\documentclass[preprint,pre,11pt,onecolumn,nobibnotes]{revtex4}%
\usepackage{amsfonts}
\usepackage{amsmath}
\usepackage{amssymb}
\usepackage{graphicx}%
\providecommand{\U}[1]{\protect\rule{.1in}{.1in}}

\begin{document}
\title{Brownian Model Theory of Nonequilibrium Liquid Structure and Hydrodynamics of
Strong Binary Electrolyte Solutions in an External Field}
\author{Byung Chan Eu and Hui Xu}
\affiliation{Department of Chemistry, McGill University, 801 Sherbrooke St. West, Montreal,
QC H3A 0B8, Canada }
\author{Kyunil Rah}
\affiliation{IT and Electronic Materials R \& D, LG\ Chem Research Park, 104-1 Moonji-dong,
Yuseung-gu, Daejeon 305-380, Korea}

\begin{abstract}
In this paper, on the basis of the Onsager--Wilson theory of strong binary
electrolyte solutions we completely work out the solutions of the governing
equations (Onsager--Fuoss equations and Poisson equations) for nonequilibrium
pair correlation functions and ionic potentials and the solutions for the
Stokes equation for the velocity and pressure in the case of strong binary
electrolyte solutions under the influence of an external electric field of
arbitrary strength. The solutions are calculated in the configuration space as
functions of coordinates and reduced field strength. Thus the axial and
transversal components of the velocity and the accompanying nonequilibrium
pressure are explicitly obtained. Computation of velocity profiles makes it
possible to visualize the movement and distortion of ion atmosphere under the
influence of an external electric field. In particular, it facilitates
tracking the movement of the center $\left(  x_{c},0\right)  $ of the ion
atmoshphere along the $x$ axis, as the field strength increases. Thus it is
possible to imagine a spherical ion atmosphere with its center displaced to
$\left(  x_{c},0\right)  $ from the origin. On the basis of this picture we
are able to formulate a computation-based procedure to unambinguously select
the values of $x$ and $r$ in the electrophoretic factor for $\xi>0$ and
thereby calculate the ionic conductance. This procedure facilitates to
overcome the mathematical divergence difficulty inherent to the method used by
Wilson in his unpublished dissertation on the ionic conductance theory
(namely, the Onsager--Wilson theory) for strong binary electrolytes. We
thereby define divergence-free electrophoretic and relaxation time factors
which would enable us to calculate equivalent conductance of strong binary
electrolytes subjected to an external electric field in excellent agreement
with experiment. We also investigate the nature of approximations that yield
Wilson's result from the exact divergence-free electrophoretic and relaxation
time coefficients. In the sequels, the results obtained in this work are
applied to study ionic conductivity and nonequilibrium pressure effects in
electrolyte solutions.

\end{abstract}
\date[Date:
\today
]{}
\startpage{1}
\endpage{102}
\maketitle

\section{Introduction}

Linear and nonlinear transport processes and nonequilibrium phenomena in
dilute non-ionic (neutral) fluids have been known adequately treatable by
means of singlet distribution functions obeying, for example, the Boltzmann
equations and related kinetic equations\cite{chapman} for singlet distribution
functions. Relying on the experience gained from the theories of neutral
dilute fluids, theories of ionized gases\cite{mason}, plasmas, and charge
carriers in semiconductors\cite{ting,nag,landsberg} often rely on singlet
distribution functions obeying Boltzmann-like kinetic equations and their
suitable modifications. However, since ions in ionized fluids interact through
long-ranged Coulombic interactions, even if the ionized species are dilute in
concentration, their spatial correlations are significant, lingering on to
manifest their effects even in the infinitely dilute regime of concentration
as the thermodynamic properties (e.g., activity coefficients) of ionic
solutions demonstrate. Therefore it would be very important to find a way, and
learn, to incorporate long-range correlations into the theory of
nonequilibrium phenomena and transport processes in ionized fluids and therein
lies the significance of the limiting theory of conductivity in ionized
liquids in the external field of arbitrary strength described in this work.

Interestingly, in the subject fields of nonlinear phenomena in ionic liquids,
the Wien effect\cite{wien} was one of the earliest experimental examples that
exhibited a marked nonlinear deviation from the Coulombic law of conduction
and, as such, it attracted considerable attention theoretically and
experimentally. Being a nonlinear effect in ionic conductance which shows a
strongly nonlinear, non-Coulombic field-dependence of ionic conductance, the
phenomenon was studied actively in physical chemistry until several decades
ago to understand ionic solutions and their physical
properties\cite{harned,accascina}. Recently, there appears to be a revival of
experimental studies on Wien effect and related aspects in ionic conductance
of ionic liquids in the presence of high external electric
fields\cite{daily,shabanov,friedman}. There are other many fascinating aspects
of physical properties of ionic liquids recently being studied actively and
reported in the recent literature\cite{castner,ionic}, although they are
mostly in the field of equilibrium phenomena. In the present series of work,
we are interested in nonlinear transport processes and, in particular,
learning about the theories of the Wien effect on ionic conductance in
electrolyte solutions in order to gain insights and theoretical approaches to
treat the currently studied properties of ionic fluids. As a first step to
this aim, we will study strong binary (symmetric) electrolytes because of the
relative simplicity of the subject matter. More complicated systems of
asymmetric electrolytes, in which the charges in a molecule are asymmetric,
will be treated in the sequels\cite{asymmetric,asymmetric-2} to this work in preparation.

The ideas\cite{onsager} of physical mechanisms underlying the Wien effect,
which might also encompass nonlinear phenomena in general in ionic fluids,
proceed as follows. It is founded on the idea of ion atmosphere in Debye's
theory\cite{debye} of electrolyte solutions. According to his theory, ion
atmosphere is formed around ions in the solution, which is spherically
symmetric if the ions are spherical and the system is in equilibrium. When the
external electric field is applied to the ionic fluid, the ions of opposite
charges begin to move in directions opposite to each other. Thus the basic
physical mechanisms involved in the ionic movements under the external field
are believed to be due to a distortion of the spherically symmetric ion
atmosphere into a non-spherical form and its subsequent tendency to relax to a
spherically symmetric form. The former effect gives rise to the
electrophoretic effect and the latter to the relaxation time effect. It should
be emphasized here that the aforementioned effects are on the ionic
atmosphere, but not on the ion of attention situated at the center of ion atmosphere.

This idea can be translated into a qualitative mathematical form as given
below: In experiments, we measure migration of ions and accompanying flow of
medium. If the external electric field is denoted $\mathbf{X}$, the force
$\mathbf{k}_{j}$ on ion $j$ of charge $e_{j}$ is then given by%
\begin{equation}
\mathbf{k}_{j}=e_{j}\mathbf{X\quad}\left(  j=1,\cdots,s\right)  . \label{k1}%
\end{equation}
Since the ion of charge $e_{j}$ in the solution creates an ion atmosphere of
charge $-e_{j}$, which is distributed in the ion atmosphere to balance the
charge in the solution, and this atmosphere is subjected to a force of
$-e_{j}\mathbf{X}$. This force tends to move the ion atmosphere in the
direction of force $-e_{j}\mathbf{X}$, while the central ion $j$ of atmosphere
is carried by force $e_{j}\mathbf{X}$ in the medium in the direction opposite
to the motion of ion atmosphere in order to balance the momentum. The velocity
of the \textit{countercurrent} generated thereby may be readily calculated if
it is assumed that the entire countercharge $-e_{j}$ of the atmosphere is
distributed in a spherical shell of radius $\kappa^{-1}$, where $\kappa^{-1}$
is the Debye radius of ion atmosphere from the central ion, and that the
motion of this sphere of radius $\kappa^{-1}$ surrounding the central charge
is governed by the Stokes law\cite{stokes,landau,batchelor,bird} holding for
the motion of a sphere in a viscous fluid. Thus, this velocity of the
countercurrent is estimated to be
\begin{equation}
\Delta\mathbf{v}_{j}=-\frac{\mathbf{k}_{j}\kappa}{6\pi\eta_{0}}=-\frac
{e_{j}\mathbf{X}\kappa}{6\pi\eta_{0}}, \label{k2}%
\end{equation}
where $\Delta\mathbf{v}_{j}$ is the velocity of the shell of radius
$\kappa^{-1}$ and $\eta_{0}$ is the viscosity of the medium. We are thus led
to the result that the medium in the interior of the shell travels with this
velocity, and that the central ion migrates against a collective current of
the medium in the shell. The deduction of this expression qualitatively
elucidates the most important part of the effect of electrophoresis. Clearly,
this effect has to do with hydrodynamic motion of the solvent around the
center ion enclosed by the ion atmosphere of radius $\kappa^{-1}$ that moves
against the former. One may therefore quantify this qualitative description by
means of a hydrodynamic method using the Navier--Stokes
equation\cite{landau,batchelor,bird}, but the Navier--Stokes equation requires
\textit{a local body-force---}local mean external force---as an input. This
local body-force cannot be obtained through a purely phenomenological
consideration, but, for example, must be calculated by means of statistical
mechanics combined with classical electrodynamics. Before proceeding to the
remaining effect, it is important to point out that Eq. (\ref{k2}) gives the
velocity of a physical object of radius $\kappa^{-1}$ (i.e., the radius of ion
atmosphere) in the direction of $\mathbf{X}$.\ 

The second effect, that is, the relaxation time effect, is seen as follows: If
the central ion possessed no atmosphere, it would simply migrate with a
velocity $\mathbf{k}_{j}/\zeta_{j}$, where $\zeta_{j}$ is the friction
constant, but owing to its ion atmosphere, the ion is subjected to a net
force, $\mathbf{k}_{j}-\Delta\mathbf{k}_{j}$, where $\Delta\mathbf{k}_{j}$ is
the force arising from the dissymmetry of the ion atmosphere created by the
movement of the ions in the external field, and hence it will move, relative
to its environment, with a velocity of a magnitude, $\left(  \mathbf{k}%
_{j}-\Delta\mathbf{k}_{j}\right)  /\zeta_{j}$. This $\Delta\mathbf{k}_{j}$ is
due to the effect arising from the relaxation of the asymmetric ion atmosphere.

Consequently, the net velocity $\mathbf{v}_{j}$ of ion $j$ is given by%
\begin{equation}
\mathbf{v}_{j}=\frac{\left(  e_{j}\mathbf{X}-\Delta\mathbf{k}_{j}\right)
}{\zeta_{j}}-\frac{e_{j}\mathbf{X}\kappa}{6\pi\eta_{0}}. \label{k4}%
\end{equation}
Here $\Delta\mathbf{k}_{j}\zeta_{j}^{-1}$ represents the relaxation time
effect on relaxation to a spherically symmetric form of the distorted ion
atmosphere, and the last term the electrophoretic effect.

The aforementioned two effects making up the velocity given in Eq. (\ref{k4})
are believed to underlie in charge conduction in electrolytic solutions. In
fact, the mobility of ions induced by an external electric field can be
calculated on the basis of the aforementioned two effects, for example, by
using Eq. (\ref{k4}).

As we can see from this heuristic discussion, the aforementioned two effects
require the velocity of the fluid (medium), which obviously obeys the
hydrodynamic equations for the system subjected to an external electric field.
Since such velocity solutions can be obtained from the Stokes equation, more
generally, Navier--Stokes equation, we may apply the solutions thereof to
calculate the charge conductance and the countercurrent of the medium to learn
the mode of charge conductance in electrolyte solutions subjected to an
external field. The hydrodynamic equations, however, contain external
body-forces, which in the present case are the external electric field. The
external electric field or body-force is generally local and depends on the
local distribution of charges. The local charge distributions require
molecular distributions in the system and a statistical mechanical theory for
them---a molecular theory.

To answer this question, Onsager\cite{fuoss} with Fuoss formulated a formal
framework of theory in which a Fokker--Planck-type differential equations for
nonequilibrium pair distribution functions are derived on the assumption of a
Brownian motion model for ions in a continuous medium of dielectric constant
$D$ and viscosity $\eta_{0}$. We will refer to these differential equations
for pair correlation functions as the Onsager--Fuoss (OF) equations
henceforth. They are coupled to the Poisson equations\cite{jackson} of
classical electrodynamics for the ionic potentials. These two coupled systems
of differential equations will be referred to as the governing equations in
the present work. The governing equations were applied to study the ionic
conductance of binary strong electrolytes in an external electric field by
Wilson in his dissertation\cite{wilson}. This theory will be referred to as
the Onsager--Wilson (OW) theory. Wilson solved the governing equations and
obtained analytic formulas for the electrophoretic and relaxation time
coefficients and the equivalent ionic conductance qualitatively displaying the
Wien effect in the regime of strong electric fields. Unfortunately, his
dissertation has never been published in public domain, but only important
results, such as the electrophoretic and relaxation time coefficients, had
been excerpted in the well-known monograph\cite{harned} by Harned and Owen on
electro-physical chemistry. Tantalized by the possibility of the utility of
the theory for recent experimental results for ionic fluids and charge carrier
mobilities in semiconductors referred to earlier, we have thoroughly examined
the OW theory to learn the details of it. \textit{Surprisingly, we have
discovered that the velocity solution of the Stokes (hydrodynamic) equation in
the OW theory can give rise to a divergent result rendering into question the
electrophoretic coefficient calculated by Wilson's procedure described in his
dissertation\cite{wilson}.} We believe that the basic framework of governing
equations---the OF equations and Poisson equations---should be correct, but
the way the solutions are evaluated by him may be called into question.
Therefore, it is our principal aim of this work to analyze the solutions of
the governing equations in the case of binary strong electrolytes in an
external electric field and obtain physically reasonable and thus acceptable
theoretical results that can be made use of to study experimental data on
conductivity and other transport phenomena in the high field regime.

This paper is organized as follows. In Sec. II, we present the governing
differential equations, which consist of the OF equations for the ionic pair
distribution functions and the Poisson equations for the potentials of ionic
interaction. We note that Kirkwood \cite{kirkwood} also derived a similar
equation for non-ionic liquids in his kinetic theory of liquids. One (BCE) of
the present authors also derived \cite{eu87} the OF equations from the
generalized Boltzmann equation.\cite{eu92,eu98} Since Wilson's
dissertation\cite{wilson} has not been published anywhere in a journal, the
governing equations and their solutions are discussed to the extent that the
present paper can be followed intelligibly.

In Sec. III, the solutions of the governing equations---the pair distribution
functions and potentials of ionic interaction---are presented in the case of a
strong binary electrolyte solution subjected to an external field. These
solutions are given in one-dimensional Fourier transforms in an axially
symmetric coordinate system, namely, a cylindrical coordinate system whose
axial coordinate is parallel to the applied external electric field. The
Fourier transform is with respect to the axial coordinate. The distribution
functions obtained are nonequilibrium pair distribution functions which
describe the nonequilibrium ionic liquid structure, and the nonequilibrium
ionic potentials of interaction in the external field. Since they should be of
considerable interest to help us learn about the nonequilibrium ionic liquid
properties we study the solutions of the governing equations in detail and
obtain, especially, their spatial profiles, indicating how ions and their
nonequilibrium part of the potentials are distributed in the external electric
field. It should be noted that the distribution functions are the
nonequilibrium corrections to the Boltzmann distribution function predicted by
the Debye--H\"{u}ckel theory\cite{debye} of electrolytes, and similarly for
the potentials.

In Sec. IV, we then discuss the solutions of the Stokes equation, which
replaces the Navier--Stokes equation in the case of incompressible fluids that
we assume the ionic solution of interest is. Solving the Stokes equation, we
obtain the axial and transversal velocity components as well as the
nonequilibrium pressure from the solutions of the Stokes equation. We present
the solution procedure for the Stokes equation in detail, because, firstly,
Wilson's thesis contains only the symmetric part of the solution, leaving out
the antisymmetric part that turns out to be comparable to the former in
magnitude and, secondly, we believe that the solution procedure of the Stokes
equations, which combines statistical mechanics and hydrodynamics in a rather
intriguing manner, appears to be very much worth learning, especially, if one
is interested in nonequilibrium theories of ionic liquids in an external
electric field. In this section we also discuss the connection with the
electrophoretic and relaxation time coefficients originally obtained by
Wilson, who evaluated them at the position of the center ion of ion
atmosphere, namely, at the coordinate origin. This discussion would show that
one of his integrals evaluated at the coordinate origin is divergent.
Therefore we evaluate explicitly the solutions to explore a way to make the OW
theory of ionic conductance unencumbered by such a divergence difficulty.

In Sec. IV, we also compute numerically the spatial profiles of the axial
velocity, and study them to guide us to avoid the divergence difficulty
mentioned in connection with Wilson's result\ and choose the optimum position
coordinates at which to calculate the relaxation time and electrophoretic
coefficients. To this aim we have either evaluated analytically or reduced to
one-dimensional quadratures, by means of contour integration methods, the
Fourier transform integrals making up the solutions of the Stokes equations
obtained earlier before computing their spatial profiles. The contour
integration methods are described in Appendix A. Since they, however, do not
cover the entire coordinate space owing to the condition imposed by Jordan's
lemma\cite{whittaker} on applicability of contour integration methods
involving integrations along a circle of infinite radius, the integrals must
be numerically computed outside the region where the aforementioned condition
is violated. The details of the condition are discussed in Sec. IV and also in
Appendix A. These numerical studies reveal the manner in which the ions flow
subject to the applied external electric field provide insight into the
behavior of the velocity and valuable clues to formulate an empirical rule to
select the position parameters ($x,r$) in the electrophoretic factor, so that
a physically sensible and non-divergent electrophoretic coefficient and the
corresponding relaxation time coefficient can be defined and ionic conductance
correctly predicted. This problem is addressed in the companion paper. Sec. V
is for discussion and concluding remark.

\section{Governing Equations}

Let $\overline{\mathbf{r}}_{j}$ denote the position vector of ion $j$ in a
fixed coordinate system and $\overline{\mathbf{r}}_{ji}$ the relative
coordinate of ion $i$ from ion $j$:%
\begin{equation}
\overline{\mathbf{r}}\equiv\overline{\mathbf{r}}_{ji}=\overline{\mathbf{r}%
}_{i}-\overline{\mathbf{r}}_{j}=\overline{\mathbf{r}}_{j}-\overline
{\mathbf{r}}_{i}=-\overline{\mathbf{r}}_{ij}\equiv-\overline{\mathbf{r}},
\label{1}%
\end{equation}
and let $f_{ji}\left(  \overline{\mathbf{r}}_{j},\overline{\mathbf{r}}%
_{ij}\right)  $ denote the concentration of ion $i$ in the atmosphere of ion
$j$ located at position $\overline{\mathbf{r}}_{j}$---in other words, the
distribution function to find ion $i$ at distance $\overline{\mathbf{r}}_{ij}$
from ion $j$ located at $\overline{\mathbf{r}}_{j}$. At equilibrium it is
given by the Boltzmann distribution function times the density of ion $j$. Let
us also denote by $\mathbf{v}_{ji}$ the velocity of ion $i$ in the
neighborhood of ion $j$. Therefore this velocity also depends on positions of
ions $i$ and $j$ in the following manner:%
\begin{equation}
\mathbf{v}_{ji}=\mathbf{v}_{ji}\left(  \overline{\mathbf{r}}_{j}%
,\overline{\mathbf{r}}_{ij}\right)  ,\qquad\mathbf{v}_{ij}=\mathbf{v}%
_{ij}\left(  \overline{\mathbf{r}}_{i},\overline{\mathbf{r}}_{ji}\right)  .
\label{2}%
\end{equation}
The equation of continuity for ion pair $(j,i)$ is then given by%
\begin{equation}
-\frac{\partial f_{ji}\left(  \overline{\mathbf{r}}_{j},\overline{\mathbf{r}%
}_{ij}\right)  }{\partial t}=\mathbf{\nabla}_{j}\cdot\left(  \mathbf{v}%
_{ij}f_{ij}\right)  +\mathbf{\nabla}_{i}\cdot\left(  \mathbf{v}_{ji}%
f_{ji}\right)  =-\frac{\partial f_{ij}\left(  \overline{\mathbf{r}}%
_{i},\overline{\mathbf{r}}_{ji}\right)  }{\partial t}, \label{2ec}%
\end{equation}
where $\nabla_{j}=\partial/\partial\overline{\mathbf{r}}_{j}$. Hence, at a
steady state $\partial f_{ji}/\partial t=0$ the steady-state equation of
continuity is given by%
\begin{equation}
\mathbf{\nabla}_{j}\cdot\left(  \mathbf{v}_{ij}f_{ij}\right)  +\mathbf{\nabla
}_{i}\cdot\left(  \mathbf{v}_{ji}f_{ji}\right)  =0. \label{2st}%
\end{equation}
Assuming that the ions, being randomly bombarded by molecules of the
continuous medium (solvent) of dielectric constant $D$ and viscosity $\eta
_{0}$, move randomly, namely, execute random Brownian motions, in the presence
of an applied external field, the velocities $\mathbf{v}_{ji}$ and
$\mathbf{v}_{ij}$ may be assumed given by the Brownian motion model%
\begin{align}
\mathbf{v}_{ji}  &  =\mathbf{V}(\overline{\mathbf{r}}_{i})+\omega
_{i}(\mathbf{K}_{ji}-k_{B}T\mathbf{\nabla}_{i}\ln f_{ji},\label{2vji}\\
\mathbf{v}_{ij}  &  =\mathbf{V}(\overline{\mathbf{r}}_{j})+\omega
_{j}(\mathbf{K}_{ij}-k_{B}T\mathbf{\nabla}_{j}\ln f_{ij}, \label{2vij}%
\end{align}
where $\mathbf{V}(\overline{\mathbf{r}}_{k})$ is the velocity of solution at
position $\overline{\mathbf{r}}_{k}$ ($k=i,j$); $\omega_{k}$ is the inverse of
the friction coefficient $\zeta_{k}$ of ion $k$, which is related to the
diffusion coefficient $D_{k}$ of ion $k$ of charge $e_{k}$ in the medium of
viscosity $\eta_{0}$%
\begin{equation}
D_{k}=k_{B}T\omega_{k}=\frac{k_{B}T}{\zeta_{k}}. \label{2di}%
\end{equation}
Here $k_{B}$ is the Boltzmann constant and $T$ the absolute temperature;
$\mathbf{K}_{ji}$ is the total force acting on ion $i$. We assume that forces
on ions $\mathbf{K}_{ji}$ are linear with respect to charge numbers%
\begin{equation}
\mathbf{K}_{ji}=\mathbf{k}_{i}-e_{i}\mathbf{\nabla}_{i}\psi_{j}\left(
\overline{\mathbf{r}}_{j},\overline{\mathbf{r}}_{ij}\right)  , \label{2Kij}%
\end{equation}
so that the superposition principle of fields is preserved. Here
$\mathbf{k}_{i}$ is the applied external force on ion $i$. Under the
assumptions for $\mathbf{v}_{ji}$ and for $\mathbf{K}_{ji}$ stated earlier,
the steady-state equation of continuity (\ref{2st}) becomes a coupled set of
differential equations\cite{fuoss} satisfied by ion pair distribution
functions $f_{ji}\left(  \overline{\mathbf{r}}_{j},\overline{\mathbf{r}}%
_{ij}\right)  $ of the ionic liquid:%
\begin{align}
k_{B}T\left(  \omega_{i}+\omega_{j}\right)  \mathbf{\nabla\cdot\nabla}%
f_{ji}\left(  \overline{\mathbf{r}}\right)  +\left(  \omega_{j}e_{j}%
-\omega_{i}e_{i}\right)  \mathbf{X\cdot\nabla}f_{ji}\left(  \overline
{\mathbf{r}}\right)   & \nonumber\\
+n_{i}n_{j}\left[  e_{i}\omega_{i}\mathbf{\nabla\cdot\nabla}\psi_{j}\left(
\overline{\mathbf{r}}\right)  +e_{j}\omega_{j}\mathbf{\nabla\cdot\nabla}%
\psi_{i}\left(  -\overline{\mathbf{r}}\right)  \right]   &  =0,\label{3}\\
\left(  i,j=1,2,\cdots,s\right)  .  & \nonumber
\end{align}
We will call this set of differential equations the Onsager--Fuoss (OF)
equations. Here for simplicity of notation we have omitted the first position
variables in the distribution functions and potentials and typeset them as
follows: $f_{ji}(\overline{\mathbf{r}})\equiv f_{ji}\left(  \overline
{\mathbf{r}}_{j},\overline{\mathbf{r}}_{ij}\right)  ,$ etc. and $\psi
_{j}\left(  \overline{\mathbf{r}}\right)  \equiv\psi_{j}\left(  \overline
{\mathbf{r}}_{j},\overline{\mathbf{r}}\right)  $ and $\psi_{i}\left(
-\overline{\mathbf{r}}\right)  \equiv\psi_{i}\left(  \overline{\mathbf{r}}%
_{i},-\overline{\mathbf{r}}\right)  $. In fact, for Eq. (\ref{3}) the
coordinate origin may be regarded as fixed on position of ion $j$. These are
Fokker--Planck-type equations for $f_{ji}\left(  \overline{\mathbf{r}}\right)
$ and $\psi_{j}\left(  \overline{\mathbf{r}}\right)  $ and $\psi_{i}\left(
-\overline{\mathbf{r}}\right)  $. In Eq. (\ref{3}), $n_{i}$ is density of ion
$i$ and $\mathbf{X}$ is the external (electric) field. The potentials
$\psi_{k}\left(  k=i,j\right)  $ appearing in this set of differential
equations, Eq. (\ref{3}), obey the Poisson equations of classical
electrodynamics\cite{jackson},%
\begin{equation}
\mathbf{\nabla\cdot\nabla}\psi_{j}\left(  \overline{\mathbf{r}}\right)
=-\frac{4\pi}{Dn_{j}}\sum_{i=1}^{s}e_{i}f_{ji}(\overline{\mathbf{r}}).
\label{4}%
\end{equation}
The two sets (\ref{3}) and (\ref{4}) are coupled to each other and will be
henceforth referred to as the governing equations in this work.

\subsection{Boundary Conditions}

The two sets of equations (\ref{3}) and (\ref{4}) are subject to the boundary
conditions stated below.

\subsubsection{\textbf{No Flux Conditions}}

The number of ions, leaving and entering the interior, $\Omega$, of a surface
$S$ should be balanced, because no ions are created or destroyed. Therefore
$f_{ji}(\overline{\mathbf{r}})\left[  \mathbf{v}_{ji}\left(  \overline
{\mathbf{r}}\right)  -\mathbf{v}_{ij}\left(  -\overline{\mathbf{r}}\right)
\right]  $ is sourceless in $\Omega$. This fact may be expressed as%
\begin{equation}
\int_{S}dSf_{ji}(\overline{\mathbf{r}})\left\{  \mathbf{e}_{n}\cdot\left[
\mathbf{v}_{ji}\left(  \overline{\mathbf{r}}\right)  -\mathbf{v}_{ij}\left(
-\overline{\mathbf{r}}\right)  \right]  \right\}  =\int_{\Omega}%
d\Omega\mathbf{\nabla}\cdot\left\{  f_{ji}(\overline{\mathbf{r}})\left[
\mathbf{v}_{ji}\left(  \overline{\mathbf{r}}\right)  -\mathbf{v}_{ij}\left(
-\overline{\mathbf{r}}\right)  \right]  \right\}  =0, \label{6}%
\end{equation}
where $\mathbf{e}_{n}$ is the vector normal to the surface $S$. This will be
henceforth called no flux condition.

\subsubsection{\textbf{Boundary Conditions on Potentials}}

If the charge $e_{j}$ is within $\Omega$, we obtain
\begin{equation}
\lim_{\Omega\rightarrow0}\int_{\Omega}d\Omega\varrho_{j}\left(  \overline
{\mathbf{r}}\right)  =e_{j}\delta, \label{7c}%
\end{equation}
where $\varrho_{j}\left(  \overline{\mathbf{r}}\right)  $ is the charge
density. Therefore, the space charge within $\Omega$ must be such that%
\begin{equation}
-\frac{D}{4\pi}\int_{\Omega}d\Omega\nabla^{2}\psi_{j}\left(  \overline
{\mathbf{r}}\right)  =\int_{\Omega}d\Omega\varrho_{j}\left(  \overline
{\mathbf{r}}\right)  , \label{7}%
\end{equation}
or alternatively%
\begin{equation}
\lim_{\Omega\rightarrow0}\int_{S}d\mathbf{S\cdot\nabla}\psi_{j}\left(
\overline{\mathbf{r}}\right)  =-\frac{4\pi e_{j}}{D}\delta, \label{7b}%
\end{equation}
for the boundary condition on the ionic potentials. Here%
\[
\delta=\left\{
\begin{array}
[c]{l}%
1\quad\text{if ion }j\text{ is located at }\overline{\mathbf{r}}=0\\
0\quad\text{otherwise}%
\end{array}
\right.  .
\]
The boundary condition (\ref{7b}) corresponds to the fact that the charge
$e_{j}$ at the origin (i.e., at the center of the ion atmosphere) must balance
the net charge of the rest of the ion atmosphere.

\subsection{Symmetric and Antisymmetric Parts of Governing Equations}

Since it is convenient to work with dimensionless variables, we first reduce
position variable $\overline{\mathbf{r}}$ with respect to the Debye parameter
$\kappa$%
\begin{equation}
\mathbf{r}=\kappa\overline{\mathbf{r}}\mathbf{,} \label{8r}%
\end{equation}
where the Debye parameter is defined by%
\begin{equation}
\kappa=\sqrt{\frac{4\pi\Gamma_{0}}{Dk_{B}T}};\qquad\Gamma_{0}=\sum_{k=1}%
^{s}n_{k}e_{k}^{2}. \label{8k}%
\end{equation}

The distribution functions and potentials change the sign of argument if
particle indices $j$ and $i$ are interchanged. Therefore they are expected to
consist of symmetric and antisymmetric components. Consequently, it is
convenient to distinguish the symmetric and antisymmetric components
$f_{ji}^{\pm}\left(  \mathbf{r}\right)  $ and $\psi_{j}^{\pm}\left(
\mathbf{r}\right)  $, etc. of distribution functions and potentials. They are
defined in reduced forms as follows:%
\begin{align}
\frac{\kappa n_{j}n_{i}e_{j}e_{i}}{Dk_{B}T}f_{ji}^{+}\left(  \mathbf{r}%
\right)   &  =\frac{1}{2}\left[  f_{ji}(\mathbf{r})+f_{ji}(-\mathbf{r}%
)\right]  -n_{i}n_{j},\label{8a}\\
\frac{\kappa n_{j}n_{i}e_{j}e_{i}}{Dk_{B}T}f_{ji}^{-}\left(  \mathbf{r}%
\right)   &  =\frac{1}{2}\left[  f_{ji}(\mathbf{r})-f_{ji}(-\mathbf{r}%
)\right]  ,\nonumber\\
\frac{\kappa e_{j}}{D}\psi_{j}^{\pm}\left(  \mathbf{r}\right)   &  =\frac
{1}{2}\left[  \psi_{j}\left(  \mathbf{r}\right)  \pm\psi_{j}\left(
-\mathbf{r}\right)  \right]  . \label{8b}%
\end{align}
Since the distribution functions tend to $n_{i}n_{j}$ as $r=\left\vert
\mathbf{r}\right\vert $ tends to infinity, $f_{ji}^{\pm}\left(  \mathbf{r}%
\right)  $ vanishes as $r\rightarrow\infty$. According to the definitions
(\ref{8a}) and (\ref{8b}), the following symmetry properties can be deduced
for them as the ion positions are interchanged:%
\begin{align}
f_{ji}^{+}\left(  -\mathbf{r}\right)   &  =f_{ji}^{+}\left(  \mathbf{r}%
\right)  =f_{ij}^{+}\left(  -\mathbf{r}\right)  ,\quad f_{ji}^{-}\left(
-\mathbf{r}\right)  =-f_{ji}^{-}\left(  \mathbf{r}\right)  =-f_{ij}^{-}\left(
-\mathbf{r}\right)  ,\label{9a}\\
\psi_{j}^{+}\left(  -\mathbf{r}\right)   &  =\psi_{j}^{+}\left(
\mathbf{r}\right)  ,\quad\psi_{j}^{-}\left(  -\mathbf{r}\right)  =-\psi
_{j}^{-}\left(  \mathbf{r}\right)  . \label{9b}%
\end{align}
The differential equations of the symmetric and antisymmetric components of
$f_{ji}$ and $\psi_{j}$ in Eqs. (\ref{3}) and (\ref{4}) then can be separated
as follows:%
\begin{align}
\mathbf{\nabla}^{2}f_{ji}^{+}\left(  \mathbf{r}\right)  -\left(  \omega
_{ij}\mu_{i}+\omega_{ji}\mu_{j}\right)  f_{ji}^{+}\left(  \mathbf{r}\right)
-\omega_{ij}\mu_{j}f_{jj}^{+}\left(  \mathbf{r}\right)  -\omega_{ji}\mu
_{i}f_{ii}^{+}\left(  \mathbf{r}\right)  +\mu_{ji}^{\prime}\mathbf{\nabla}%
_{x}f_{ji}^{-}\left(  \mathbf{r}\right)   &  =0,\nonumber\\
\mathbf{\nabla}^{2}f_{jj}^{+}\left(  \mathbf{r}\right)  -\left[  \mu_{i}%
f_{ji}^{+}\left(  \mathbf{r}\right)  +\mu_{j}f_{jj}^{+}\left(  \mathbf{r}%
\right)  \right]   &  =0,\nonumber\\
\mathbf{\nabla}^{2}f_{ii}^{+}\left(  \mathbf{r}\right)  -\left[  \mu_{i}%
f_{ii}^{+}\left(  \mathbf{r}\right)  +\mu_{j}f_{ji}^{+}\left(  \mathbf{r}%
\right)  \right]   &  =0,\label{G1}\\
\mathbf{\nabla}^{2}f_{ji}^{-}\left(  \mathbf{r}\right)  -\left(  \omega
_{ij}\mu_{i}+\omega_{ji}\mu_{j}\right)  f_{ji}^{-}\left(  \mathbf{r}\right)
-\omega_{ij}\mu_{j}f_{jj}^{-}\left(  \mathbf{r}\right)  +\omega_{ji}\mu
_{i}f_{ii}^{-}\left(  \mathbf{r}\right)  +\mu_{ji}^{\prime}\mathbf{\nabla}%
_{x}f_{ji}^{+}\left(  \mathbf{r}\right)   &  =0,\nonumber\\
\mathbf{\nabla}^{2}f_{jj}^{-}\left(  \mathbf{r}\right)   &  =0,\nonumber\\
\mathbf{\nabla}^{2}f_{ii}^{-}\left(  \mathbf{r}\right)   &  =0,\nonumber
\end{align}%
\begin{align}
\mathbf{\nabla}^{2}\psi_{j}^{+}\left(  \mathbf{r}\right)   &  =-\left[
\mu_{i}f_{ji}^{+}\left(  \mathbf{r}\right)  +\mu_{j}f_{jj}^{+}\left(
\mathbf{r}\right)  \right]  ,\nonumber\\
\mathbf{\nabla}^{2}\psi_{i}^{+}\left(  \mathbf{r}\right)   &  =-\left[
\mu_{i}f_{ii}^{+}\left(  \mathbf{r}\right)  +\mu_{j}f_{ji}^{+}\left(
\mathbf{r}\right)  \right]  ,\nonumber\\
\mathbf{\nabla}^{2}\psi_{j}^{-}\left(  \mathbf{r}\right)   &  =-\left[
\mu_{i}f_{ji}^{-}\left(  \mathbf{r}\right)  +\mu_{j}f_{jj}^{-}\left(
\mathbf{r}\right)  \right]  ,\label{G2}\\
\mathbf{\nabla}^{2}\psi_{i}^{-}\left(  \mathbf{r}\right)   &  =-\left[
\mu_{i}f_{ii}^{-}\left(  \mathbf{r}\right)  +\mu_{j}f_{ji}^{-}\left(
\mathbf{r}\right)  \right]  ,\nonumber
\end{align}
where $\mathbf{\nabla}^{2}$ now stands for Laplacian operator of reduced
variable $\mathbf{r}$,
\begin{align}
\mu_{i}  &  =\frac{\nu_{i}z_{i}^{2}}{\nu_{i}z_{i}^{2}+\nu_{j}z_{j}^{2}}%
,\qquad\omega_{ji}=\frac{\omega_{i}}{\omega_{i}+\omega_{j}},\label{10}\\
\mu_{ji}^{\prime}  &  =\frac{X}{\kappa k_{B}T}\frac{\left(  \omega_{j}%
e_{j}-\omega_{i}e_{i}\right)  }{\omega_{j}+\omega_{i}}\quad\left(  X=\text{
external electric field strength}\right)  , \label{10a}%
\end{align}
with $\nu_{i}$ denoting the stoichiometric coefficient of ion $i$ in the
($j,i$) electrolyte and $z_{k}$ the charge number of ion $k$: $e_{k}%
=ez_{k}\quad\left(  k=j,i\right)  $. Henceforth the indices $j$ and $i$ refer
to ions of the binary electrolyte ($j,i$), but also may dually refer to other
ions belonging to species $j$ or $i$. This notational device prevents
proliferation of subscripts distinguishing ionic particles. The 10
differential equations of Eqs. (\ref{G1}) and (\ref{G2}) will be referred to
as the governing equations,\ the set (\ref{G1}) as the Onsager--Fuoss (OF)
equations, and the set (\ref{G2}) as the Poisson equations. The solutions of
the governing equations provide the information on the nonequilibrium ionic
liquid structure and ionic potentials of the electrolyte solutions subjected
to an external electric field of arbitrary strength. A theory of transport
processes in ionic solutions can be developed by making use of them.

\section{Nonequilibrium Ionic Liquid Structure and Potentials of Binary
Electrolytes}

\subsection{Complete Solutions of the Governing Equations}

We now limit our study to strong binary electrolyte solutions as in the theory
of Wilson\cite{wilson} and Onsager. If the electrolyte is binary and strong,
then $\left\vert e_{j}\right\vert =\left\vert e_{i}\right\vert =ez$ with
$z=\left\vert z_{i}\right\vert =\left\vert z_{j}\right\vert $, and
\[
\mu_{i}=\mu_{j}=\frac{1}{2},\qquad\mu_{ji}^{\prime}=\frac{zeX}{\kappa k_{B}%
T}\equiv\xi.
\]
Wilson in his unpublished PhD thesis\cite{wilson} obtained formal solutions of
the governing equations in the forms of Fourier transforms under the
assumption that $\omega_{i}=\omega_{j}$, which means $\omega_{ji}=\omega
_{ij}=\frac{1}{2}$; that is, the diffusivities of the constituent ions are
equal. (As it will turn out, the difference in the diffusivities has only a
minor effect that can be ignored without much effect on the solutions.) And
therewith he formulated a theory of ionic conductance of binary electrolytes
under the influence of external electric field. However, Wilson's thesis
unfortunately has not been published in a journal in public domain, nor have
the nonequilibrium ionic liquid structures and accompanying potentials been
explicitly evaluated and studied. In fact, neither were the velocity profiles
completely calculated in the full configuration space since he limited the
study to the velocity of the center ion of the ion atmosphere located at the
coordinate origin in his calculation of the electrophoretic effect. Moreover,
the particular velocity formula made use of by Wilson had a divergence
difficulty at the origin, but he argued it away on the ground that the
divergent term would not contribute to the ionic conductance. We will show his
argument was mathematically groundless and would not hold true. For these
reasons, in this work we will first evaluate the velocity formulas explicitly
by applying analytic methods or methods of contour integrations or numerical
computation methods for wide ranges of position coordinates, and then will
explore a way to overcome or get around the divergence difficulty. The results
obtained thereby for the nonequilibrium ionic liquid structure and potentials
as well as the velocity profiles would be principal contributions of the
present work, which are not available in the literature on ionic liquids at
present. In the subsequent paper\cite{eurah2}, the solutions of the present
paper will be applied to study the Wien effect on equivalent ionic conductance
as a function of the applied field strength.

Since Wilson's dissertation is not only not readily accessible as mentioned
earlier, but also his solution procedure is difficult to follow, on the basis
of our understanding of his solution procedure we will reconstruct the
solutions for the governing equations (\ref{G1}) and (\ref{G2}). The solution
procedure presented below is not exactly the same as his except in spirit, but
most of the final results agree with his in the main. Under the assumptions on
$\omega_{ji}=\omega_{ij}$ mentioned earlier, the governing equations
(\ref{G1}) and (\ref{G2}) are given as follows:%
\begin{align}
\mathbf{\nabla}^{2}f_{ji}^{+}\left(  \mathbf{r}\right)  -\frac{1}{2}f_{ji}%
^{+}\left(  \mathbf{r}\right)  -\frac{1}{4}f_{jj}^{+}\left(  \mathbf{r}%
\right)  -\frac{1}{4}f_{ii}^{+}\left(  \mathbf{r}\right)  +\xi\nabla_{x}%
f_{ji}^{-}\left(  \mathbf{r}\right)   &  =0,\nonumber\\
\mathbf{\nabla}^{2}f_{jj}^{+}\left(  \mathbf{r}\right)  -\frac{1}{2}\left[
f_{ji}^{+}\left(  \mathbf{r}\right)  +f_{jj}^{+}\left(  \mathbf{r}\right)
\right]   &  =0,\nonumber\\
\mathbf{\nabla}^{2}f_{ii}^{+}\left(  \mathbf{r}\right)  -\frac{1}{2}\left[
f_{ii}^{+}\left(  \mathbf{r}\right)  +f_{ji}^{+}\left(  \mathbf{r}\right)
\right]   &  =0,\label{G3}\\
\mathbf{\nabla}^{2}f_{ji}^{-}\left(  \mathbf{r}\right)  -\frac{1}{2}f_{ji}%
^{-}\left(  \mathbf{r}\right)  -\frac{1}{4}f_{jj}^{-}\left(  \mathbf{r}%
\right)  +\frac{1}{4}f_{ii}^{-}\left(  \mathbf{r}\right)  +\xi\nabla_{x}%
f_{ji}^{+}\left(  \mathbf{r}\right)   &  =0,\nonumber\\
\mathbf{\nabla}^{2}f_{jj}^{-}\left(  \mathbf{r}\right)   &  =0,\nonumber\\
\mathbf{\nabla}^{2}f_{ii}^{-}\left(  \mathbf{r}\right)   &  =0,\nonumber
\end{align}%
\begin{align}
\mathbf{\nabla}^{2}\psi_{j}^{+}\left(  \mathbf{r}\right)   &  =-\frac{1}%
{2}\left[  f_{ji}^{+}\left(  \mathbf{r}\right)  +f_{jj}^{+}\left(
\mathbf{r}\right)  \right]  ,\nonumber\\
\mathbf{\nabla}^{2}\psi_{i}^{+}\left(  \mathbf{r}\right)   &  =-\frac{1}%
{2}\left[  f_{ii}^{+}\left(  \mathbf{r}\right)  +f_{ji}^{+}\left(
\mathbf{r}\right)  \right]  ,\nonumber\\
\mathbf{\nabla}^{2}\psi_{j}^{-}\left(  \mathbf{r}\right)   &  =-\frac{1}%
{2}\left[  f_{ji}^{-}\left(  \mathbf{r}\right)  +f_{jj}^{-}\left(
\mathbf{r}\right)  \right]  ,\label{G4}\\
\mathbf{\nabla}^{2}\psi_{i}^{-}\left(  \mathbf{r}\right)   &  =-\frac{1}%
{2}\left[  f_{ii}^{-}\left(  \mathbf{r}\right)  +f_{ji}^{-}\left(
\mathbf{r}\right)  \right]  .\nonumber
\end{align}

Owing to the fact that the solution of the Laplace equation is constant, the
solutions of the fifth and sixth equations of the set (\ref{G3}) are
constant:
\[
f_{jj}^{-}\left(  \mathbf{r}\right)  =C_{j},\qquad f_{ii}^{-}\left(
\mathbf{r}\right)  =C_{i},
\]
but by the boundary conditions that they must vanish as $r\rightarrow\infty$.
Therefore, the constants $C_{j}$ and $C_{i}$ must be equal to zero. Hence
\begin{equation}
f_{jj}^{-}\left(  \mathbf{r}\right)  =0,\qquad f_{ii}^{-}\left(
\mathbf{r}\right)  =0. \label{11}%
\end{equation}
Consequently, the governing equations reduce to the following 8 differential
equations:%
\begin{align}
\left(  \mathbf{\nabla}^{2}-\frac{1}{2}\right)  f_{ji}^{+}\left(
\mathbf{r}\right)  -\frac{1}{4}\left[  f_{jj}^{+}\left(  \mathbf{r}\right)
+f_{ii}^{+}\left(  \mathbf{r}\right)  \right]  +\xi\nabla_{x}f_{ji}^{-}\left(
\mathbf{r}\right)   &  =0,\label{12a}\\
\left(  \mathbf{\nabla}^{2}-\frac{1}{2}\right)  f_{jj}^{+}\left(
\mathbf{r}\right)  -\frac{1}{2}f_{ji}^{+}\left(  \mathbf{r}\right)   &
=0,\label{12b}\\
\left(  \mathbf{\nabla}^{2}-\frac{1}{2}\right)  f_{ii}^{+}\left(
\mathbf{r}\right)  -\frac{1}{2}f_{ji}^{+}\left(  \mathbf{r}\right)   &
=0,\label{12c}\\
\left(  \mathbf{\nabla}^{2}-\frac{1}{2}\right)  f_{ji}^{-}\left(
\mathbf{r}\right)  +\xi\nabla_{x}f_{ji}^{+}\left(  \mathbf{r}\right)   &  =0,
\label{12d}%
\end{align}%
\begin{align}
\mathbf{\nabla}^{2}\psi_{j}^{+}\left(  \mathbf{r}\right)   &  =-\frac{1}%
{2}\left[  f_{ji}^{+}\left(  \mathbf{r}\right)  +f_{jj}^{+}\left(
\mathbf{r}\right)  \right]  ,\label{13a}\\
\mathbf{\nabla}^{2}\psi_{i}^{+}\left(  \mathbf{r}\right)   &  =-\frac{1}%
{2}\left[  f_{ii}^{+}\left(  \mathbf{r}\right)  +f_{ji}^{+}\left(
\mathbf{r}\right)  \right]  ,\label{13b}\\
\mathbf{\nabla}^{2}\psi_{j}^{-}\left(  \mathbf{r}\right)   &  =-\frac{1}%
{2}f_{ji}^{-}\left(  \mathbf{r}\right)  ,\label{13c}\\
\mathbf{\nabla}^{2}\psi_{i}^{-}\left(  \mathbf{r}\right)   &  =-\frac{1}%
{2}f_{ji}^{-}\left(  \mathbf{r}\right)  . \label{13d}%
\end{align}
These two sets, (\ref{12a})--(\ref{12d}) and (\ref{13a})--(\ref{13d}), suggest
that having obtained the solutions of the first set (\ref{12a})--(\ref{12d})
we can look for the solutions of the second set (\ref{13a})--(\ref{13d}),
inhomogeneous differential equations . We will follow this strategy by
applying the method of Fourier transform.

Since there exists an axial symmetry present in the system owing to the fact
that a uniform external electric field is applied in a direction, we choose a
cylindrical coordinate system whose axial coordinate axis is parallel to the
external field direction. The cylindrical coordinates will be denoted
($x,\rho,\theta$) where $x$ is the axial coordinate, $\rho$ the radial
coordinate transversal to the axis $x$, and $\theta$ the azimuthal angle; see
Fig. 1. Then the distribution functions and potentials have axial symmetry
around the $x$ axis, and hence they are independent of angle $\theta$. Now
Fourier transforms are taken with respect to the axial coordinate $x$:%
\begin{align}
\left(
\begin{array}
[c]{c}%
f_{ji}^{+}\left(  x,\rho\right)  \smallskip\\
f_{ji}^{-}\left(  x,\rho\right)
\end{array}
\right)   &  =\frac{2}{\pi}\int_{0}^{\infty}d\alpha\left(
\begin{array}
[c]{c}%
\cos\left(  \alpha x\right)  \widehat{f}_{ji}^{+}\left(  \alpha,\rho\right)
\smallskip\\
\sin\left(  \alpha x\right)  \widehat{f}_{ji}^{-}\left(  \alpha,\rho\right)
\end{array}
\right)  ,\qquad etc.\label{14a}\\
\left(
\begin{array}
[c]{c}%
\psi_{j}^{+}\left(  x,\rho\right)  \smallskip\\
\psi_{j}^{-}\left(  x,\rho\right)
\end{array}
\right)   &  =\frac{2}{\pi}\int_{0}^{\infty}d\alpha\left(
\begin{array}
[c]{c}%
\cos\left(  \alpha x\right)  \widehat{\psi}_{j}^{+}\left(  \alpha,\rho\right)
\smallskip\\
\sin\left(  \alpha x\right)  \widehat{\psi}_{j}^{-}\left(  \alpha,\rho\right)
\end{array}
\right)  ,\qquad etc. \label{14b}%
\end{align}
Here $\alpha$ is a dimensionless wave number in units of $\kappa$. When
Fourier transformed in this manner, the governing equations (\ref{12a}%
)--(\ref{12d}) and (\ref{13a})--(\ref{13d}) become sets of coupled
second-order ordinary differential equations with respect to the reduced
radial coordinate $\rho$ (perpendicular to the $x$ axis) given below:%
\begin{align}
\left(  D_{\rho}^{2}-\frac{1}{2}\right)  \widehat{f}_{ji}^{+}\left(
\alpha,\rho\right)  -\frac{1}{4}\left[  \widehat{f}_{jj}^{+}\left(
\alpha,\rho\right)  +\widehat{f}_{ii}^{+}\left(  \alpha,\rho\right)  \right]
+\alpha\xi\widehat{f}_{ji}^{-}\left(  \alpha,\rho\right)  =0,  & \label{15a}\\
\left(  D_{\rho}^{2}-\frac{1}{2}\right)  \widehat{f}_{jj}^{+}\left(
\alpha,\rho\right)  -\frac{1}{2}\widehat{f}_{ji}^{+}\left(  \alpha
,\rho\right)  =0,  & \label{15b}\\
\left(  D_{\rho}^{2}-\frac{1}{2}\right)  \widehat{f}_{ii}^{+}\left(
\alpha,\rho\right)  -\frac{1}{2}\widehat{f}_{ji}^{+}\left(  \alpha
,\rho\right)  =0,  & \label{15c}\\
\left(  D_{\rho}^{2}-\frac{1}{2}\right)  \widehat{f}_{ji}^{-}\left(
\alpha,\rho\right)  -\alpha\xi\widehat{f}_{ji}^{+}\left(  \alpha,\rho\right)
=0,  &  \label{15d}%
\end{align}%
\begin{align}
D_{\rho}^{2}\widehat{\psi}_{j}^{+}\left(  \mathbf{r}\right)   &  =-\frac{1}%
{2}\left[  \widehat{f}_{ji}^{+}\left(  \mathbf{r}\right)  +\widehat{f}%
_{jj}^{+}\left(  \mathbf{r}\right)  \right]  ,\label{16a}\\
D_{\rho}^{2}\widehat{\psi}_{i}^{+}\left(  \mathbf{r}\right)   &  =-\frac{1}%
{2}\left[  \widehat{f}_{ii}^{+}\left(  \mathbf{r}\right)  +\widehat{f}%
_{ji}^{+}\left(  \mathbf{r}\right)  \right]  ,\label{16b}\\
D_{\rho}^{2}\widehat{\psi}_{j}^{-}\left(  \mathbf{r}\right)   &  =-\frac{1}%
{2}\widehat{f}_{ji}^{-}\left(  \mathbf{r}\right)  ,\label{16c}\\
D_{\rho}^{2}\widehat{\psi}_{i}^{-}\left(  \mathbf{r}\right)   &  =-\frac{1}%
{2}\widehat{f}_{ji}^{-}\left(  \mathbf{r}\right)  . \label{16d}%
\end{align}
Here symbol $D_{\rho}^{2}$ is defined by the differential operator%
\begin{equation}
D_{\rho}^{2}=\frac{1}{\rho}\frac{d}{d\rho}\rho\frac{d}{d\rho}-\alpha^{2}.
\label{17}%
\end{equation}
Because the zeroth-order Bessel function $K_{0}\left(  \alpha\rho\right)  $ of
second kind is an irregular solution of the differential
equation\cite{watson,abramowitz}%
\begin{equation}
D_{\rho}^{2}K_{0}\left(  \alpha\rho\right)  =\left(  \frac{1}{\rho}\frac
{d}{d\rho}\rho\frac{d}{d\rho}-\alpha^{2}\right)  K_{0}\left(  \alpha
\rho\right)  =0, \label{18}%
\end{equation}
the coupled inhomogeneous differential equations (\ref{15a})--\ref{15d}) are
expected to be solved by linear combinations of zeroth-order Bessel functions
but of different arguments $\lambda_{k}\rho$, where $\lambda_{k}$ $\left(
k=1,2,3,4\right)  $ are characteristic values of the differential equation
system. Unfortunately, two of the characteristic values turn out to be
degenerate. Therefore it is not possible to apply the method of linear algebra
to solve the system in the conventional manner in which the solutions are
expanded in characteristic vectors. This difficulty is overcome if Eqs.
(\ref{15a})--(\ref{15d}) are solved in the following manner.

Operating $\left(  D_{\rho}^{2}-\frac{1}{2}\right)  $ on Eq. (\ref{15a}) and
eliminating resulting $\left(  D_{\rho}^{2}-\frac{1}{2}\right)  \widehat
{f}_{jj}^{+}$, $\left(  D_{\rho}^{2}-\frac{1}{2}\right)  \widehat{f}_{ii}^{+}%
$, and $\left(  D_{\rho}^{2}-\frac{1}{2}\right)  \widehat{f}_{ji}^{-}$ using
Eqs. (\ref{15b})--(\ref{15d}), we obtain the fourth-order differential
equation%
\begin{equation}
\left(  D_{\rho}^{2}-\frac{1}{2}+\frac{1}{2}R\right)  \left(  D_{\rho}%
^{2}-\frac{1}{2}-\frac{1}{2}R\right)  \widehat{f}_{ji}^{+}\left(  \alpha
,\rho\right)  =0, \label{19}%
\end{equation}
where%
\begin{equation}
R=\sqrt{1-4\left(  \alpha\xi\right)  ^{2}}. \label{20}%
\end{equation}
This fourth-order differential equations can be solved by Bessel function
$K_{0}\left(  \lambda_{1}\rho\right)  $ and $K_{0}\left(  \lambda_{2}%
\rho\right)  $, where $\lambda_{1}$ and $\lambda_{2}$ are two characteristic
values
\begin{equation}
\lambda_{1}=\sqrt{\frac{1}{2}+\alpha^{2}+\frac{1}{2}R},\quad\lambda_{2}%
=\sqrt{\frac{1}{2}+\alpha^{2}-\frac{1}{2}R}. \label{21}%
\end{equation}
These are non-degenerate. Therefore the general solution for Eq. (\ref{19})
may be written as a linear combination of the Bessel functions%
\begin{equation}
\widehat{f}_{ji}^{+}\left(  \alpha,\rho\right)  =A_{1}K_{0}\left(  \lambda
_{1}\rho\right)  +A_{2}K_{0}\left(  \lambda_{2}\rho\right)  , \label{22}%
\end{equation}
where $A_{1}$ and $A_{2}$ are constant coefficients that must be determined by
the boundary conditions, Eqs. (\ref{6}) and (\ref{7b}), or the equivalent
conditions, for the symmetric and antisymmetric parts. Note that this solution
satisfies the boundary condition as $\rho\rightarrow\infty$ since the Bessel
functions $K_{0}\left(  \lambda_{k}\rho\right)  $ vanish at $\rho=\infty$.
Upon substituting this expansion into Eqs. (\ref{15a})--(\ref{15d}) we obtain%
\begin{align}
-\frac{1}{4}\widehat{f}_{jj}^{+}\left(  \alpha,\rho\right)  -\frac{1}%
{4}\widehat{f}_{ii}^{+}\left(  \alpha,\rho\right)  +\alpha\xi\widehat{f}%
_{ji}^{-}\left(  \alpha,\rho\right)   &  =-\frac{1}{2}RA_{1}K_{0}\left(
\lambda_{1}\rho\right)  +\frac{1}{2}RA_{2}K_{0}\left(  \lambda_{2}\rho\right)
,\label{23a}\\
\left(  D_{\rho}^{2}-\frac{1}{2}\right)  \widehat{f}_{jj}^{+}\left(
\alpha,\rho\right)   &  =\frac{1}{2}A_{1}K_{0}\left(  \lambda_{1}\rho\right)
+\frac{1}{2}A_{2}K_{0}\left(  \lambda_{2}\rho\right)  ,\label{23b}\\
\left(  D_{\rho}^{2}-\frac{1}{2}\right)  \widehat{f}_{ii}^{+}\left(
\alpha,\rho\right)   &  =\frac{1}{2}A_{1}K_{0}\left(  \lambda_{1}\rho\right)
+\frac{1}{2}A_{2}K_{0}\left(  \lambda_{2}\rho\right)  ,\label{23c}\\
\left(  D_{\rho}^{2}-\frac{1}{2}\right)  \widehat{f}_{ji}^{-}\left(
\alpha,\rho\right)   &  =\alpha\xi A_{1}K_{0}\left(  \lambda_{1}\rho\right)
+\alpha\xi A_{2}K_{0}\left(  \lambda_{2}\rho\right)  . \label{23d}%
\end{align}
This\ inhomogeneous set may be also solved by expansion. Let%
\begin{align}
\widehat{f}_{jj}^{+}\left(  \alpha,\rho\right)   &  =B_{1}K_{0}\left(
\lambda_{1}\rho\right)  +B_{2}K_{0}\left(  \lambda_{2}\rho\right)  +B_{3}%
K_{0}\left(  \lambda_{3}\rho\right)  ,\nonumber\\
\widehat{f}_{ii}^{+}\left(  \alpha,\rho\right)   &  =C_{1}K_{0}\left(
\lambda_{1}\rho\right)  +C_{2}K_{0}\left(  \lambda_{2}\rho\right)  +C_{3}%
K_{0}\left(  \lambda_{3}\rho\right)  ,\label{24a}\\
\widehat{f}_{ji}^{-}\left(  \alpha,\rho\right)   &  =D_{1}K_{0}\left(
\lambda_{1}\rho\right)  +D_{2}K_{0}\left(  \lambda_{2}\rho\right)  +D_{3}%
K_{0}\left(  \lambda_{3}\rho\right)  ,\nonumber
\end{align}
where $B_{k}$, $C_{k}$, and $D_{k}$ are expansion coefficients to be
determined and $\lambda$ is the degenerate characteristic value to be
determined self-consistently. Inserting these expansions into Eqs.
(\ref{23a})--(\ref{23d}) we find relations between the coefficients and also
the as-yet undetermined characteristic value $\lambda$. We find%
\begin{equation}
\lambda=\sqrt{\alpha^{2}+\frac{1}{2}}\equiv\lambda_{3} \label{24}%
\end{equation}
which is the degenerate third characteristic value of the governing
OF\ equations for binary electrolytes. It is independent of the external field
strength $X$ or $\xi$ unlike $\lambda_{1}$ and $\lambda_{2}$. The relations
between the coefficients are also obtained as follows:%
\begin{align}
B_{1}  &  =\frac{2\alpha\xi}{R}A_{1},\qquad B_{2}=-\frac{2\alpha\xi}{R}%
A_{2},\nonumber\\
C_{1}  &  =\frac{1}{R}A_{1},\qquad C_{2}=-\frac{1}{R}A_{2},\label{25}\\
D_{1}  &  =\frac{1}{R}A_{1},\qquad D_{2}=-\frac{1}{R}A_{2}.\nonumber
\end{align}
Thus the distribution functions $\widehat{f}_{ji}^{+}$, $\widehat{f}_{jj}^{+}%
$, $\widehat{f}_{ii}^{+}$, $\widehat{f}_{ji}^{-}$ are given as linear
combinations of Bessel functions $K_{0}(\lambda_{k}\rho)$ ($k=1,2,3$):%
\begin{align}
\widehat{f}_{ji}^{+}\left(  \alpha,\rho\right)   &  =A_{1}K_{0}\left(
\lambda_{1}\rho\right)  +A_{2}K_{0}\left(  \lambda_{2}\rho\right)
,\label{25a}\\
\widehat{f}_{jj}^{+}\left(  \alpha,\rho\right)   &  =\frac{2\alpha\xi}{R}%
A_{1}K_{0}\left(  \lambda_{1}\rho\right)  -\frac{2\alpha\xi}{R}A_{2}%
K_{0}\left(  \lambda_{2}\rho\right)  +B_{3}K_{0}\left(  \lambda\rho\right)
,\label{25b}\\
\widehat{f}_{ii}^{+}\left(  \alpha,\rho\right)   &  =\frac{1}{R}A_{1}%
K_{0}\left(  \lambda_{1}\rho\right)  -\frac{1}{R}A_{2}K_{0}\left(  \lambda
_{2}\rho\right)  +C_{3}K_{0}\left(  \lambda\rho\right)  ,\label{25c}\\
\widehat{f}_{ji}^{-}\left(  \alpha,\rho\right)   &  =\frac{1}{R}A_{1}%
K_{0}\left(  \lambda_{1}\rho\right)  -\frac{1}{R}A_{2}K_{0}\left(  \lambda
_{2}\rho\right)  +D_{3}K_{0}\left(  \lambda\rho\right)  . \label{25d}%
\end{align}
The solutions of Poisson equations (\ref{16a})--(\ref{16d}) can be similarly
obtained as linear combinations of Bessel functions $K_{0}(\lambda_{k}\rho)$:%
\begin{align}
\widehat{\psi}_{j}^{+}\left(  \alpha,\rho\right)   &  =-\frac{1}{R}A_{1}%
K_{0}\left(  \lambda_{1}\rho\right)  +\frac{1}{R}A_{2}K_{0}\left(  \lambda
_{2}\rho\right)  -C_{3}K_{0}\left(  \lambda_{3}\rho\right)  ,\label{26a}\\
\widehat{\psi}_{i}^{+}\left(  \alpha,\rho\right)   &  =-\frac{1}{R}A_{1}%
K_{0}\left(  \lambda_{1}\rho\right)  +\frac{1}{R}A_{2}K_{0}\left(  \lambda
_{2}\rho\right)  -\left(  4\alpha\xi B_{3}-C_{3}\right)  K_{0}\left(
\lambda_{3}\rho\right)  ,\label{26b}\\
\widehat{\psi}_{j}^{-}\left(  \alpha,\rho\right)   &  =-\frac{2\alpha\xi
}{R\left(  1+R\right)  }A_{1}K_{0}\left(  \lambda_{1}\rho\right)
+\frac{2\alpha\xi}{R\left(  1-R\right)  }A_{2}K_{0}\left(  \lambda_{2}%
\rho\right)  -B_{3}K_{0}\left(  \lambda_{3}\rho\right)  ,\label{26c}\\
\widehat{\psi}_{i}^{-}\left(  \alpha,\rho\right)   &  =-\frac{2\alpha\xi
}{R\left(  1+R\right)  }A_{1}K_{0}\left(  \lambda_{1}\rho\right)
+\frac{2\alpha\xi}{R\left(  1-R\right)  }A_{2}K_{0}\left(  \lambda_{2}%
\rho\right)  -B_{3}K_{0}\left(  \lambda_{3}\rho\right)  . \label{26d}%
\end{align}
The coefficients $A_{1}$, $A_{2}$, $B_{3}$, and $C_{3}$ in these expansions
are determined by imposing the boundary conditions (\ref{6}) and (\ref{7b}),
which for the symmetric and antisymmetric parts become
\begin{align}
\lim_{\rho\rightarrow0}\left(  \rho\frac{d}{d\rho}\widehat{f}_{ji}^{\pm}%
+\frac{1}{2}\rho\frac{d}{d\rho}\widehat{\psi}_{j}^{\pm}+\frac{1}{2}\rho
\frac{d}{d\rho}\widehat{\psi}_{i}^{\pm}\right)   &  =0,\label{27a}\\
\lim_{\rho\rightarrow0}\left(  \rho\frac{d}{d\rho}\widehat{f}_{kk}^{+}%
+\rho\frac{d}{d\rho}\widehat{\psi}_{k}^{+}\right)   &  =0\quad(k=j,i),
\label{27b}%
\end{align}%
\begin{align}
\lim_{\rho\rightarrow0}\rho\frac{d}{d\rho}\widehat{\psi}_{k}^{+}\left(
\rho\right)   &  =-\delta\qquad(k=i,j),\label{28a}\\
\lim_{\rho\rightarrow0}\rho\frac{d}{d\rho}\widehat{\psi}_{k}^{-}\left(
\rho\right)   &  =0\qquad(k=i,j). \label{28b}%
\end{align}
We note that the behavior of $K_{0}(\lambda_{k}\rho)$ near $\rho=0$ has the
property%
\begin{equation}
\lim_{\rho\rightarrow0}\rho\frac{d}{d\rho}K_{0}\left(  \lambda_{k}\rho\right)
=-\lim_{\rho\rightarrow0}\rho\frac{d}{d\rho}\ln\rho=-1. \label{29}%
\end{equation}
Imposing the boundary conditions, we obtain the linear algebraic relations of
coefficients, which can be easily solved upon reducing them to independent
linear equations. To save the space we simply present the final results only:%
\begin{align}
A_{1}  &  =-\frac{\left(  R+1\right)  }{2R},\qquad A_{2}=-\frac{\left(
R-1\right)  }{2R},\label{30a}\\
B_{1}  &  =-\frac{\left(  R+1\right)  \alpha\xi}{R^{2}},\qquad B_{2}%
=-\frac{\left(  1-R\right)  \alpha\xi}{R^{2}},\qquad B_{3}=\frac{2\alpha\xi
}{R^{2}},\label{30b}\\
C_{1}  &  =-\frac{\left(  R+1\right)  }{2R^{2}},\qquad C_{2}=\frac{\left(
R-1\right)  }{2R^{2}},\qquad C_{3}=\frac{16\left(  \alpha\xi\right)  ^{2}%
}{R^{2}},\label{30c}\\
D_{1}  &  =-\frac{\left(  R+1\right)  }{2R^{2}},\qquad D_{2}=\frac{\left(
R-1\right)  }{2R^{2}},\qquad D_{3}=\frac{4\left(  \alpha\xi\right)  ^{2}%
}{R^{2}}. \label{30d}%
\end{align}
The Fourier components $\widehat{f}_{ji}^{\pm}$ and $\widehat{\psi}_{k}^{+}$
in Eqs. (\ref{31fa}) and (\ref{31fb}) are finally given by
\begin{align}
\widehat{f}_{ji}^{+}\left(  \alpha,\rho\right)   &  =-\frac{1}{2R}\left[
\left(  R+1\right)  K_{0}\left(  \lambda_{1}\rho\right)  +\left(  R-1\right)
K_{0}\left(  \lambda_{2}\rho\right)  \right]  ,\label{31a}\\
\widehat{f}_{jj}^{+}\left(  \alpha,\rho\right)   &  =-\frac{\xi\alpha}{R^{2}%
}\left[  \left(  R+1\right)  K_{0}\left(  \lambda_{1}\rho\right)  +\left(
R-1\right)  K_{0}\left(  \lambda_{2}\rho\right)  -2K_{0}\left(  \lambda
\rho\right)  \right]  ,\label{31jj}\\
\widehat{f}_{ii}^{+}\left(  \alpha,\rho\right)   &  =-\frac{1}{2R^{2}}\left[
\left(  R+1\right)  K_{0}\left(  \lambda_{1}\rho\right)  -\left(  R-1\right)
K_{0}\left(  \lambda_{2}\rho\right)  -8\left(  \alpha\xi\right)  ^{2}%
K_{0}\left(  \lambda\rho\right)  \right] \label{31ii}\\
\widehat{f}_{ji}^{-}\left(  \alpha,\rho\right)   &  =-\frac{\xi\alpha}{R^{2}%
}\left[  \left(  R+1\right)  K_{0}\left(  \lambda_{1}\rho\right)  +\left(
1-R\right)  K_{0}\left(  \lambda_{2}\rho\right)  -2K_{0}\left(  \lambda
_{3}\rho\right)  \right]  ,\label{31c}\\
\widehat{\psi}_{k}^{+}\left(  \alpha,\rho\right)   &  =\frac{1}{2R^{2}}\left[
\left(  R+1\right)  K_{0}\left(  \lambda_{1}\rho\right)  +\left(  1-R\right)
K_{0}\left(  \lambda_{2}\rho\right)  -8\left(  \alpha\xi\right)  ^{2}%
K_{0}\left(  \lambda_{3}\rho\right)  \right]  ,\label{31d}\\
\widehat{\psi}_{k}^{-}\left(  \alpha,\rho\right)   &  =\frac{\xi\alpha}{R^{2}%
}\left[  K_{0}\left(  \lambda_{1}\rho\right)  +K_{0}\left(  \lambda_{2}%
\rho\right)  -2K_{0}\left(  \lambda_{3}\rho\right)  \right]  \quad\left(
k=j,i\right)  . \label{31e}%
\end{align}
We summarize the Fourier transforms of the solutions for the distribution
functions and potentials we set out to find:%
\begin{align}
f_{ji}\left(  \mathbf{r}\right)  -n_{i}n_{j}  &  =-\frac{2\kappa^{2}n_{j}%
n_{i}e_{j}e_{i}}{D\pi k_{B}T}\int_{0}^{\infty}d\alpha\frac{\cos\left(  \alpha
x\right)  }{R}\left[  \left(  R+1\right)  K_{0}\left(  \lambda_{1}\rho\right)
+\left(  R-1\right)  K_{0}\left(  \lambda_{2}\rho\right)  \right] \nonumber\\
&  \quad-\frac{2\kappa^{2}n_{j}n_{i}e_{j}e_{i}\xi}{D\pi k_{B}T}\int
_{0}^{\infty}d\alpha\frac{\alpha\sin\left(  \alpha x\right)  }{R^{2}}%
\times\label{31fa}\\
&  \qquad\qquad\qquad\qquad\quad\left[  \left(  R+1\right)  K_{0}\left(
\lambda_{1}\rho\right)  +\left(  1-R\right)  K_{0}\left(  \lambda_{2}%
\rho\right)  -2K_{0}\left(  \lambda_{3}\rho\right)  \right]  ,\nonumber\\
\psi_{k}\left(  \mathbf{r}\right)   &  =\frac{e_{k}}{2D}\int_{0}^{\infty
}d\alpha\frac{\cos\left(  \alpha x\right)  }{R^{2}}\times\nonumber\\
&  \qquad\qquad\left[  \left(  R+1\right)  K_{0}\left(  \lambda_{1}%
\rho\right)  +\left(  1-R\right)  K_{0}\left(  \lambda_{2}\rho\right)
-8\left(  \alpha\xi\right)  ^{2}K_{0}\left(  \lambda_{3}\rho\right)  \right]
\nonumber\\
&  \quad+\frac{e_{k}\xi}{D}\int_{0}^{\infty}d\alpha\frac{\alpha\sin\left(
\alpha x\right)  }{R^{2}}\left[  K_{0}\left(  \lambda_{1}\rho\right)
+K_{0}\left(  \lambda_{2}\rho\right)  -2K_{0}\left(  \lambda_{3}\rho\right)
\right]  \quad\left(  k=j,i\right)  , \label{31fb}%
\end{align}
and similarly for $f_{jj}\left(  x,\rho\right)  $ and $f_{ii}\left(
x,\rho\right)  $ to $f_{ji}\left(  x,\rho\right)  $ in Eq. (\ref{31fa}). It
should be noted that the variables in the integrals are reduced variables in
the units of the Debye parameter $\kappa$; see Eq. (\ref{32a}) below.

The solutions presented in Eqs. (\ref{31fa}) and (\ref{31fb}) are the
nonequilibrium parts of the pair distribution functions and potentials in the
ionic liquid in the external field $X$ (or $\xi$ in reduced form) at arbitrary
strength. Therefore they represent the nonequilibrium ionic liquid structure
and potentials when the ions are moving subjected to the external field at a
steady-state condition. If the full potential is desired, $\psi_{k}\left(
\alpha,\rho\right)  $ must be combined with the equilibrium Debye
potential---the Yukawa-type potential. Therefore it would be of great interest
to see how the nonequilibrium liquid structure and potentials vary with
respect to spatial positions and the field strength. We will investigate these
aspects (i.e., profiles) in the following.

\subsection{Evaluation of Nonequilibrium Ionic Liquid Structure and
Potentials}

The Fourier integrals in Eqs. (\ref{31fa}) and (\ref{31fb}) contain three
parameters, position coordinates $x$ and $\rho$ and the reduce field strength
$\xi$. Although looking complicated, they can be evaluated analytically in the
region where the transversal (radial) coordinate $\rho$ satisfies a certain
condition with respect to the axial coordinate $x$, as will be stated more
precisely later; see Eq. (\ref{33}) below. In the rest of the $\left(
x,\rho\right)  $ plane where the condition is not met, they can be evaluated
numerically, provided that the singular behavior of the integrands is properly
handled by using the method of principal values used for singular
integrals\cite{singular}.

For the evaluation of the integrals, it is convenient to scale further the
variables as follows:%
\begin{equation}
t=\sqrt{2}\alpha,\quad r=\rho/\sqrt{2},\quad\widehat{x}=x/\sqrt{2},\quad
\omega_{k}=\sqrt{2}\lambda_{k}. \label{32a}%
\end{equation}
Thus%
\begin{equation}
\omega_{1}=\sqrt{1+t^{2}+\sqrt{1-2\xi^{2}t^{2}}},\quad\omega_{2}=\sqrt
{1+t^{2}-\sqrt{1-2\xi^{2}t^{2}}},\quad\omega_{3}=\sqrt{1+t^{2}}. \label{32b}%
\end{equation}
It is also convenient to define
\begin{align}
\overline{\omega}_{1}  &  =\sqrt{1-y^{2}+\sqrt{1+2\xi^{2}y^{2}}}%
,\quad\overline{\omega}_{2}=\sqrt{1-y^{2}-\sqrt{1+2\xi^{2}y^{2}}}\nonumber\\
\overline{\omega}_{3}  &  =\sqrt{1-y^{2}},\quad\overline{\omega}=\frac
{\sqrt{1+2\xi^{2}}}{\sqrt{2}\xi}. \label{32c}%
\end{align}
Notice that $\overline{\omega}_{k}=\omega_{k}|_{t=iy}$ ($k=1,\cdots$) with the
complex variable $t$ taken along the imaginary axis. As shown in Appendix A,
if the condition\cite{whittaker}
\begin{equation}
\frac{\widehat{x}}{r}+\frac{\operatorname{Re}\omega_{k}\left(  t\right)
}{\operatorname{Im}t}>0 \label{33}%
\end{equation}
for $\widehat{x},r>0$ in the complex plane of variable $t$, the integrals in
Eqs. (\ref{31fa}) and (\ref{31fb}) can be evaluated analytically or reduced to
simple one-dimensional quadratures, if methods of contour integration are
employed. As a matter of fact, the one-dimensional quadratures thus obtained
can be analytically evaluated term by term in series if the series
representation for Bessel functions $I_{0}(z)$ is used. Condition (\ref{33})
means that the region in question is roughly within a conical domain
surrounding the $x$ axis. Outside this region the integrals must be computed
numerically by applying methods of principal integrations for singular
integrals\cite{singular}. In this exterior region the integrals vanish
uniformly as $\widehat{x},r\rightarrow\infty$.

In his dissertation\cite{wilson}, Wilson did not evaluated either $f_{ji}$ or
$\psi_{k}$, but only the axial velocity at the special position of
$\widehat{x}=r=0$, namely, the coordinate origin. Henceforth for notational
brevity the reduced variable $\widehat{x}$ will be simply typeset $x$ without
the caret $\widehat{}$ .

The integrals in Eqs. (\ref{31fa}) and (\ref{31fb}), reduced as described
above, are evaluated by means of the contour integration methods described in
Appendix A. They are given by the expressions
\begin{align}
f_{ji}\left(  \pm\mathbf{r}\right)   &  =n^{2}-\frac{\kappa zn^{2}e^{2}}{8\pi
Dk_{B}T}\left\{  \int_{0}^{\sqrt{2\left(  1+\xi^{2}\right)  }}dye^{-yx}\left(
1+\frac{1}{\sqrt{1+2\xi^{2}y^{2}}}\right)  I_{0}\left(  \overline{\omega}%
_{1}r\right)  \right. \nonumber\\
&  \qquad\qquad\qquad\qquad\pm\xi\int_{0}^{\sqrt{2\left(  1+\xi^{2}\right)  }%
}dy\frac{e^{-xy}y\left(  1+\sqrt{1+2\xi^{2}y^{2}}\right)  }{\left(  1+2\xi
^{2}y^{2}\right)  }I_{0}\left(  \overline{\omega}_{1}r\right) \nonumber\\
&  \qquad\qquad\qquad\qquad\left.  \mp2\xi\int_{0}^{1}dy\frac{ye^{-xy}}%
{1+2\xi^{2}y^{2}}I_{0}\left(  \overline{\omega}_{3}r\right)  \right\}  ,
\label{34f}%
\end{align}%
\begin{align}
\psi_{j}\left(  \pm\mathbf{r}\right)   &  =-\psi_{i}\left(  \mp\mathbf{r}%
\right) \nonumber\\
&  =-\frac{z\kappa e}{8\pi\sqrt{2}D}\left\{  \left[  \int_{0}^{\sqrt{2\left(
1+\xi^{2}\right)  }}dy\frac{e^{-xy}\left(  1+\sqrt{1+2\xi^{2}y^{2}}\right)
}{1+2\xi^{2}y^{2}}I_{0}\left(  \overline{\omega}_{1}r\right)  \right.  \right.
\nonumber\\
&  \qquad\qquad\qquad\left.  -4\xi^{2}\int_{0}^{1}dy\frac{e^{-xy}y^{2}}%
{1+2\xi^{2}y^{2}}I_{0}\left(  \overline{\omega}_{3}r\right)  \right]
\nonumber\\
&  \qquad\qquad\qquad\left.  \pm\sqrt{2}\xi\left[  \int_{0}^{\sqrt{2\left(
1+\xi^{2}\right)  }}dy\frac{ye^{-xy}I_{0}\left(  \overline{\omega}%
_{1}r\right)  }{1+2\xi^{2}y^{2}}-2\int_{0}^{1}dy\frac{ye^{-xy}I_{0}\left(
\overline{\omega}_{3}r\right)  }{1+2\xi^{2}y^{2}}\right]  \right\}  .
\label{34p}%
\end{align}
Here $I_{0}\left(  \overline{\omega}_{k}r\right)  $ $\left(  k=1,3\right)  $
are the regular Bessel functions of zeroth order of second kind. In these
expressions the range of position variables $x$ and $r$ must be such that
$\sqrt{2\left(  1+\xi^{2}\right)  }x>r$ for the integrals involving
$\overline{\omega}_{1}$, and $x>r$ for the integrals involving $\overline
{\omega}_{3}$. These conditions, related to Condition (\ref{33}) stemming from
Jordan's lemma\cite{whittaker} on contour integrals, ensure the boundary
conditions for the distribution functions and the potentials, which vanish as
$x$ and $r$ tend to infinity.

The results presented in Eqs. (\ref{34f}) and (\ref{34p}) for the reduced
nonequilibrium part of pair distribution function $\Delta\overline{f}_{ji}%
\ $and the reduced nonequilibrium part of ionic potential $\Delta\psi_{j}$,
respectively, defined by%
\begin{equation}
\Delta\overline{f}_{ji}=\frac{\pi Dk_{B}T}{\sqrt{2}\kappa ze^{2}}\left(
f_{ji}-n^{2}\right)  ,\quad\Delta\overline{\psi}_{j}=\frac{\sqrt{2}\pi
D}{\kappa ze}\left(  \psi_{j}-\psi_{j}^{0}\right)  \label{34del}%
\end{equation}
are graphically depicted in the case of $\xi=1$ in Figs. 2--3 to give
pictorial representations for the nonequilibrium parts of the ionic liquid
structure and the mean ionic potential in the Brownian motion model. They
vanish as $x$ and $r$ increase to infinity from a finite value at the origin.
The choice \ of the value of the reduced field strength $\xi$ is arbitrary; it
could be as large as desired.

Fig. 2 displays an important feature most distinguishable from the equilibrium
pair distribution function for ion pair $\left(  j,i\right)  $ that should be
spherically symmetric and peaked at the coordinate origin $\left(  x,r\right)
=\left(  0,0\right)  $. Instead, the nonequilibrium part of the pair
distribution function $\Delta\overline{f}_{ji}\left(  x,r,\xi\right)  $ at
$\xi>0$ has a peak displaced from the coordinate origin. This means that the
spherical symmetry originally present at equilibrium (i.e., at $\xi=0$) not
only has been destroyed with its peak position displaced to a point $\left(
x,r\right)  \neq\left(  0,0\right)  $ from the coordinate origin, but also the
ion atmosphere is no longer spherically symmetric if $\xi>0$. This means that
the center of ion atmosphere has also been displaced by the the external field
along the $x$ axis. As a matter of fact, the present exact solutions of the
governing equations provide the details of the state of distortion of the
spherical ionic atmosphere and its migration as $\xi$ increases from $\xi=0$.
We will see in the next section how this mode of distortion in the ion
atmosphere is further modified in a manner of feedback process by the
hydrodynamic motion of medium induced by the motions of ions under the
influence of the external field. Fig. 3 for $\Delta\overline{\psi}_{j}$
illustrates the molecular cause for the distortion of the spherical ionic
atmosphere through the nonequilibrium change in the ionic potentials.

If the series representations for the Bessel functions $I_{0}\left(
\overline{\omega}_{1}r\right)  $ and $I_{0}\left(  \overline{\omega}%
_{3}r\right)  $ are used, the integrals can be evaluated in terms of
elementary functions of $\xi$, $x$, and $r$, but since these series converges
slowly, such series representations would have a limited practical value for
precise evaluation of integrals. Nevertheless, such representation might be of
some use for some theoretical study. Eqs. (\ref{34f}) and (\ref{34p}) contain
the information on the nonequilibrium ionic liquid structure and the mean
potentials for the ionic liquid subjected to the external electric
field.\textbf{ }They are new results for ionic solutions in an external field
examined here. The distribution functions and ionic potentials could be made
use of to develop a theory of transport processes in binary electrolyte
solutions. In this sense, they would be potentially very useful, especially,
for calculating transport coefficients of the ionic solution in the electric field.

\section{Hydrodynamic Equation and Flow Profiles}

In the conventional ionic conductance experiments the flow velocity of the
medium is usually not large. Therefore flow may be regarded as laminar. Under
this condition the nonlinear inertial term can be neglected in the
Navier--Stokes equation. Moreover, the liquid may be considered incompressible
to a good approximation. Under these conditions the Navier--Stokes equation
becomes the Stokes equation \cite{landau,batchelor,bird} for an incompressible
fluid. We therefore use the Stokes equation to calculate the flow velocity of
the medium around the moving ions pulled by the external field. It may be
helpful to point out that the flow field generated would be schematically
reminiscent of the flow field around a moving object submerged in a medium.

We assume that there are no body-forces other than an applied electric field.
However, because ions are strongly correlated by long-range Coulomb forces and
also interacting with the applied external electric field, it is necessary to
calculate the mean local electric field. For the purpose of calculating it we
may use the solutions of the OF equations and the Poisson equations presented
in the previous section. Therefore the mean local electric field is expected
to depend on the spatial position and the external field strength $\xi$.

\subsection{\textbf{Local Electric Field}}

Since the field is aligned along the $x$ axis and the charge density is given
by the Poisson equation, the local force due to the field $X$ on charge
density $\varrho$ is given by%
\begin{equation}
F_{x}=-\frac{DX}{4\pi}\nabla^{2}\psi_{j}\left(  \mathbf{r}\right)  .
\label{35a}%
\end{equation}
Since $\psi_{j}\left(  \mathbf{r}\right)  $ is given by
\begin{align}
\psi_{j}\left(  \mathbf{r}\right)   &  =\frac{\kappa e_{j}}{\pi D}\int
_{0}^{\infty}d\alpha\frac{\cos\left(  \alpha x\right)  }{R^{2}}\left[  \left(
R+1\right)  K_{0}\left(  \lambda_{1}\rho\right)  \right. \nonumber\\
&  \qquad\qquad\qquad\qquad\qquad\left.  +\left(  1-R\right)  K_{0}\left(
\lambda_{2}\rho\right)  -2\left(  1-R^{2}\right)  K_{0}\left(  \lambda_{3}%
\rho\right)  \right] \nonumber\\
&  \quad+\frac{8\kappa e_{j}\xi}{D}\int_{0}^{\infty}d\alpha\sin\left(  \alpha
x\right)  \frac{\alpha}{R^{2}}\left[  K_{0}\left(  \lambda_{1}\rho\right)
+K_{0}\left(  \lambda_{2}\rho\right)  -2K_{0}\left(  \lambda_{3}\rho\right)
\right]  , \label{36}%
\end{align}
we obtain the mean local body-force
\begin{align}
F_{x}  &  =-\frac{Xe_{j}\kappa^{3}}{2\pi^{2}}\int_{0}^{\infty}d\alpha
\cos\left(  \alpha x\right)  \left[  \frac{\left(  R+1\right)  \left(
\lambda_{1}^{2}-\alpha^{2}\right)  }{2R^{2}}K_{0}\left(  \lambda_{1}%
\rho\right)  \right. \nonumber\\
&  \qquad\qquad\qquad\left.  +\frac{\left(  1-R\right)  \left(  \lambda
_{2}^{2}-\alpha^{2}\right)  }{2R^{2}}K_{0}\left(  \lambda_{2}\rho\right)
-\frac{\left(  1-R^{2}\right)  \left(  \lambda_{3}^{2}-\alpha^{2}\right)
}{R^{2}}K_{0}\left(  \lambda_{3}\rho\right)  \right] \nonumber\\
&  \quad-\frac{Xe_{j}\kappa^{3}\xi}{2\pi^{2}}\int_{0}^{\infty}d\alpha
\sin\left(  \alpha x\right)  \left[  \frac{\alpha\left(  \lambda_{1}%
^{2}-\alpha^{2}\right)  }{R^{2}}K_{0}\left(  \lambda_{1}\rho\right)  \right.
\nonumber\\
&  \qquad\qquad\qquad\left.  +\frac{\alpha\left(  \lambda_{2}^{2}-\alpha
^{2}\right)  }{R^{2}}K_{0}\left(  \lambda_{2}\rho\right)  -2\frac
{\alpha\left(  \lambda_{3}^{2}-\alpha^{2}\right)  }{R^{2}}K_{0}\left(
\lambda_{3}\rho\right)  \right]  . \label{37F}%
\end{align}
This expression shows that the external force $e_{j}X$ is dressed up by the
long-range correlations between the ions interacting through Coulomb forces
and the interaction of ions and ion atmosphere with the external field. The
effects of long-range correlations are described by the governing equations,
and their feedback effect manifests itself in the form of dressed external
force. This aspect is an important characteristic of the present theory of
ionic solutions not usually seen in theories of charge carrier mobilities and
their transport processes in recent literatures \cite{ting,nag,landsberg}.

It is convenient to write Eq. (\ref{37F}) in a compact form to solve the
Stokes equation:%
\begin{equation}
F_{x}=\frac{Xe_{j}\kappa^{3}}{2\pi^{2}}\sum_{l=1}^{3}\left[  C_{l}\cos\left(
\alpha x\right)  K_{0}\left(  \lambda_{l}\rho\right)  +S_{l}\sin\left(  \alpha
x\right)  K_{0}\left(  \lambda_{l}\rho\right)  \right]  , \label{37a}%
\end{equation}
where integral operators $C_{l}$ and $S_{l}$ are defined by%
\begin{align}
C_{l}  &  =-\int_{0}^{\infty}d\alpha\left\{
\begin{array}
[c]{c}%
\frac{\left(  1+R\right)  \left(  \lambda_{1}^{2}-\alpha^{2}\right)  }{2R^{2}%
}\quad\text{for }l=1\medskip\\
\frac{\left(  1-R\right)  \left(  \lambda_{2}^{2}-\alpha^{2}\right)  }{2R^{2}%
}\quad\text{for }l=2\medskip\\
-\frac{\left(  1-R^{2}\right)  \left(  \lambda_{3}^{2}-\alpha^{2}\right)
}{R^{2}}\quad\text{for }l=2
\end{array}
\right.  ,\label{37c}\\
S_{l}  &  =-\int_{0}^{\infty}d\alpha\left\{
\begin{array}
[c]{c}%
\frac{\xi\alpha\left(  \lambda_{1}^{2}-\alpha^{2}\right)  }{R^{2}}%
\quad\text{for }l=1\medskip\\
\frac{\xi\alpha\left(  \lambda_{2}^{2}-\alpha^{2}\right)  }{R^{2}}%
\quad\text{for }l=2\medskip\\
-\frac{2\xi\alpha\left(  \lambda_{3}^{2}-\alpha^{2}\right)  }{R^{2}}%
\quad\text{for }l=3
\end{array}
\right.  . \label{37s}%
\end{align}
This mean local force $F_{x}$ is an input for the Stokes equation of the flow
problem under consideration.

\subsection{\textbf{Stokes Equation and its Equivalent Form}}

At an arbitrary Reynolds number the steady Navier--Stokes
equation\cite{landau} must be used:
\begin{equation}
\overline{\rho}\mathbf{v\cdot\nabla v}-\eta_{0}\nabla^{2}\mathbf{v-}\eta
_{b}\mathbf{\nabla}\left(  \mathbf{\nabla\cdot v}\right)  =-\mathbf{\nabla
}p+\mathbf{F,} \label{NS}%
\end{equation}
where $\overline{\rho}$ is the fluid density, $\eta_{0}$ is the shear
viscosity of the electrolyte solution, $\eta_{b}$ is its bulk viscosity, $p$
is the pressure, and $\mathbf{F}$ is the body (external) force density. For an
incompressible fluid $\mathbf{\nabla\cdot v=\,}0$. For a fluid undergoing
laminar flow of low Reynolds number (typically Re = $O(10^{-6})$ at the field
gradient of $1$ kVolt/m in aqueous solution) the inertial term can be
neglected. Thus the Navier--Stokes equation for velocity $\mathbf{v}$ becomes
the Stokes equations for divergenceless flow
\begin{align}
-\eta_{0}\nabla^{2}\mathbf{v}  &  =-\mathbf{\nabla}p+\mathbf{F,}\label{NSa}\\
\mathbf{\nabla\cdot v\,}  &  \mathbf{=\,}0. \label{NSb}%
\end{align}
Note that the presence of an external field makes the pressure nonuniform in
space. As is well known, if $curl$ of Eq. (\ref{NSa}) is taken, the
$\mathbf{\nabla}p$ term vanishes and Eq. (\ref{NSa}) takes the form%
\begin{equation}
\eta_{0}\mathbf{\nabla\times\nabla\times\nabla}\times\mathbf{v=\nabla}%
\times\mathbf{F.} \label{38a}%
\end{equation}

Since $\mathbf{\nabla\times\nabla}\times\mathbf{v=\nabla}\left(
\mathbf{\nabla\cdot v\,}\right)  -\nabla^{2}\mathbf{v}$ by vector algebra, the
two equations (\ref{NSa}) and (\ref{NSb}) may be combined into a single
equation%
\begin{equation}
\eta_{0}\mathbf{\nabla\times\nabla}\times\mathbf{v=}-\mathbf{\nabla
}p+\mathbf{F.} \label{38b}%
\end{equation}
For the present problem $\mathbf{F=}%
\mbox{\boldmath$\delta$}%
_{x}F_{x}$, where $%
\mbox{\boldmath$\delta$}%
_{x}$ is the unit vector along the $x$ axis.

To solve Eq. (\ref{38a}) for $\mathbf{v}$, we observe $\mathbf{\nabla\cdot
v\,}\mathbf{=\,}0$, which means that there exists an axial vector $\mathbf{A}$
such that $\mathbf{v\,}\mathbf{=\nabla\times A}$, where $\mathbf{A}$ must
depend on position vector $\mathbf{r}$ and field vector $\mathbf{X}$, both of
which are ordinary vectors. Thus we may transform the solution $\mathbf{v}$ of
Eq. (\ref{38b}) into the form
\begin{equation}
\mathbf{v=\nabla\times\nabla}\times\mathbf{a}+\mathbf{v}^{0}\mathbf{,}
\label{38c}%
\end{equation}
where $\mathbf{v}^{0}$ is a constant satisfying the appropriate boundary
conditions of the velocity. Since $\mathbf{a\rightarrow\,}0$ and also
$\mathbf{v}$ should vanish as $\left\vert \mathbf{r}\right\vert \rightarrow
\infty$, it follows $\mathbf{v}^{0}=0$. Thus we will set $\mathbf{v}^{0}=0$ henceforth.

In the first step to formally solve Eq. (\ref{38b}), substitute Eq.
(\ref{38c}) with $\mathbf{v}^{0}=0$ into Eq. (\ref{38b}) to obtain the
equation%
\begin{equation}
\eta_{0}\mathbf{\nabla\times\nabla}\times\mathbf{\nabla\times\nabla}%
\times\mathbf{a=}-\mathbf{\nabla}p+\mathbf{F.} \label{39s}%
\end{equation}
By the identities of vector algebra%
\begin{align}
\mathbf{\nabla\times\nabla}\times\mathbf{a\,}  &  \mathbf{=\nabla}\left(
\operatorname{div}\mathbf{a}\right)  \mathbf{-\nabla}^{2}\mathbf{a,}%
\label{L1}\\
\mathbf{\nabla\times\nabla\times\nabla}\left(  \mathbf{\nabla\cdot a}\right)
&  =0, \label{L2}%
\end{align}
and%
\begin{equation}
\mathbf{\nabla\times\nabla}\times\left(  \nabla^{2}\mathbf{a}\right)
=\mathbf{\nabla}\left(  \nabla^{2}\operatorname{div}\mathbf{a}\right)
-\mathbf{\nabla}^{2}\left(  \nabla^{2}\mathbf{a}\right)  , \label{L3}%
\end{equation}
it follows that%
\begin{align}
\mathbf{\nabla\times\nabla\times\nabla\times\nabla\times a}  &
=\mathbf{\nabla\times\nabla\times\nabla}\left(  \mathbf{\nabla\cdot a}\right)
\,\mathbf{-\,\nabla\times\nabla\times}\left(  \nabla^{2}\mathbf{a}\right)
\nonumber\\
&  =-\mathbf{\nabla\times\nabla\times}\left(  \nabla^{2}\mathbf{a}\right)  .
\label{L4}%
\end{align}
Upon using Eq. (\ref{L4}) in Eq. (\ref{39s}) and substituting the result into
Eq. (\ref{38b}), we obtain a fourth-order differential equation of vector
$\mathbf{a}$:%
\begin{equation}
\eta_{0}\mathbf{\nabla}^{2}\mathbf{\nabla}^{2}\mathbf{a-F=\nabla}\left(
\eta_{0}\nabla^{2}\operatorname{div}\mathbf{a-}p\right)  . \label{39f}%
\end{equation}
This equation is equivalent to Eq. (\ref{NSa}) or the Stokes equations.
Because the left and right hand sides of Eq. (\ref{39f}) are of two different
kinds of vectors the equation may be separated into two equations:%
\begin{align}
\eta_{0}\mathbf{\nabla}^{2}\mathbf{\nabla}^{2}\mathbf{a}  &  =\mathbf{F,}%
\label{40}\\
p  &  =p_{0}+\eta_{0}\mathbf{\nabla}^{2}\operatorname{div}\mathbf{a.}\nonumber
\end{align}
The solution of the Stokes equations is now reduced to that of Eq. (\ref{40}),
a fourth-order differential equation with $\mathbf{F}$ given by the solutions
of the OF equations and the Poisson equations---namely, the governing
equations. In summary, we have for the velocity and pressure the expressions
\begin{align}
\mathbf{v}  &  =\mathbf{\nabla\times\nabla\times a+v}^{0}=\mathbf{\nabla
\times\nabla\times a,}\label{41}\\
p  &  =p_{0}+\eta_{0}\mathbf{\nabla}^{2}\left(  \mathbf{\nabla\cdot a}\right)
. \label{42}%
\end{align}
Vector $\mathbf{a}$ is determined by solving Eq. (\ref{40}) in terms of the
local force density given by Eq. (\ref{37F}) or (\ref{37a}), a compact
abbreviation of the former. In Eq. (\ref{25}) $p_{0}$ is a homogeneous
pressure uniform in space, that is, the equilibrium pressure consisting of the
osmotic pressure of the solution. This equilibrium pressure must be either
supplied phenomenologically by using thermodynamics or from the statistical
mechanics of the electrolyte solution. Therefore, given the solution for
vector $\mathbf{a}$, both velocity and pressure can be determined from the
Stokes equation.

To solve Eq. (\ref{40}) for $\mathbf{a}$, substitute Eq. (\ref{37a}) into the
former, which then reads%
\begin{equation}
\mathbf{\nabla}^{2}\left(  \mathbf{\nabla}^{2}\mathbf{a}\right)  =\frac
{Xe_{j}\kappa^{3}}{2\pi^{2}\eta_{0}}%
\mbox{\boldmath$\delta$}%
_{x}\left[  C_{l}\cos\left(  \alpha x\right)  K_{0}(\lambda_{l}\rho)+S_{l}%
\sin\left(  \alpha x\right)  K_{0}(\lambda_{l}\rho)\right]  , \label{26}%
\end{equation}
where the repeated index $l$ means a sum over $l=1,2,3$. Since Eq. (\ref{26})
suggests that $\mathbf{\nabla}^{2}\mathbf{a}$ must be a linear combination of
the Bessel functions in the right hand side of the equation, recalling Eqs.
(\ref{17}) and (\ref{18}) we find
\begin{equation}
\mathbf{\nabla}^{2}\mathbf{a=\,}\frac{Xe_{j}\kappa}{2\pi^{2}\eta_{0}}%
\mbox{\boldmath$\delta$}%
_{x}\left[  C_{l}\frac{\cos\left(  \alpha x\right)  K_{0}(\lambda_{l}\rho
)}{\lambda_{l}^{2}-\alpha^{2}}+S_{l}\frac{\sin\left(  \alpha x\right)
K_{0}(\lambda_{l}\rho)}{\lambda_{l}^{2}-\alpha^{2}}\right]  +%
\mbox{\boldmath$\delta$}%
_{x}A^{\ast}, \label{28}%
\end{equation}
where $A^{\ast}$ is the homogeneous solution obeying the equation%
\begin{equation}
\mathbf{\nabla}^{2}\left(  \mathbf{\nabla}^{2}\mathbf{A}^{\ast}\right)  =0
\label{homo}%
\end{equation}
with $\mathbf{A}^{\ast}=%
\mbox{\boldmath$\delta$}%
_{x}A^{\ast}$. The solution $\mathbf{A}^{\ast}$ must satisfy the boundary
conditions at infinite $\rho$. Thus we choose
\begin{equation}
\mathbf{A}^{\ast}=-\frac{Xe_{j}\kappa}{2\pi^{2}\eta_{0}}%
\mbox{\boldmath$\delta$}%
_{x}\left[  C_{l}\frac{\cos\left(  \alpha x\right)  K_{0}(\alpha\rho)}%
{\lambda_{l}^{2}-\alpha^{2}}+S_{l}\frac{\sin\left(  \alpha x\right)
K_{0}(\alpha\rho)}{\lambda_{l}^{2}-\alpha^{2}}\right]  , \label{30}%
\end{equation}
since this satisfies Eq. (\ref{homo}). Therefore we obtain the equation%
\begin{align}
\mathbf{\nabla}^{2}\mathbf{a}  &  \mathbf{=\,}\frac{Xe_{j}\kappa}{2\pi^{2}%
\eta_{0}}%
\mbox{\boldmath$\delta$}%
_{x}\times\nonumber\\
&  \qquad\left\{  C_{l}\frac{\cos\left(  \alpha x\right)  \left[
K_{0}(\lambda_{l}\rho)-K_{0}(\alpha\rho)\right]  }{\lambda_{l}^{2}-\alpha^{2}%
}+S_{l}\frac{\sin\left(  \alpha x\right)  \left[  K_{0}(\lambda_{l}\rho
)-K_{0}(\alpha\rho)\right]  }{\lambda_{l}^{2}-\alpha^{2}}\right\}  .
\label{31}%
\end{align}
Since the solution of this inhomogeneous second-order differential equation
must be a linear combination of the Bessel functions making up the
inhomogeneous term on the right, it is sought in the form%
\begin{align}
\mathbf{a\,}  &  \mathbf{=\,}\frac{Xe_{j}}{2\pi^{2}\eta_{0}\kappa}%
\mbox{\boldmath$\delta$}%
_{x}\left[  C_{l}\frac{\cos\left(  \alpha x\right)  }{\lambda_{l}^{2}%
-\alpha^{2}}+S_{l}\frac{\sin\left(  \alpha x\right)  }{\lambda_{l}^{2}%
-\alpha^{2}}\right]  \times\nonumber\\
&  \qquad\qquad\left\{  b_{1}\left[  K_{0}(\lambda_{l}\rho)-K_{0}(\alpha
\rho)\right]  +b_{2}\left[  K_{0}(\beta_{l}\rho)-K_{0}(\alpha\rho)\right]
\right\}  , \label{h2}%
\end{align}
where $b_{1}$, $b_{2}$, and $\beta_{l}$ are constants determined as follows:
On inserting this expansion into Eq. (\ref{31}) we find%
\begin{equation}
\left[  b_{1}\left(  \lambda_{l}^{2}-\alpha^{2}\right)  -1\right]
K_{0}(\lambda_{l}\rho)+b_{2}\left(  \beta_{l}^{2}-\alpha^{2}\right)
K_{0}(\beta_{l}\rho)+K_{0}(\alpha\rho)=0. \label{h2b}%
\end{equation}
The expansion coefficients $b_{1}$ and $b_{2}$ and the parameter $\beta_{l}$
are determined below. Since Bessel functions $K_{0}(\lambda_{l}\rho)$,
$K_{0}(\beta_{l}\rho)$, and $K_{0}(\alpha\rho)$ not only do not vanish
everywhere in $\rho$, but also their arguments are arbitrary, we may choose
$b_{1}$ and $b_{2}$ such that%
\begin{equation}
b_{1}=\frac{1}{\lambda_{l}^{2}-\alpha^{2}} \label{h3}%
\end{equation}
and%
\begin{equation}
\lim_{\beta_{l}\rightarrow\alpha}b_{2}\left(  \beta_{l}^{2}-\alpha^{2}\right)
K_{0}(\beta_{l}\rho)=-K_{0}(\alpha\rho). \label{h4}%
\end{equation}
Then Eq. (\ref{h2b}) is satisfied and hence Eq. (\ref{h2}) is a solution of
Eq. (\ref{31}). Eq. (\ref{h4}) therefore implies%
\begin{equation}
b_{2}=-\frac{1}{\left(  \beta_{l}^{2}-\alpha^{2}\right)  }. \label{h5}%
\end{equation}
Finally, we obtain for the solution of Eq. (\ref{31})%
\begin{align}
\mathbf{a}  &  =\frac{Xe_{j}}{2\pi^{2}\eta_{0}\kappa}%
\mbox{\boldmath$\delta$}%
_{x}\left[  C_{l}\cos\left(  \alpha x\right)  +S_{l}\sin\left(  \alpha
x\right)  \right]  \times\nonumber\\
&  \qquad\qquad\qquad\left\{  \frac{\left[  K_{0}(\lambda_{l}\rho
)-K_{0}(\alpha\rho)\right]  }{\left(  \lambda_{l}^{2}-\alpha^{2}\right)  ^{2}%
}+\frac{\alpha\rho K_{1}(\alpha\rho)}{2\left(  \lambda_{l}^{2}-\alpha
^{2}\right)  \alpha^{2}}\right\}  . \label{32}%
\end{align}
For the solution (\ref{32}) for $\mathbf{a}$, we have used Eq. (\ref{L1}) and
the recurrence relations of the Bessel functions\cite{watson,abramowitz}%
\begin{align}
\frac{d}{dz}K_{0}(z)  &  =-K_{1}(z),\nonumber\\
\frac{d}{dz}K_{1}\left(  z\right)   &  =-K_{0}\left(  z\right)  -\frac{1}%
{z}K_{1}\left(  z\right)  ,\label{37}\\
\left(  \frac{1}{z}\frac{d}{dz}z\frac{d}{dz}-1\right)  zK_{1}\left(  z\right)
&  =-2K_{0}\left(  z\right)  ,\nonumber
\end{align}
as well as $\operatorname{div}\mathbf{a=\,}\partial a_{x}/\partial x$ owing to
the fact that $\mathbf{F=}%
\mbox{\boldmath$\delta$}%
_{x}F_{x}$\ and hence $a_{\rho}=a_{\theta}=0$\ identically.

\subsection{\textbf{Fourier Transform Solution for the Axial Velocity}}

It is now possible to obtain the Fourier transform solution for the axial
component of the velocity. Since%
\[
\left(  \mathbf{\nabla\times\nabla}\times\mathbf{a}\right)  _{x}%
\;\mathbf{=\nabla}_{x}\left(  \operatorname{div}\mathbf{a}\right)
-\mathbf{\nabla}^{2}a_{x},
\]
by using the formulas for $\mathbf{\nabla}_{x}\left(  \operatorname{div}%
\mathbf{a}\right)  $ and $\mathbf{\nabla}^{2}a_{x}$ it follows from Eq.
(\ref{41}) the Fourier transform solution for the axial velocity component for
all values of $x$ and $\rho$:%
\begin{align}
\mathbf{v}_{x}\left(  x,\rho;\xi\right)   &  =\mathbf{-}\frac{Xe_{j}}{2\pi
^{2}\eta_{0}\kappa}\sum_{l=1}^{3}C_{l}\cos\left(  \alpha x\right)  \left\{
\frac{\lambda_{l}^{2}\left[  K_{0}(\lambda_{l}\rho)-K_{0}(\alpha\rho)\right]
}{\left(  \lambda_{l}^{2}-\alpha^{2}\right)  ^{2}}+\frac{\alpha\rho
K_{1}(\alpha\rho)}{2\left(  \lambda_{l}^{2}-\alpha^{2}\right)  }\right\}
\nonumber\\
&  \quad\,-\frac{Xe_{j}}{2\pi^{2}\eta_{0}\kappa}\sum_{l=1}^{3}S_{l}\sin\left(
\alpha x\right)  \left\{  \frac{\lambda_{l}^{2}\left[  K_{0}(\lambda_{l}%
\rho)-K_{0}(\alpha\rho)\right]  }{\left(  \lambda_{l}^{2}-\alpha^{2}\right)
^{2}}+\frac{\alpha\rho K_{1}(\alpha\rho)}{2\left(  \lambda_{l}^{2}-\alpha
^{2}\right)  }\right\}  . \label{38}%
\end{align}
Here we now have restored the summation sign over index $l$. For a more
explicit expression the sum over $l$ may be expanded. This formula does not
exactly agree with Wilson's expression\cite{wilson} for the axial velocity
because of some missing terms and typographical errors in his formula.

The Fourier transform integrals in Eq. (\ref{38}) may be expressed by using
the reduced variables defined in Eqs. (\ref{32a}) and (\ref{32b}) to cast them
into as simple forms as possible. We will also define the reduced velocity%
\begin{equation}
\widehat{\mathbf{v}}=\left(  2\sqrt{2}\pi^{2}\eta_{0}/zeX\kappa\right)
\mathbf{v}. \label{Rv1}%
\end{equation}
Then the axial velocity $\mathbf{v}_{x}\left(  x,\rho;\xi\right)  $ is given
by%
\begin{equation}
\mathbf{v}_{x}\left(  x,\rho;\xi\right)  =\frac{zeX\kappa}{2\sqrt{2}\pi
^{2}\eta_{0}}\widehat{\mathbf{v}}_{x}\left(  x,r,\xi\right)  , \label{Rv}%
\end{equation}
where the reduced axial velocity is now given by components made up of cosine
and sine Fourier transforms:
\begin{equation}
\widehat{\mathbf{v}}_{x}\left(  x,r,\xi\right)  =\frac{1}{2}K_{c}^{B}%
+\frac{\xi}{\sqrt{2}}K_{s}^{B}-K_{4}^{c}+\frac{1}{2}rK_{5}^{c}-\frac{1}%
{\sqrt{2}\xi}K_{4}^{s}. \label{Kvx}%
\end{equation}
Various components in Eq. (\ref{Kvx}) are defined by the Fourier transforms%
\begin{equation}
K_{c}^{B}\left(  x,r,\xi\right)  =\int_{0}^{\infty}dt\frac{\cos\left(
xt\right)  }{\left(  1-2\xi^{2}t^{2}\right)  }\left[  \omega_{1}^{2}%
K_{0}(\omega_{1}r)+\omega_{2}^{2}K_{0}(\omega_{2}r)-4\xi^{2}t^{2}\omega
_{3}^{2}K_{0}(\omega_{3}r)\right]  , \label{KcB}%
\end{equation}%
\begin{align}
K_{s}^{B}\left(  x,r,\xi\right)   &  =\int_{0}^{\infty}dt\frac{t\sin\left(
xt\right)  }{\left(  1-2\xi^{2}t^{2}\right)  }\times\nonumber\\
&  \qquad\left[  \frac{\omega_{1}^{2}K_{0}(\omega_{1}r)}{\left(
1+\sqrt{1-2\xi^{2}t^{2}}\right)  }+\frac{\omega_{2}^{2}K_{0}(\omega_{2}%
r)}{\left(  1-\sqrt{1-2\xi^{2}t^{2}}\right)  }-2\omega_{3}^{2}K_{0}(\omega
_{3}r)\right]  , \label{KsB}%
\end{align}%
\begin{align}
K_{4}^{c}  &  =\int_{0}^{\infty}dt\cos\left(  xt\right)  \omega_{3}^{2}%
K_{0}(rt),\label{Kc4a}\\
K_{5}^{c}  &  =\int_{0}^{\infty}dt\cos\left(  xt\right)  tK_{1}%
(rt),\label{Kc5a}\\
K_{4}^{s}  &  =\int_{0}^{\infty}dtt\sin\left(  xt\right)  K_{0}(rt).
\label{Ks4}%
\end{align}

Integrals $K_{4}^{c}$, $K_{5}^{c}$, and $K_{4}^{s}$ can be evaluated
analytically in closed algebraic forms. On the other hand, the Brownian motion
part of the integrals $K_{c}^{B}\left(  x,r,\xi\right)  $ and $K_{s}%
^{B}\left(  x,r,\xi\right)  $ can be evaluated by methods of contour
integration in the region satisfying Condition (\ref{33}) required by the
Jordan lemma\cite{whittaker} for the contour integrals. Outside the region,
they are computed by using straightforward numerical integration methods
employing a method of principal values.

\subsubsection{Evaluation of Integrals $K_{4}^{c}$, $K_{5}^{c}$, and
$K_{4}^{s}$}

All the integrals appearing in the expression for $\widehat{\mathbf{v}}%
_{x}\left(  x,r,\xi\right)  $ do not appear simple at first glance.
Presumably, for this reason Wilson evaluated the integrals for the case of
$x=r=0$ only. However, the integrals $K_{4}^{c}$, $K_{5}^{c}$, and $K_{4}^{s}$
are indeed amenable to analytic evaluations in closed form. We explicitly
illustrate the method by using $K_{4}^{c}$ as an example. Other integrals can
be evaluated similarly.

\paragraph{$K_{4}^{c}$}

On substitution of the integral representation\cite{watson} of the Bessel
function $K_{\nu}(rt)$ of integer order%
\begin{equation}
K_{\nu}(z)=\int_{0}^{\infty}dse^{-z\cosh s}\cosh\left(  \nu s\right)
\qquad\left(  \left\vert \arg z\right\vert <\frac{\pi}{2}\right)  ,
\label{40IR}%
\end{equation}
the integral $K_{4}^{c}$ can be written as
\[
K_{4}^{c}=\frac{1}{2}\int_{0}^{\infty}dt\int_{0}^{\infty}ds\left(
1+t^{2}\right)  \left[  e^{-t\left(  r\cosh s-ix\right)  }+e^{-t\left(  r\cosh
s+ix\right)  }\right]  .
\]
It is legitimate to interchange the order of integrals. Then the integration
over $t$ is trivial; changing variable to $z=\sinh s$, we obtain elementary
integrals with respect to $z$, which can be easily integrated:
\begin{equation}
K_{4}^{c}=\frac{\pi}{2\left(  x^{2}+r^{2}\right)  ^{1/2}}-\frac{\pi\left(
2x^{2}-r^{2}\right)  }{2\left(  r^{2}+x^{2}\right)  ^{5/2}}. \label{43}%
\end{equation}
It reminds us of Coulombic and dipole contributions, which are purely mechanical.

\paragraph{$K_{5}^{c}$}

Upon using the integral representation of $K_{1}(rt)$ and the same procedure
as for integral $K_{4}^{c}$, we obtain $K_{5}^{c}$,%
\begin{equation}
K_{5}^{c}=\frac{\pi r}{2\left(  x^{2}+r^{2}\right)  ^{\frac{3}{2}}}.
\label{44}%
\end{equation}
This integral appears in Wilson's formulation as the divergence-causing term.
We will return to it again when we compare the present result with
Wilson's\cite{wilson} in more detail.

\paragraph{$K_{4}^{s}$}

This integral also can be evaluated in the same manner as for $K_{4}^{c}$. We
obtain
\begin{equation}
K_{4}^{s}=\frac{\pi x}{2\left(  x^{2}+r^{2}\right)  ^{\frac{3}{2}}}.
\label{45}%
\end{equation}
The three integrals $K_{4}^{c}$, $K_{5}^{c}$, and $K_{4}^{s}$ make up purely
mechanical contributions to the axial velocity $\mathbf{v}_{x}$. They may be
interpreted as either Coulombic or dipolar contributions of the ion
atmosphere, which acts as if it is a dipole toward the external field. The
collection of the three integrals evaluated up to this point will be
collectively referred to as a mechanical velocity $\left(  \widehat
{\mathbf{v}}_{x}\right)  _{\text{me}}$, which is a countercurrent induced by
Coulomb and dipole interactions of ion atmosphere interacting with the applied
external field:
\begin{align}
\left(  \widehat{\mathbf{v}}_{x}\right)  _{\text{me}}  &  \equiv-K_{4}%
^{c}+\frac{r}{2}K_{5}^{c}-\frac{1}{\sqrt{2}\xi}K_{4}^{s}\nonumber\\
&  =-\frac{\pi}{2\sqrt{2}\xi}\frac{x}{\left(  x^{2}+r^{2}\right)  ^{\frac
{3}{2}}}\nonumber\\
&  \quad\,-\frac{\pi}{2\left(  x^{2}+r^{2}\right)  ^{1/2}}+\frac{\pi r^{2}%
}{4\left(  x^{2}+r^{2}\right)  ^{\frac{3}{2}}}+\frac{\pi\left(  2x^{2}%
-r^{2}\right)  }{2\left(  x^{2}+r^{2}\right)  ^{5/2}}. \label{vacm}%
\end{align}
This contribution of $\left(  \widehat{\mathbf{v}}_{x}\right)  _{\text{me}}$
to $\widehat{\mathbf{v}}_{x}$ represents \textit{the fully deterministic part
of the hydrodynamic velocity} that is not associated with the Brownian motion
of particles giving rise to the dissipative part of the local body-force. In
fact, one of these terms [i.e., the first term in the second equality of Eq.
(\ref{vacm})], when inserted into the velocity formula (\ref{Rv}), becomes
field-independent and, consequently, does not contribute to the mobility or
electrophoretic coefficient. Moreover, $\left(  \widehat{\mathbf{v}}%
_{x}\right)  _{\text{me}}$ is negatively divergent at the coordinate origin,
and its manner of divergence is clearly direction-dependent, that is,
depending on whether the zero of $x$ or $r$ is approached first. Note that
when converted to the axial velocity in real units, the last three terms in
$\left(  \widehat{\mathbf{v}}_{x}\right)  _{\text{me}}$ in Eq. (\ref{vacm})
are proportional to the reduced field strength $\xi$.

\subsubsection{\textbf{Evaluation of }$K_{c}^{B}\left(  x,r,\xi\right)  $ and
$K_{s}^{B}\left(  x,r,\xi\right)  $\textbf{ Arising from Brownian Motions}}

The remaining integrals (\ref{KcB}) and (\ref{KsB}) can be calculated by means
of contour integration methods described in Appendix A. We collect them in the
form%
\begin{equation}
\frac{1}{2}K_{c}^{B}+\frac{\xi}{\sqrt{2}}K_{s}^{B}=-\frac{\pi}{4}\left[
\mathfrak{C}_{1}\left(  x,r,\xi\right)  -\mathfrak{S}_{1}\left(
x,r,\xi\right)  \right]  -\frac{\pi}{2}\left[  \mathfrak{C}_{2}\left(
x,r,\xi\right)  -\mathfrak{S}_{2}\left(  x,r,\xi\right)  \right]  ,
\label{vbr}%
\end{equation}
where
\begin{align}
\mathfrak{C}_{1}\left(  x,r;\xi\right)   &  =\int_{0}^{\sqrt{2\left(
1+\xi^{2}\right)  }}dy\frac{\left(  1-y^{2}+\sqrt{1+2\xi^{2}y^{2}}\right)
}{1+2\xi^{2}y^{2}}e^{-xy}I_{0}(\overline{\omega}_{1}r),\label{vC1}\\
\mathfrak{C}_{2}\left(  x,r;\xi\right)   &  =\int_{0}^{1}dy\frac{2\xi^{2}%
y^{2}\left(  1-y^{2}\right)  }{1+2\xi^{2}y^{2}}e^{-xy}I_{0}(\overline{\omega
}_{3}r),\label{vC2}\\
\mathfrak{S}_{1}\left(  x,r;\xi\right)   &  =\int_{0}^{\sqrt{2\left(
1+\xi^{2}\right)  }}dy\frac{\sqrt{2}\xi y\left(  1-y^{2}+\sqrt{1+2\xi^{2}%
y^{2}}\right)  }{\left(  1+2\xi^{2}y^{2}\right)  \left(  1+\sqrt{1+2\xi
^{2}y^{2}}\right)  }e^{-xy}I_{0}(\overline{\omega}_{1}r),\label{vC3}\\
\mathfrak{S}_{2}\left(  x,r;\xi\right)   &  =\int_{0}^{1}dy\frac{\sqrt{2}\xi
y\left(  1-y^{2}\right)  }{1+2\xi^{2}y^{2}}e^{-xy}I_{0}(\overline{\omega}%
_{3}r). \label{vC4}%
\end{align}
The parameters ($x,r$) in integrals (\ref{vC1}) and (\ref{vC3}) are subject to
the condition%
\begin{equation}
x\sqrt{1+\xi^{2}}>r \label{vca}%
\end{equation}
and integrals (\ref{vC2}) and (\ref{vC4}) to the condition
\begin{equation}
x>r, \label{vcb}%
\end{equation}
both of which arise from the condition to satisfy the Jordan
lemma\cite{whittaker} on the contour integrals involving an infinite
semicircle in the complex plane; see contours in Figs. 11--13 in Appendix A:%
\begin{equation}
x/r+\operatorname{Re}\omega_{l}\left(  t\right)  /\operatorname{Im}%
t>0\quad\left(  t=\text{ complex};\;l=1,2,3\right)  . \label{vc}%
\end{equation}
These conditions also have been mentioned in connection with the pair
distribution functions and potentials in Sec. II. If these conditions are not
met, the contour integration methods cannot be applied because the integrals
along curve $C_{\infty}$ of infinite radius diverge. In the region of $(x,r)$
plane not satisfying these conditions (i.e., exterior to the region) the
integrals must be evaluated numerically by using the method of principal
values\cite{singular}. Note that as in the contour integration methods used
for Eqs. (\ref{vC1})--(\ref{vC4}) the contributions from the singular points
cancel in the end, leaving only the principal value parts. It is also
important to note the sine transform terms make significant contributions
comparable in magnitude to the cosine transform terms, as will be found later
in the numerical analysis. On the other hand, if $x$ were set equal to zero,
the sine integral would identically vanish and thus have made no contribution
to the velocity. Consequently, the final velocity values would be different
depending on whether setting $x$ and $r$ equal to zero before or after
integration. This subtle, but important point should be kept in mind when we
handle this kind of integrals or the result obtained could be misleading.

In summary for the axial velocity, we obtain
\begin{align}
\mathbf{v}_{x}\left(  x,r,\xi\right)   &  =-\frac{\kappa^{2}k_{B}T}{8\pi
\eta_{0}}\frac{x}{\left(  x^{2}+r^{2}\right)  ^{3/2}}\nonumber\\
&  \quad\,-\frac{ze\kappa X}{4\sqrt{2}\pi\eta_{0}}\left[  \frac{1}{\left(
x^{2}+r^{2}\right)  ^{\frac{1}{2}}}-\frac{r^{2}}{2\left(  x^{2}+r^{2}\right)
^{\frac{3}{2}}}-\frac{2x^{2}-r^{2}}{\left(  x^{2}+r^{2}\right)  ^{\frac{5}{2}%
}}\!\right] \nonumber\\
&  \quad\,-\frac{ze\kappa X}{4\sqrt{2}\pi\eta_{0}}\left[  \mathfrak{C}%
_{1}\left(  x,r,\xi\right)  +2\mathfrak{C}_{2}\left(  x,r,\xi\right)
-\mathfrak{S}_{1}\left(  x,r,\xi\right)  -2\mathfrak{S}_{2}\left(
x,r,\xi\right)  \right]  . \label{vax}%
\end{align}

The axial velocity obtained here contains a term independent of the external
field---i.e., the first term on the right, whereas the rest of terms are led
by terms proportional to $X$ (or $\xi$ in reduced units); they are in fact
rather complicated functions of the field strength $\xi$. Physically, the
velocity calculated from the Stokes equation represents the flow profile of
the countercurrent induced by the moving center ion and its ion atmosphere in
response to the external electric field.

\subsubsection{Electrophoretic Factor}

The mobility of ions in the $x$ direction is associated with the
field-dependent terms of the axial velocity, and the mobility or
electrophoretic coefficient can be defined as the coefficient in the axial
velocity vs. electric field according to the thermodynamic force-flux
relations in thermodynamics of irreversible processes\cite{mazur, haase}.
Therefore, according to the usual practice in the theory of ionic
conductance\cite{harned} within the framework of irreversible thermodynamics,
we define the electrophoretic factor $\mathfrak{f}(x,r;\xi)$ as follows:%
\begin{equation}
\mathbf{v}_{x}\left(  x,r,\xi\right)  =-\frac{\kappa^{2}k_{B}T}{8\pi\eta_{0}%
}\frac{x}{\left(  x^{2}+r^{2}\right)  ^{\frac{3}{2}}}-\frac{zeX\kappa}%
{6\sqrt{2}\pi\eta_{0}}\mathfrak{f}(x,r;\xi). \label{ephc}%
\end{equation}
In fact, the factor $\mathfrak{f}(x,r;\xi)$ is generally dependent on position
coordinates $x$ and $r$ as well as $\xi$. Note that the second term on the
right of Eq. (\ref{ephc}) is reminiscent of the velocity formula in Eq.
(\ref{k4}), which was obtained by a heuristic argument on the basis of the
Stokes law in contrast to the hydrodynamic derivation of Eq. (\ref{ephc}).
Then upon comparison with the axial velocity formula (\ref{vax}) the
electrophoretic factor $\mathfrak{f}(x,r;\xi)$ is identified with the
expression%
\begin{align}
\mathfrak{f}(x,r;\xi)  &  =\frac{3}{2\left(  x^{2}+r^{2}\right)  ^{1/2}}%
-\frac{3r^{2}}{4\left(  r^{2}+x^{2}\right)  ^{3/2}}-\frac{3\left(
2x^{2}-r^{2}\right)  }{2\left(  r^{2}+x^{2}\right)  ^{5/2}}\nonumber\\
&  +\frac{3}{2}\left[  \mathfrak{C}_{1}\left(  x,r,\xi\right)  +2\mathfrak{C}%
_{2}\left(  x,r,\xi\right)  -\mathfrak{S}_{1}\left(  x,r,\xi\right)
-2\mathfrak{S}_{2}\left(  x,r,\xi\right)  \right]  . \label{fephc}%
\end{align}
Since it generally depends on coordinates as does the axial velocity, the
factor $\mathfrak{f}(x,r;\xi)$, in fact, describes the electrophoretic profile
in ($x,r$) plane as is evident from the figure shown below.

\subsubsection{Numerical Evaluation of the Axial Velocity}

Since it is important to learn about the axial velocity profiles we have
plotted them in the $\left(  x,r\right)  $ plane in the case of $\xi=1$. In
the region satisfying Conditions (\ref{vc}) the formula given in Eq.
(\ref{vax}) is used with the integrals (\ref{vC1})--(\ref{vC4}), and in the
exterior to the region defined by the conditions the velocity integrals for
the Brownian motion contributions---i.e., Eqs. (\ref{KcB}) and (\ref{KsB}%
)---are calculated by applying methods of principal integration because the
integrals have singularities on the real axis. Thus computed axial velocity
profiles are summarized in Figs. 4--6.

In Fig. 4 the axial velocity is plotted in ($x,r$) plane in 3D with the
vertical axis indicating the magnitude (color coded) of the axial velocity. It
is seen negative in a semi-elliptic region enclosing the $r$ axis beginning
from $r=0$ (yellow-green color) as predicted by Formula (\ref{vax}), it being
negative principally because of the mechanical part of the axial velocity
$\left(  \widehat{\mathbf{v}}_{x}\right)  _{\text{me}}$, which becomes
dominant over the Brownian motion contributions---the last group of terms in
Eq. (\ref{vax}). According to Fig. 4, the maximum of the axial velocity in the
positive $x$ direction (dark red region) is located in the neighborhood of the
coordinate origin, but displaced from the origin $(x,r)=(0,0)$. The axial
velocity decreases gradually and eventually vanishes as $x$ and $r$ values
increase to infinity. To gain a better idea of the electrophoretic factor
$\mathfrak{f}\left(  x,r,\xi\right)  $ it is plotted 3-dimensionally in
($x,r$) plane in Fig. 5 with the magnitude in the similar color coding used
for Fig. 4. Its shape is rather similar to the velocity profile in Fig. 4, but
its sign is opposite to that of $\widehat{\mathbf{v}}_{x}$ owing to the way it
is defined. To have a better idea of the behavior of the electrophoretic
factor we have plotted the projection of the level curves of the
$\mathfrak{f}\left(  x,r,\xi\right)  $ surface onto the ($x,r)$ plane in Fig.
6. It displays two sets of roughly elliptical contours, one with the major
axis on the $x$ axis and the other with the major axis on the $r$ axis
excluding the coordinate origin. The former set of contours corresponds to the
negative portion of $\mathfrak{f}\left(  x,r,\xi\right)  $, whereas the latter
to the positive portion of $\mathfrak{f}\left(  x,r,\xi\right)  $ but
transversal to the $x$ axis. The outermost level curve denoted $C_{p}$ in fact
represents the locus of zero of $\mathfrak{f}\left(  x,r,\xi\right)  $, that
is, $\mathfrak{f}\left(  x,r,\xi\right)  =0$. These two sets of
quasi-elliptical contours, and particularly, curve $C_{p}$ (i.e., the
quasi-ellipse above the $x$ axis) indicates how the spherical ion atmosphere
at equilibrium with its center located at the coordinate origin when $\xi$ was
equal to zero drifts away from the origin along the $x$ axis and the spherical
form is, at the same time, distorted to a non-spherical (quasi-elliptical)
form with its center at $x>0$ as the external field strength increases---i.e.,
a nonequilibrium state. For example, in the present reduced variables
employed, the equilibrium radius of the ion atmosphere $\left(  \xi=0\right)
$ is $\left(  1/\sqrt{2}\right)  \kappa^{-1}\simeq0.7\kappa^{-1}$ with the
center at the coordinate origin, but if $\xi=1$, not only the center of the
quasi-ellipse has migrated to $x\simeq0.6\kappa^{-1}$ and the curve $C_{p}$ is
no longer spherical with the major axis reduced to approximately
$1.2\kappa^{-1}=2\left(  0.6\kappa^{-1}\right)  \ $instead of $2\left(
1/\sqrt{2}\right)  \kappa^{-1}\simeq1.4\kappa^{-1}$ in the case of $\xi=0$.
This trend persists with increasing $\xi$. This behavior is numerically
examined in Fig. 7, where the position of the center $(x_{c},0)$ of the
quasi-ellipse is plotted against $\xi$. It gradually and significantly
diminishes with increasing $\xi$ after having reached a maximum. Since this
position $\left(  x_{c},0\right)  $ is at the center of displaced ion
atmosphere that is simultaneously distorted by the external field it is
natural to choose $x$ in $\mathfrak{f}\left(  x,r,\xi\right)  $ with the $x$
coordinate of the center $x_{c}$ of the quasi-ellipse as the center of the ion
atmosphere at $\xi$. Since the electrophoretic coefficient may be regarded as
the force on the imaginary spherical ion atmosphere with its center at
$(x_{c},0)$, then it is reasonable to choose $r$ in $\mathfrak{f}\left(
x,r,\xi\right)  $ with $r=x_{c}$. With this choice of the $x$ and $r$ values
in the electrophoretic factor $\mathfrak{f}\left(  x,r,\xi\right)  $ we have
verified that the electrophoretic coefficient thus calculated invariably
produces the correctly behaved equivalent ionic conductance over a wide range
of the external field strength, provided that the relaxation time coefficient
[see Eqs. (\ref{GC7})\ and (\ref{GC8}) below] is calculated with the same set
of $\left(  x,r\right)  $. Thus, in this manner we have been able to formulate
a procedure based on computation result for selecting the position parameters
$\left(  x,r\right)  $ in the electrophoretic and relaxation time factors and
therewith the ionic conductance unambiguously. We now state this procedure as
follows: \textit{The values of the coordinates }$x$\textit{ and }$r$\textit{
in the electrophoretic factor }$f\left(  x,r,\xi\right)  $\textit{ are
selected to be the }$x$\textit{ coordinate of the center of the quasi-elliptic
level curve }$C_{P}$\textit{ and the corresponding value for }$r$\textit{ of
the imaginary spherical ion atmosphere }$r_{c}=x_{c}$\textit{ centered at
}$\left(  x_{c},0\right)  $\textit{. The relaxation time factor is similarly
calculated. }In retrospect, this procedure---which may be called a
rule---seems natural since the center of the ion atmosphere drifts along the
$x$ axis as $\xi$ increases and the electrophoretic coefficient must be
reckoned with respect to the center of ion located at the $\left(
x_{c},0\right)  $ of the spherical ion atmosphere of radius $x_{c}$, namely,
$\left(  1/\sqrt{2}\right)  \kappa^{-1}$ in the actual units, which means
$r_{c}=x_{c}$.

With this identification of the coordinate parameters $\left(  x,r\right)  $
in the electrophoretic and relaxation time factors the electrophoretic and
relaxation time coefficients are rendered unambiguous and unique. They are
also divergence-free because the center of the displaced and distorted ion
atmosphere does not occur at the coordinate origin for all values of $\xi$ and
the OW theory becomes free from the divergence difficulty inherent to Wilson's
procedure of selecting $x=r=0$.

\subsubsection{Comparison with Wilson's Result for the Electrophoretic
Coefficient}

Having defined the electrophoretic factor based on the full formula
(\ref{vax}) for the axial velocity obtained from the Stokes equation, we
investigate how Wilson's result for the electrophoretic coefficient can be
recovered. He observed that since the ion of interest in conductance
experiment is the center ion of the ion atmosphere, which is located at the
coordinate origin, the axial velocity must be considered at $x=r=0$. He then
noticed that the Fourier transform integrals comprising the axial velocity
could be analytically evaluated at $x=r=0$, because in the Bessel function
$K_{0}(z)$ represented in power series as\cite{watson, abramowitz}%
\begin{equation}
K_{0}\left(  z\right)  =-\left[  \ln\left(  \frac{1}{2}z\right)
+\gamma\right]  \sum_{k=0}^{\infty}\frac{\left(  \frac{z}{2}\right)  ^{2k}%
}{\left(  k!\right)  ^{2}}+\frac{z^{2}}{2^{2}\left(  1!\right)  ^{2}}%
+\frac{\left(  1+\frac{1}{2}\right)  z^{4}}{2^{4}\left(  2!\right)  ^{2}%
}+\frac{\left(  1+\frac{1}{2}+\frac{1}{3}\right)  z^{4}}{2^{6}\left(
3!\right)  ^{2}}+\cdots, \label{Bess}%
\end{equation}
where $\gamma$ is Euler's constant, if $x=r=0$, only the leading term of
$K_{0}\left(  z\right)  $ contributes. Therefore, at $x=r=0$ Formula
(\ref{Rv}) for the axial velocity can be written as a sum of simple integrals%
\begin{align}
\widehat{v}_{x}\left(  0,0;\xi\right)   &  =-\frac{1}{2}\int_{0}^{\infty
}dt\frac{1}{\left(  1-2\xi^{2}t^{2}\right)  }\times\nonumber\\
&  \qquad\qquad\left[  \omega_{1}^{2}\ln\left(  \frac{2\omega_{1}}{t}\right)
+\omega_{2}^{2}\ln\left(  \frac{2\omega_{2}}{t}\right)  -2\left(
1-R^{2}\right)  \omega_{3}^{2}\ln\left(  \frac{2\omega_{3}}{t}\right)
\right]  +\int_{0}^{\infty}dt\frac{1}{2}. \label{vw}%
\end{align}
The logarithmic integrals can be exactly evaluated by means of contour
integrations by using contours similar to Figs. 10--12 in Appendix A. (Note,
however, his contours used are not exactly the same as Figs. 10--12 we have
employed in Appendix A except for the locations of simple poles and branch
cuts.) With so evaluated integrals and the electrophoretic coefficient defined
by the relation\cite{wilson,harned}%
\begin{equation}
v_{x}\left(  0,0;\xi\right)  =-\frac{zeX\kappa}{6\sqrt{2}\pi\eta_{0}}f(\xi),
\label{vwf}%
\end{equation}
the electrophoretic coefficient $f(\xi)$ could be shown given by the
expression%
\begin{align}
f\left(  \xi\right)   &  =1+\frac{3}{4\sqrt{2}\xi^{3}}\left\{  2\xi^{2}%
\sinh^{-1}\xi+\sqrt{2}\xi-\xi\sqrt{1+\xi^{2}}\right. \nonumber\\
&  \quad\left.  -\left(  1+2\xi^{2}\right)  \tan^{-1}\left(  \sqrt{2}%
\xi\right)  +\left(  1+2\xi^{2}\right)  \tan^{-1}\left(  \frac{\xi}%
{\sqrt{1+\xi^{2}}}\right)  \right\}  , \label{Wf}%
\end{align}
provided that the last integral in Eq. (\ref{vw}) is ignored. For this formula
for $f\left(  \xi\right)  $ we have used the identities:
\[
\frac{1}{2}\tan^{-1}\left(  2\xi\sqrt{1+\xi^{2}}\right)  =\tan^{-1}\left(
\xi/\sqrt{1+\xi^{2}}\right)  ;\quad\sinh^{-1}\left(  2\xi\sqrt{1+\xi^{2}%
}\right)  =2\sinh^{-1}\xi.
\]
As a matter of fact, for the last integral in Eq. (\ref{vw}) for $\widehat
{v}_{x}\left(  0,0;\xi\right)  $ Wilson\cite{wilson} argued that the integral
of $\frac{1}{2}$ contributes nothing to the electrophoretic coefficient
because its contour integral vanishes. This argument is fallacious because
although the contour integral in question certainly vanishes, it is composed
of two integrals which are manifestly infinite, but opposite in sign:%
\begin{equation}
\int_{C}\frac{1}{2}dz=\int_{-\infty}^{\infty}\frac{1}{2}dx+\int_{C_{\infty}%
}\frac{1}{2}dz=0, \label{Ic}%
\end{equation}
As a matter of fact, according to the analysis leading to Eq. (\ref{vax}) the
last integral in Eq. (\ref{vw}) originates from the integral $K_{5}^{c}$,
which we have already evaluated analytically for all values of $x$ and $r$,
and it is equal to zero at $r=0$ only if $x\neq0$, as is obvious from the
following consideration:
\begin{equation}
K_{5}^{c}=\frac{\pi r}{2\left(  x^{2}+r^{2}\right)  ^{\frac{3}{2}}}=\left\{
\begin{array}
[c]{c}%
0\quad\text{as }r\rightarrow0\text{, }x\neq0\\
\frac{\pi}{2r}\quad\text{for}\;x\rightarrow0\text{, }r\neq0
\end{array}
\right.  . \label{fvb}%
\end{equation}
However, if $x$ and $r$ simultaneously tend to $0$ at the same rate,
$K_{5}^{c}$ is manifestly divergent. Therefore rigorously speaking, Wilson's
electrophoretic coefficient cannot be defined upon evaluation of $v_{x}\left(
0,0;\xi\right)  $ with preset values of $x=r=0$ unless we simply abandon the
divergent term. It now appears that his procedure of setting $x=r=0$ in the
velocity integrals before evaluating the integrals is the cause for the
divergence difficulty to obtain a finite electrophoretic coefficient, or the
position $x=r=0$ should not have been taken in the electrophoretic coefficient
defined through the thermodynamic force--flux relation for mobility or the
Stokes law. This divergence difficulty and our desire to obtain physically
sensible mobility coefficient was the principal motivation that we have
evaluated and examined the velocity profiles in the $\left(  x,r\right)  $
plane to understand how the velocity varies in space and to find out what
would be the most probable or reasonable velocity that should be used to
calculate ionic conductance if the Wilson--Onsager theory of conduction is
adopted as the theory to rely on. We believe that OW theory is a correct
approach to the ionic conduction problem, but the solutions must be evaluated
more carefully for a wider range of ($x,r$), because the ion atmosphere, and
therefore its center, migrates under the influence of an external field.

We now would like to show in what manner the Wilson formula for $f\left(
\xi\right)  $ would emerge from Eq. (\ref{fephc}). First of all, the
potentially divergent mechanical contribution $\left(  \widehat{\mathbf{v}%
}_{x}\right)  _{\text{me}}$ should be ignored to obtain a finite numerical
value for $\mathfrak{f}(x,r;\xi)$ at $x=r=0$, although neglecting $\left(
\widehat{\mathbf{v}}_{x}\right)  _{\text{me}}$ would result in a significant
error to the axial velocity, and integrals $\mathfrak{C}_{1}\left(
x,r,\xi\right)  $, $\cdots$, $\mathfrak{S}_{2}\left(  x,r,\xi\right)  $ should
be approximated as follows. First, let the Brownian motion contributions be
denoted by
\[
\widehat{\mathbf{v}}_{\text{Brown}}\left(  x,r,\xi\right)  =\mathfrak{C}%
_{1}\left(  x,r,\xi\right)  +2\mathfrak{C}_{2}\left(  x,r,\xi\right)
-\mathfrak{S}_{1}\left(  x,r,\xi\right)  -2\mathfrak{S}_{2}\left(
x,r,\xi\right)  .
\]
Then if the Bessel functions $I_{0}(z)$ in the integrals for $\mathfrak{C}%
_{1}\left(  x,r,\xi\right)  $, etc. are expanded in series%
\begin{subequations}
\begin{equation}
I_{0}(z)=\sum_{k=0}^{\infty}\frac{\left(  \frac{1}{4}z^{2}\right)  ^{k}%
}{k!\Gamma\left(  k+1\right)  }, \label{BI}%
\end{equation}
they can be evaluated analytically in closed form (at least, for quite a few
leading order terms) at $x=0$. Especially, at $x=0$ the zeroth-order term,
namely, the $k=0$ term in Eq. (\ref{BI}), gives rise to exactly the same
$f\left(  \xi\right)  $ as in Wilson's, Eq. (\ref{Wf}):%
\end{subequations}
\begin{equation}
\widehat{\mathbf{v}}_{\text{Brown}}\left(  0,0,\xi\right)  =-\frac{zeX\kappa
}{6\sqrt{2}\pi\eta_{0}}\mathfrak{f}_{\text{Brown}}(\xi), \label{BF}%
\end{equation}
where $\mathfrak{f}_{\text{Brown}}(\xi)$ is then given by the expression
\begin{align}
\mathfrak{f}_{\text{Brown}}(\xi)  &  =\frac{3}{2}\int_{0}^{\sqrt{2\left(
1+\xi^{2}\right)  }}dy\frac{\left(  1-y^{2}+\sqrt{1+2\xi^{2}y^{2}}\right)
}{1+2\xi^{2}y^{2}}+3\int_{0}^{1}dy\frac{2\xi^{2}y^{2}\left(  1-y^{2}\right)
}{1+2\xi^{2}y^{2}}\nonumber\\
&  -\frac{3}{2}\int_{0}^{\sqrt{2\left(  1+\xi^{2}\right)  }}dy\frac{\sqrt
{2}\xi y\left(  1-y^{2}+\sqrt{1+2\xi^{2}y^{2}}\right)  }{\left(  1+2\xi
^{2}y^{2}\right)  \left(  1+\sqrt{1+2\xi^{2}y^{2}}\right)  }-3\int_{0}%
^{1}dy\frac{\sqrt{2}\xi y\left(  1-y^{2}\right)  }{1+2\xi^{2}y^{2}},
\label{Bf}%
\end{align}
which can be shown identical with $f(\xi)$ in Eq. (\ref{Wf}). This process of
arriving at $f(\xi)$ from $\mathfrak{f}(x,r;\xi)$, as a matter of fact,
indicates that the electrophoretic coefficient $f(\xi)$ obtained by
Wilson\cite{wilson} must be regarded as an approximation to the more precisely
defined exact electrophoretic factor (or mobility coefficient) $\mathfrak{f}%
(x,r;\xi)$ through the mobility\cite{mazur} of ions on the basis of the
irreversible thermodynamic force--flux relation between the external electric
field and flow velocity. Recall that for $f(\xi)$ it is necessary to leave out
$\left(  \widehat{\mathbf{v}}_{x}\right)  _{\text{me}}$ from the axial
velocity $\mathbf{v}_{x}\left(  x,r,\xi\right)  $ in Eq. (\ref{vax}). It is of
course necessary also to leave out the field independent term---the first term
on the right in Eq. (\ref{vax})---for both $f(\xi)$ and $\mathfrak{f}%
(x,r;\xi)$ because the term has nothing to do with the mobility of ions in the
external electric field.

The axial velocity profiles presented in Eq. (\ref{vax}) arise from the
presence of ion atmosphere and its interaction with the center ion itself and
the external electric field. We must recognize that dynamics of ions in a
solution and their interactions with the external field is not like that of an
isolated single ion in the external field. Moreover, the center ion of the ion
atmosphere does not directly contribute to the ionic conduction because of the
countercurrent of the medium produced by the ion atmosphere, and the
electrophoretic coefficient must be appropriately calculated taking this fact
and the interaction of ion atmosphere with the external field into account.
Therefore the position coordinate values should be taken with those of a point
other than the coordinate origin, preferably, exterior to the curve $C_{p}$,
to calculate the electrophoretic coefficient because the center ion of the ion
atmosphere moves with $\xi$ increasing; see Fig. 6 and the rule for choosing
$(x,r)$ in $\mathfrak{f}(x,r;\xi)$ proposed. In this regard, recall that
$\mathfrak{f}(x,r;\xi)=0$ on $C_{p}$.

\subsection{\textbf{Fourier Transform Solution for the Transversal Velocity}}

By using the relation%
\begin{equation}
\left(  \mathbf{\nabla\times\nabla}\times\mathbf{a}\right)  _{\rho
}\mathbf{=\nabla}_{\rho}\left(  \operatorname{div}\mathbf{a}\right)
-\nabla^{2}\mathbf{a}_{\rho}=\mathbf{\nabla}_{\rho}\left(  \operatorname{div}%
\mathbf{a}\right)  \label{39}%
\end{equation}
in the case of $a_{\rho}=a_{\theta}=0$ we find the transversal velocity
component in the form
\begin{equation}
\widehat{\mathbf{v}}_{\rho}\left(  x,r,\xi\right)  =\frac{\xi}{2\sqrt{2}}%
J_{c}^{B}-\frac{1}{2}J_{s}^{B}-\frac{1}{\sqrt{2}\xi}J_{4}^{c}+J_{4}^{s}%
-\frac{1}{2}rJ_{5}^{s}, \label{rvr}%
\end{equation}
where $\widehat{\mathbf{v}}_{\rho}\left(  x,r,\xi\right)  $ is the reduced
transversal velocity defined by Eq. (\ref{Rv1}),
\begin{align}
J_{c}^{B}  &  =\int_{0}^{\infty}dt\frac{\cos\left(  tx\right)  }{\left(
1-2\xi^{2}t^{2}\right)  }\left[  \frac{t^{2}\omega_{1}K_{1}(\omega_{1}%
r)}{\left(  1+\sqrt{1-2\xi^{2}t^{2}}\right)  }\right. \nonumber\\
&  \qquad\qquad\qquad\qquad\qquad\left.  +\frac{t^{2}\omega_{1}K_{1}%
(\omega_{2}r)}{\left(  1-\sqrt{1-2\xi^{2}t^{2}}\right)  }-2t^{2}\omega
_{3}K_{1}(\omega_{3}r)\right]  ,\label{J1c}\\
J_{s}^{B}  &  =\int_{0}^{\infty}dt\frac{\sin\left(  tx\right)  }{\left(
1-2\xi^{2}t^{2}\right)  }\left[  t\omega_{1}K_{1}(\omega_{1}r)+t\omega
_{2}K_{1}(\omega_{2}r)-4\xi^{2}t^{3}\omega_{3}K_{1}(\omega_{3}r)\right]  ,
\label{J1s}%
\end{align}%
\begin{align}
J_{4}^{c}  &  =\int_{0}^{\infty}dt\cos\left(  tx\right)  tK_{1}%
(tr),\label{J4c}\\
J_{4}^{s}  &  =\int_{0}^{\infty}dt\sin\left(  tx\right)  t^{2}K_{1}%
(rt),\label{Js4}\\
J_{5}^{s}  &  =\int_{0}^{\infty}dt\sin\left(  tx\right)  tK_{0}(rt).
\label{Js5}%
\end{align}
The integrals $J_{4}^{c}$, $J_{4}^{s}$, and $J_{5}^{s}$ are analytically
evaluated by using the integral representations of the Bessel functions in the
same manner as for $K_{4}^{c}$, $K_{4}^{s}$, and $K_{5}^{c}$:
\begin{align}
J_{4}^{c}  &  =\frac{\pi r}{2\left(  x^{2}+r^{2}\right)  ^{\frac{3}{2}}%
},\label{J4ce}\\
J_{4}^{s}  &  =-\frac{3\pi x\left(  x^{4}-x^{2}r^{2}-r^{4}\right)  }%
{2r^{3}\left(  x^{2}+r^{2}\right)  ^{\frac{5}{2}}},\label{J4se}\\
J_{5}^{s}  &  =\frac{\pi x}{2\left(  x^{2}+r^{2}\right)  ^{3/2}}. \label{J5se}%
\end{align}
The integrals $J_{c}^{B}$ and $J_{s}^{B}$\ can be evaluated by using the
contour integration methods similarly for the integrals $K_{c}^{B}$ and
$K_{s}^{B}$. Jordan's lemma gives rise to the same conditions as Inequalities
(\ref{vc}).\ In the region outside the validity of Ineq. (\ref{vc}) the method
of principal value integration is numerically employed.

In summary, we obtain the transversal velocity component $\mathbf{v}_{\rho
}\left(  x,r,\xi\right)  $ in the form%
\begin{align}
\mathbf{v}_{\rho}\left(  x,r,\xi\right)   &  =-\frac{\kappa^{2}k_{B}T}%
{8\pi\eta_{0}}\frac{r}{\left(  x^{2}+r^{2}\right)  ^{\frac{3}{2}}}\nonumber\\
&  -\frac{\kappa zeX}{4\sqrt{2}\pi\eta_{0}}\left[  \frac{xr}{2\left(
x^{2}+r^{2}\right)  ^{3/2}}+\frac{3x\left(  x^{4}-x^{2}r^{2}-r^{4}\right)
}{r^{3}\left(  x^{2}+r^{2}\right)  ^{\frac{5}{2}}}\right] \nonumber\\
&  +\frac{\kappa zeX}{8\sqrt{2}\pi\eta_{0}}\left[  \mathfrak{C}_{3}\left(
x,r,\xi\right)  +\mathfrak{S}_{3}\left(  x,r,\xi\right)  +4\mathfrak{C}%
_{4}\left(  x,r,\xi\right)  -4\mathfrak{S}_{4}\left(  x,r,\xi\right)  \right]
, \label{S5}%
\end{align}
where%
\begin{align}
\mathfrak{C}_{3}\left(  x,r,\xi\right)   &  =\int_{0}^{\sqrt{2\left(
1+\xi^{2}\right)  }}dy\frac{y\sqrt{1-y^{2}+\sqrt{1+2\xi^{2}y^{2}}}}{\left(
1+2\xi^{2}y^{2}\right)  }e^{-xy}I_{1}(\overline{\omega}_{1}r),\label{C3}\\
\mathfrak{C}_{4}\left(  x,r,\xi\right)   &  =\int_{0}^{1}dy\frac{\xi^{2}%
y^{3}\sqrt{1-y^{2}}}{1+2\xi^{2}y^{2}}e^{-xy}I_{1}(\overline{\omega}%
_{3}r),\label{C4}\\
\mathfrak{S}_{3}\left(  x,r,\xi\right)   &  =\int_{0}^{\sqrt{2\left(
1+\xi^{2}\right)  }}dy\frac{\xi y^{2}\sqrt{1-y^{2}+\sqrt{1+2\xi^{2}y^{2}}}%
}{\sqrt{2}\left(  1+2\xi^{2}y^{2}\right)  \left(  1+\sqrt{1+2\xi^{2}y^{2}%
}\right)  }e^{-xy}I_{1}(\overline{\omega}_{1}r),\label{S3}\\
\mathfrak{S}_{4}\left(  x,r,\xi\right)   &  =\int_{0}^{1}dy\frac{\sqrt{2}%
\xi^{3}y^{2}\sqrt{1-y^{2}}}{1+2\xi^{2}y^{2}}e^{-xy}I_{1}(\overline{\omega}%
_{3}r). \label{S4}%
\end{align}
These integrals can be analytically evaluated term by term by using the series
expansion of the Bessel function $I_{1}(z)$ or numerically by a fairly
straightforward procedure. The profiles of $\mathbf{v}_{\rho}\left(
x,r,\xi\right)  $ look quite similar to those of the axial velocity
$\mathbf{v}_{x}\left(  x,r,\xi\right)  $ in Figs. 4--6.

\subsection{\textbf{Fourier Transform Solution for Pressure}}

Since for the present system the (nonequilibrium) pressure is given by%
\begin{equation}
p=p_{0}+\eta_{0}\mathbf{\nabla}^{2}\left(  \frac{\partial a_{x}}{\partial
x}\right)  , \label{fpa}%
\end{equation}
it is easy to calculate it from Eq. (\ref{32}):%
\begin{align}
p-p_{0}  &  =\frac{zeX}{4\pi^{2}}\int_{0}^{\infty}d\alpha\frac{\alpha
\sin\left(  \alpha x\right)  }{2R^{2}}\left[  \left(  1+R\right)
K_{0}(\lambda_{1}\rho)\right. \nonumber\\
&  \qquad\qquad\qquad\qquad\qquad\left.  +\left(  1-R\right)  K_{0}%
(\lambda_{2}\rho)-2\left(  1-R^{2}\right)  K_{0}(\lambda_{3}\rho)\right]
\nonumber\\
&  \quad-\frac{zeX}{4\pi^{2}}\int_{0}^{\infty}d\alpha\alpha\sin\left(  \alpha
x\right)  K_{0}(\alpha\rho)\nonumber\\
&  \quad-\frac{zeX}{4\pi^{2}}\xi\int_{0}^{\infty}d\alpha\frac{\alpha^{2}%
\cos\left(  \alpha x\right)  }{R^{2}}\left[  K_{0}(\lambda_{1}\rho
)+K_{0}(\lambda_{2}\rho)-2K_{0}(\lambda_{3}\rho)\right]  . \label{fp}%
\end{align}
The formula presented above represents a nonequilibrium part of pressure
$\Delta p=p-p_{0}$ consistent with the velocity components obtained as the
solution of the Stokes equation for a fluid in an external electric field. It
also can be decomposed into the mechanical and Brownian motion parts as for
the axial and transversal velocity components. They can be evaluated by the
same methods as for the axial velocity, for example. With the reduced
nonequilibrium pressure $\Delta\widehat{p}$ defined by the formula%
\begin{equation}
\Delta\widehat{p}=\Delta p\left(  \frac{zeX\kappa^{2}}{4\pi^{2}}\right)  ^{-1}
\label{rdp}%
\end{equation}
we obtain the nonequilibrium pressure profile in the form%
\begin{align}
\Delta\widehat{p}  &  =-\frac{\pi x}{2\left(  x^{2}+r^{2}\right)  ^{3/2}%
}\nonumber\\
&  \quad-\frac{\pi}{4}\int_{0}^{\sqrt{2\left(  1+\xi^{2}\right)  }}%
dy\frac{e^{-xy}y\left(  1+\sqrt{2}\xi y+\sqrt{1+2\xi^{2}y^{2}}\right)
}{1+2\xi^{2}y^{2}}I_{0}(\overline{\omega}_{1}r)\nonumber\\
&  \quad+\frac{\pi}{2}\int_{0}^{1}dy\frac{e^{-xy}\sqrt{2}\xi y^{2}\left(
1-\sqrt{2}\xi y\right)  }{1+2\xi^{2}y^{2}}I_{0}(\overline{\omega}_{3}r),
\label{dpfin}%
\end{align}
This shows that $\Delta\widehat{p}$ is also singular at the origin of the
coordinates. It is significant to observe that the nonequilibrium pressure
$\Delta p$ is generally negative, that is, there is a tension that becomes
negative infinite at the origin. This implies that the nonequilibrium pressure
is compressional in the neighborhood of the origin. Moreover, it is
proportional to the field strength $X$. It seems to be a remarkable result,
probably deserving a deeper consideration, because the degree of compression
can be manipulated by the applied external electric field strength. We will
report on a further study of this nonequilibrium pressure separately
elsewhere\cite{pressure}.

\subsection{Ionic Field and Relaxation Time Effect}

Just as the velocity is induced by the mean local body-force which in turn is
produced by interaction of the ion atmosphere\cite{debye} with the external
field, the local ionic force field is modified by a feedback process of
correlations arising from the Coulomb potentials and their interaction with
the external field. Thus we may express the total electric field $X_{t}$
acting on the ion in the $x$ direction as%
\begin{equation}
X_{t}=X+\Delta X, \label{If1}%
\end{equation}
where the local contribution $\Delta X$ is the ionic field produced by the
interaction of the ion atmosphere with the external force field. If the
potential of ion $j$ in the electrolyte solution is denoted by $\psi
_{j}\left(  \mathbf{r}\right)  $ the force arising from the potential
$\psi_{j}\left(  \mathbf{r}\right)  $ is given by
\begin{equation}
e_{j}\Delta\mathbf{X}\left(  \mathbf{r}\right)  =-\mathbf{\nabla}\psi
_{j}\left(  \mathbf{r}\right)  . \label{If2}%
\end{equation}
Upon using the solution for $\psi_{j}\left(  \mathbf{r}\right)  $, Eq.
(\ref{34p}), we obtain the mean local ionic force%
\begin{align}
e_{j}\Delta X\left(  \mathbf{r}\right)   &  =\mp\frac{2e_{j}\mu^{\prime}}{\pi
D\kappa^{2}}\int_{0}^{\infty}d\alpha\alpha\cos\left(  \alpha x\right)
\frac{\alpha}{R^{2}}\left[  K_{0}\left(  \lambda_{1}\rho\right)  +K_{0}\left(
\lambda_{2}\rho\right)  -2K_{0}\left(  \lambda_{3}\rho\right)  \right]
\nonumber\\
&  \quad\ +\frac{e_{j}^{2}}{\pi D}\int_{0}^{\infty}d\alpha\alpha\frac
{\sin\left(  \alpha x\right)  }{2R^{2}}\left[  \left(  1+R\right)
K_{0}\left(  \lambda_{1}\rho\right)  +\left(  1-R\right)  K_{0}\left(
\lambda_{2}\rho\right)  \right. \nonumber\\
&  \qquad\qquad\qquad\qquad\qquad\qquad\left.  -2\left(  1-R^{2}\right)
K_{0}\left(  \lambda_{3}\rho\right)  \right]  \label{Ifex}%
\end{align}
in the notation already defined. It should be noted that the formula for
$e_{j}\Delta X\left(  \mathbf{r}\right)  $ in Eq. (\ref{Ifex}) is an exact
result, although formal. The external field dependence $\xi$ enters the theory
in a nonlinear manner through the arguments of the Bessel functions. We have
shown that the Fourier transforms such as those in Eq. (\ref{Ifex}) can be
reduced to finite quadratures consisting of regular Bessel functions of zeroth
order of second kind $I_{0}(z)$ weighted by some algebraic functions.

The integrals in the expression for $\Delta X\left(  \mathbf{r}\right)  $ were
also evaluated by Wilson in his dissertation\cite{wilson}\ for the case of
$x=r=0$ in the same manner as for the electrophoretic coefficient $f(\xi)$,
Eq. (\ref{Wf}). With the so-obtained result the relaxation time coefficient
$g\left(  \xi\right)  $ was defined by the relation
\begin{equation}
ze\Delta X\left(  0,0;\xi\right)  =-\frac{e_{j}\mu^{\prime}\kappa}{2D}g\left(
\xi\right)  =-\frac{e_{j}\kappa^{2}\xi}{2D}g\left(  \xi\right)  , \label{If12}%
\end{equation}
for which he obtained $g(\xi)$ in a simple analytic form
\begin{equation}
g(\xi)=\frac{1}{2\xi^{3}}\left[  \xi\sqrt{\left(  1+\xi^{2}\right)  }%
-\tan^{-1}\left(  \frac{\xi}{\sqrt{1+\xi^{2}}}\right)  -\sqrt{2}\xi+\tan
^{-1}\left(  \sqrt{2}\xi\right)  \right]  . \label{If13}%
\end{equation}
It is a nonlinear but well-behaved function of the reduced field strength
$\xi$; its limiting values are $g(0)=\frac{1}{3}\left(  2-\sqrt{2}\right)  $
as $\xi\rightarrow0$ and $g(\infty)=0$ as $\xi\rightarrow\infty$. We will show
presently under what condition this result, Eq. (\ref{If13}), is recovered
from the exact formula for $e_{j}\Delta X\left(  \mathbf{r}\right)  $, Eq.
(\ref{Ifex}).

To obtain a complete formula for the local ionic field from Eq. (\ref{Ifex})
the Fourier transform integrals therein must be calculated without setting
$x=r=0$ before evaluating them. For this purpose we use the same contour
integration methods described in Appendix A as for the velocity formulas. We
thereby obtain $e_{j}\Delta X\left(  \mathbf{r}\right)  $ in the form%
\begin{align}
e_{j}\Delta X\left(  \mathbf{r}\right)   &  =-\frac{e_{j}\kappa^{2}\xi}%
{2\sqrt{2}D}\left[  \int_{0}^{\sqrt{2\left(  1+\xi^{2}\right)  }}%
dy\frac{e^{-xy}y^{2}I_{0}\left(  \overline{\omega}_{1}r\right)  }{1+2\xi
^{2}y^{2}}-2\int_{0}^{1}dy\frac{e^{-xy}y^{2}I_{0}\left(  \overline{\omega}%
_{3}r\right)  }{1+2\xi^{2}y^{2}}\right] \nonumber\\
&  \quad\quad-\frac{e_{j}^{2}\kappa^{2}}{4D}\left[  \int_{0}^{\sqrt{2\left(
1+\xi^{2}\right)  }}dy\frac{e^{-xy}y\left(  1+\sqrt{1+2\xi^{2}y^{2}}\right)
I_{0}\left(  \overline{\omega}_{1}r\right)  }{1+2\xi^{2}y^{2}}\right.
\nonumber\\
&  \quad\quad\qquad\qquad\left.  +4\xi^{2}\int_{0}^{1}dy\frac{y^{3}%
e^{-xy}I_{0}\left(  \overline{\omega}_{3}r\right)  }{1+2\xi^{2}y^{2}}\right]
, \label{Nex}%
\end{align}
The variables $x$ and $r$ as well as $\xi$ and $y$ appearing in the integrals
in Eq. (\ref{Nex}) are dimensionless reduced variables defined earlier. The
integrals in Eq. (\ref{Nex}), of course, are subject to conditions deduced
from conditions (\ref{vc}) related to the Jordan lemma\cite{whittaker} for the
contour integrals. Exterior to the region of $(x,r)$ satisfying Conditions
(\ref{vc}) the method of principal values for singular integrals is used to
numerically compute the Fourier transform integrals.

The first group of terms in Eq. (\ref{Nex}) descends from the cosine transform
terms in Eq. (\ref{Nex}) whereas the second group originates from the sine
transform terms. If $x$ and $r$ are set equal to zero in Eq. (\ref{Nex}) the
first group exactly gives rise to Wilson's result, Eq. (\ref{If13}), but the
integrals in the second group do not vanish even if $x=r=0$ is taken after
their evaluation, but give rise to a field-independent term $-e_{j}\kappa
^{2}/2D$ in the limits of $x\rightarrow0$ and $r\rightarrow0$, in addition to
$\xi$-dependent terms as shown below. However, if we took $x=0$ in the sine
transform integrals in Eq. (\ref{Ifex}) there would have been no contribution
from it at all. This example, once again, manifestly demonstrates a need for
caution to take in evaluating the Fourier transforms, especially, with regard
to setting $x=r=0$: We reiterate that \textit{the values obtained of the
integrals are different, depending on whether particular parameter values,
especially, }$x=r=0$\textit{, are taken before or after evaluation of the
integrals, or even depending on the order of taking the limits }%
$x\rightarrow0$ and $r\rightarrow0$\textit{; }we have seen a similar situation
in the previous section for velocity profiles. The fact that $x$ and $r$ must
be set equal to zero to obtain Wilson's formula for $g(\xi)$ also suggests
that his formula is an approximation to the full relaxation time factor
defined below, as is $f\left(  \xi\right)  $ an approximation to the full
$\mathfrak{f}\left(  x,r;\xi\right)  $ given in Eq. (\ref{fephc}).

To define the appropriate relaxation time factor we split $e_{j}\Delta
X\left(  \mathbf{r}\right)  $ in Eq. (\ref{Nex}) into two parts%
\begin{equation}
e_{j}\Delta X\left(  \mathbf{r}\right)  =-\frac{e_{j}\xi\kappa^{2}}{2D}%
g_{c}\left(  x,r;\xi\right)  -\frac{e_{j}^{2}\kappa^{2}}{2D}g_{s}(x,r;\xi)
\label{GC1}%
\end{equation}
with the definitions%
\begin{align}
g_{c}\left(  x,r;\xi\right)   &  =\frac{1}{\sqrt{2}}\int_{0}^{\sqrt{2\left(
1+\xi^{2}\right)  }}dy\frac{e^{-xy}y^{2}I_{0}\left(  \overline{\omega}%
_{1}r\right)  }{1+2\xi^{2}y^{2}}-\sqrt{2}\int_{0}^{1}dy\frac{e^{-xy}y^{2}%
I_{0}\left(  \overline{\omega}_{3}r\right)  }{1+2\xi^{2}y^{2}},\label{GC2}\\
g_{s}(x,r;\xi)  &  =\frac{1}{2}\int_{0}^{\sqrt{2\left(  1+\xi^{2}\right)  }%
}dy\frac{e^{-xy}y\left(  1+\sqrt{1+2\xi^{2}y^{2}}\right)  I_{0}\left(
\overline{\omega}_{1}r\right)  }{1+2\xi^{2}y^{2}}\nonumber\\
&  \qquad+2\xi^{2}\int_{0}^{1}dy\frac{y^{3}e^{-xy}I_{0}\left(  \overline
{\omega}_{3}r\right)  }{1+2\xi^{2}y^{2}}. \label{GC3}%
\end{align}
The reason for the splitting made above is that the term $g_{s}(x,r;\xi)$
tends to a value independent of $\xi$ as $\xi\rightarrow0$:%
\begin{equation}
g_{s}(x,r;0)=\int_{0}^{\sqrt{2}}dye^{-xy}yI_{0}\left(  r\sqrt{2-y^{2}}\right)
\neq0, \label{GC4}%
\end{equation}
which contributes a field-independent term to $e_{j}\Delta X\left(
\mathbf{r}\right)  $:%
\begin{equation}
-\frac{e_{j}^{2}\kappa^{2}}{2D}\int_{0}^{\sqrt{2}}dye^{-xy}yI_{0}\left(
r\sqrt{2-y^{2}}\right)  , \label{GC5}%
\end{equation}
and this contribution would have nothing to do with the relaxation of ion
atmosphere. Therefore it is useful to define the field-dependent part of
$g_{s}(x,r;\xi)$ by the expression%
\begin{align}
\Delta g_{s}(x,r;\xi)  &  =\frac{1}{\xi}\left[  \frac{1}{2}\int_{0}%
^{\sqrt{2\left(  1+\xi^{2}\right)  }}dy\frac{e^{-xy}y\left(  1+\sqrt
{1+2\xi^{2}y^{2}}\right)  I_{0}\left(  \overline{\omega}_{1}r\right)  }%
{1+2\xi^{2}y^{2}}\right. \nonumber\\
&  \qquad\quad\left.  -\int_{0}^{\sqrt{2}}dye^{-xy}yI_{0}\left(
r\sqrt{2-y^{2}}\right)  \right]  +2\xi\int_{0}^{1}dy\frac{y^{3}e^{-xy}%
I_{0}\left(  \overline{\omega}_{3}r\right)  }{1+2\xi^{2}y^{2}}. \label{GC6}%
\end{align}
With this we are now able to cast $e_{j}\Delta X\left(  \mathbf{r}\right)  $
into a more appropriate form%
\begin{equation}
e_{j}\Delta X\left(  \mathbf{r}\right)  =-\frac{e_{j}^{2}\kappa^{2}}{2D}%
g_{s}(x,r;0)-\frac{e_{j}\xi\kappa^{2}}{2D}\mathfrak{g}\left(  x,r;\xi\right)
, \label{GC7}%
\end{equation}
where the relaxation time factor $\mathfrak{g}\left(  x,r;\xi\right)  $ in the
cylindrical coordinate representation is defined by the formula%
\begin{equation}
\mathfrak{g}\left(  x,r;\xi\right)  =g_{c}\left(  x,r;\xi\right)  +\Delta
g_{s}(x,r;\xi). \label{GC8}%
\end{equation}
It is easily verifiable that $\Delta g_{s}(x,r;\xi)$ is indeed a constant in
the limit of $\xi=0$ and hence $\xi\Delta g_{s}(x,r;\xi)$ vanishes as
$\xi\rightarrow0$. With this definition of relaxation time factor
$\mathfrak{g}\left(  x,r;\xi\right)  $, we are now ready to examine its
relation to Wilson's $g(\xi)$ formula.

If both $x$ and $r$ are set equal to zero in integrals in Eqs. (\ref{GC2}),
(\ref{GC5}), and (\ref{GC6}), it follows
\begin{align}
g_{c}\left(  0,0;\xi\right)   &  =\frac{1}{\sqrt{2}}\int_{0}^{\sqrt{2\left(
1+\xi^{2}\right)  }}dy\frac{y^{2}}{1+2\xi^{2}y^{2}}-\sqrt{2}\int_{0}%
^{1}dy\frac{y^{2}}{1+2\xi^{2}y^{2}}\nonumber\\
&  =\frac{1}{4\xi^{3}}\left(  2\xi\sqrt{\xi^{2}+1}-\arctan2\xi\sqrt{\xi^{2}%
+1}-2\sqrt{2}\xi+2\arctan\sqrt{2}\xi\right)  ,\label{GC9}\\
\Delta g_{s}(0,0;\xi)  &  =\frac{1}{2\xi}\left(  \int_{0}^{\sqrt{2\left(
1+\xi^{2}\right)  }}dy\frac{y\left(  1+\sqrt{1+2\xi^{2}y^{2}}\right)  }%
{1+2\xi^{2}y^{2}}-2\right)  +2\xi\int_{0}^{1}dy\frac{y^{3}}{1+2\xi^{2}y^{2}%
}\nonumber\\
&  =\frac{1}{4\xi^{3}}\left\{  \left[  \ln\left(  2\xi^{2}+1\right)  +\left(
2\xi^{2}+1\right)  -1\right]  -4\xi^{2}\right\}  +\frac{1}{4\xi^{3}}\left[
2\xi^{2}-\ln\left(  2\xi^{2}+1\right)  \right] \nonumber\\
&  =0,\label{GC10}\\
g_{s}(0,0;0)  &  =1. \label{GC11}%
\end{align}
Here $g_{c}\left(  0,0;\xi\right)  $ is identical with Formula $g(\xi)$ in Eq.
(\ref{If13}) for the relaxation time coefficient in Wilson's
method\cite{wilson}, whereas $\Delta g_{s}(0,0;\xi)$ together with
$g_{s}(0,0;0)$ represents extra terms not present in his result. We reiterate
that in Wilson's method $g_{s}(0,0;\xi)$ does not appear because the
$\sin\left(  \alpha x\right)  $ term in the sine transform integral vanishes
if $x$ is set equal to zero before evaluating the sine transform integral.
This shows under what condition Wilson's $g(\xi)$ is recoverable from the
present result for $\mathfrak{g}\left(  x,r;\xi\right)  $.

In Fig. 8, $\mathfrak{g}\left(  x,r;\xi\right)  $ computed, for example, at
$x=0.5$ and $r=0.5$ is plotted as a function of the field strength $\xi$ and
compared with Wilson's relaxation time coefficient $g\left(  \xi\right)  $ in
Eq. (\ref{If13}), the dotted curve. To better comprehend the profile of the
ionic field graphically, we plot a 3D example of $\mathfrak{g}\left(
x,r;\xi\right)  $ in the case of $\xi=1$ in Fig. 9. A combination of formulas
(\ref{GC8})--(\ref{GC11}) and the method of principal values for integration
is used to compute the relaxation time factor $\mathfrak{g}\left(
x,r;\xi\right)  $ presented in Fig. 9. As does the electrophoretic factor, it
also exhibits a singular behavior near the origin, although the details are
different from the behavior of $\mathfrak{f}\left(  x,r;\xi\right)  $ in Fig.
5. It also vanishes as $x$ and $r$ increase to infinity.

We give a short summary of this long section: The results of evaluation of
$\mathbf{v}_{x}\left(  x,r,\xi\right)  $, $\mathbf{v}_{\rho}\left(
x,r,\xi\right)  $, and $\Delta p\left(  x,r,\xi\right)  $ for all values of
$x$ and $r$ and the related electrophoretic and relaxation time factors
constitute some of the important contributions of this work to the
hydrodynamics of strong binary electrolyte solutions in the external electric
field. On extensively studying the velocity profiles we have been able to
formulate a rule for selecting the position variable ($x,r$) in the
electrophoretic and relaxation time coefficients, which are finite everywhere.
By using this rule and the profiles of velocities and nonequilibrium pressure
as well as the distribution functions and mean potentials calculated, we will
also be able to predict or deduce, in a well-defined manner, hydrodynamic
consequences to transport properties, such as conductivity, and related
nonequilibrium properties of ionic motions in the medium in an external
electric field of arbitrary strength.

\section{Discussion and \textbf{Concluding Remarks}}

In this paper, we have shown that since ions interact with each other through
long-range Coulombic interactions, ion atmosphere with ions, and both of them
with the external electric field, the correlations of particles in the ionic
liquids are quite complex and the whole body of an ionic solution collectively
and cooperatively moves subjected to the external electric field.
Consequently, even at a dilute ionic concentration the macroscopic behavior of
electrolyte solutions under an external electric field is not simple, but
rather complex and, therefore, exhibits an interesting feedback system. In
this regard, the subject matter is interesting from not only the theoretical,
but also practical standpoint to gain insights into the behavior of complex
liquids. For this reason, we believe the ideas of the OW
theory\cite{onsager,fuoss,wilson} as a theory designed to treat ionic fluid
systems in the external field are worth studying in depth for the insights
they provide for dynamical theories of ionic matter in general. However,
examining in detail the solutions of the Stokes equation for flow velocity
obtained from the solutions of the governing equations in the OW theory, we
find that the velocity formula not only had a divergence difficulty that we
have unexpectedly encountered while studying it, but also was incompletely
treated mathematically in Wilson's work\cite{wilson} because only the behavior
of the center ion at the coordinate origin was examined despite the fact that
the ion atmosphere moves in the external electric field and develops a
non-simple spatial structure. Therefore, we felt that there still remained the
task of fully implementing the theory in a mathematically satisfactory manner
to make it serve as a complete theory of ion conductivity in the nonlinear
regime of external field dependence.

To achieve the goal in mind, we have numerically studied the velocity profiles
in the configuration space over a range of external field strength and, in
particular, the movement and distortion of the ion atmosphere, as the external
electric field strength is continuously varied over a wide range. Thereby we
have numerically quantified the trajectory of the center ion of the ion
atmosphere with respect to $\xi$, but also studied the manner of its
distortion from a spherical form to a quasi-elliptical form, as the field
strength $\xi$ is varied. The general picture we obtain of the electrophoretic
factor\textit{ }$\mathfrak{f}\left(  x,r;\xi\right)  $ is as follows: within
curve $C_{P}$ it is negative whereas outside $C_{P}$ it is positive. Moreover,
$C_{P}$ is non-spherical. This implies that the ion atmosphere not only
polarizes into a negative and a positive domain (typical of a dipolar
distribution), but also the boundary curve (i.e., $C_{P}$) gets distorted to a
quasi-ellipse from a spherically symmetric form, as $\xi$ increases from zero.
On the basis of the body of numerical studies of the axial velocity profiles
we have been able to formulate a procedure by which it is sufficient to
calculate the center position of the moving ion atmosphere at every value of
$\xi$ and therewith calculate the electrophoretic and relaxation time
coefficients as functions of $\xi$.

\textit{The identified procedure is that: the electrophoretic coefficient at a
value of }$\xi$\textit{ is given by the electrophoretic factor }%
$\mathfrak{f}\left(  x,r;\xi\right)  $\textit{ evaluated at the center of the
displaced spherical ion atmosphere of radius }$x_{c}$\textit{, whose center is
located at }$\left(  x,r\right)  =\left(  x_{c},0\right)  $\textit{, the
center of the displaced quasi-elliptic curve }$C_{P}$\textit{ that is the
locus of }$\mathfrak{f}\left(  x,r;\xi\right)  =0$\textit{. Since the
spherical ion atmosphere with its center at }$\left(  x_{c},0\right)
$\textit{ has a radius }$x_{c}$\textit{, the value of }$r_{c}$\textit{ must be
equal to }$x_{c}$\textit{. Therefore the electrophoretic coefficient
}$\mathfrak{f}\left(  \xi\right)  $\textit{ is given by }$\mathfrak{f}\left(
\xi\right)  \equiv\mathfrak{f}\left(  x_{c},r_{c};\xi\right)  =\mathfrak{f}%
\left(  x_{c},x_{c};\xi\right)  $\textit{ according to this finding. Since the
center position of quasi-elliptic curve }$C_{P}$\textit{ is unique for every
}$\xi$\textit{, the electrophoretic coefficient }$\mathfrak{f}\left(
\xi\right)  $\textit{ defined is unique. The relaxation time coefficient
}$\mathfrak{g}\left(  \xi\right)  $\textit{ is then calculated by
}$\mathfrak{g}\left(  \xi\right)  \equiv\mathfrak{g}\left(  x_{c},r_{c}%
;\xi\right)  =\mathfrak{g}\left(  x_{c},x_{c};\xi\right)  $\textit{ to be
consistent with the electrophoretic coefficient defined.}

This behavior (trajectory) of the center of ion atmosphere gives rise to
non-divergent electrophoretic coefficients for all field strengths and hence
the ionic conductance based on the Fokker--Planck equations employed is now
rendered divergence-free. This is made possible by recognizing that the
electrophoretic coefficient must be calculated for the moving ionic atmosphere
with the center of the displaced ion atmosphere at $\left(  x_{c},0\right)  $
when the external electric field is applied to the system.

The set of values for $x$ and $r$\ obtained to use for $\mathfrak{f}\left(
x,r;\xi\right)  $ and $\mathfrak{g}\left(  x,r;\xi\right)  $ is $\left(
x_{c},r_{c}\right)  =\left(  x_{c},x_{c}\right)  $ which is in significant
contrast to the values $x=r=0$ taken by Wilson to evaluate the integrals that
gives rise to a divergence difficulty. In the companion paper\cite{eurah2}, we
apply this identification of $x$ and $r$ to compute the electrophoretic and
relaxation time coefficients, and calculate therewith the equivalent ionic
conductance and, in particular, the Wien effect of a binary electrolyte
solution\ in comparison with experimental data available.

The velocity profiles graphically presented also suggest a skin effect by
which the mobility of ions in solution is predominantly contributed by ions
outside the curve $C_{P}$. We have not suspected the existence of this effect
before: that the conduction currents are mostly carried by ions in the
periphery---i.e., in the shells of radius of $O\left(  \kappa^{-1}\right)
$---of ion atmosphere, but not by the center ions, as is obvious from Fig.
4--Fig. 6.

The another important mathematical question we are answering in the present
work is that variable parameters, such as $x$, $r$, and $\xi$, in the Fourier
transform solutions of the OF equations, Poisson equations, and Stokes
equation should not be set equal to zero before fully evaluating them, since
the results so obtained do not generally yield the same results as those
obtained by setting them equal to zero after their complete evaluation. They
would give identical results only if the results of the integrals are analytic
everywhere in the space of $x$, $r$, and $\xi$, but the examples we have
studied definitely show that the evaluated results are not necessarily
analytic everywhere in the aforementioned space, and as a consequence the
results of evaluations by the aforementioned two different modes can be
significantly different; that is, the results are not uniformly convergent to
the same conclusion. This should be regarded as a significant point of the
present analysis to keep in mind in the study of this line of theories for
ionic solutions.

What we have shown in this work are the exact velocity and pressure profiles
in space in a Brownian motion model, which we may apply to study other
irreversible phenomena in the binary electrolyte solutions in the electric
field than the Wien effect. Being full exact solutions without an
approximation within the framework of the Brownian motion model, not only\ do
they, at least in the low density regime, promise to provide a more complete
picture of conduction phenomena, but also the insights gained therefrom should
also help us develop theories of related transport phenomena in
systems\cite{castner,ionic,daily,shabanov,friedman} of current interest in
science and engineering, such as plasmas\cite{ting,plasma},
semiconductors\cite{nag,landsberg,benabdallah}, small systems\cite{micro-nano}%
, etc. in electromagnetic fields. In any case, they represent new results in
the subject field. In the sequels\cite{asymmetric} we will also study
asymmetric electrolyte solutions, in which charges of the cation and anion are
not symmetric, and ionic conductance under an external electric field.
Although more complicated than the present symmetric binary electrolyte
solutions, we find that a similar mathematical analysis is possible to obtain
for them. The results of the mathematical solutions will be reported in the
near future\cite{asymmetric}, together with their numerical
results\cite{asymmetric-2} in comparison with experimental data.

\bigskip

\textbf{Acknowledgment}

\textit{The present work has been supported in part by the Discovery grants
from the Natural Sciences and Engineering Research Council of Canada.}

\appendix{}

\section{Contour Integration Method}

Some of the integrals in the formulas for the distribution functions,
potentials, velocities, and pressure can be evaluated by applying the method
of contour integration which yields formulas more readily amenable to analysis
and further approximations giving rise to simple results which will make them
possible to use for the purpose of assessing the existing results on the
subject matter.

\subsection{Axial Velocity}

We consider the axial velocity first for the reason that it contains more
experimentally direct features than the nonequilibrium structure and
potentials. The integrals appearing in the formal Fourier transform solutions
in the present theory all involve the Bessel functions $K_{0}(\lambda_{l}%
\rho)$ $\left(  l=1,2,3\right)  $ of argument $\lambda_{l}\rho$ with
$\lambda_{l}$ being relatively complicated functions of the integration
variable, the wave number; see Eq. (\ref{32a}) and Eq. (\ref{32b}) and also
Eq. (\ref{32c}) for $\overline{\omega}_{l}$. In reduced variables we have
defined for the analysis they have the mathematical properties listed below.

\begin{enumerate}
\item[(1)] The zeros of the arguments of the Bessel function $K_{\nu}%
(\omega_{l}r)$ for $r\neq0$ are found to be :%
\begin{align}
t  &  =\pm i\sqrt{2\left(  1+\xi^{2}\right)  }\quad\text{for }\omega
_{1},\label{A46a}\\
t  &  =0\quad\text{for }\omega_{2},\label{B46b}\\
t  &  =\pm i\frac{\kappa}{\sqrt{2}}\quad\text{for }\omega_{3}. \label{A46c}%
\end{align}
The argument of $K_{\nu}(\omega_{1}r)$ therefore has branch points at $tr=\pm
i\sqrt{2\left(  1+\xi^{2}\right)  }r$, whereas the argument of $K_{\nu}%
(\omega_{2}r)$ has branch points at $tr=0\times r$ and $-\infty\times r$ and
the argument of $K_{\nu}(\omega_{3}r)$ branch points at $tr=\pm ir$.\newline
Thus we may insert a branch cut on the imaginary axis of complex $t$ plane
between $t=i\sqrt{2\left(  1+\xi^{2}\right)  }$ and $t=-i\sqrt{2\left(
1+\xi^{2}\right)  }$ for the integral of $K_{\nu}(\omega_{1}r)$, while a
branch cut may be inserted along the negative real axis for the integral of
$K_{\nu}(\omega_{2}r)$, and on the imaginary axis between $t=i$ and $t=-i$ for
the integral of $K_{\nu}(\omega_{3}r)$, respectively. See Figs 11--13 below.

\item[(2)] We recall that Bessel function $K_{\nu}(z)$ of complex variable $z$
is regular in $z$ plane cut along the negative real
axis\cite{watson,abramowitz}. That is, it is a multi-valued function in the
cut plane. Therefore, in the present case, $K_{0}(\omega_{1}r)$ changes
discontinuously as the branch cut $\left[  -ir\sqrt{2\left(  1+\xi^{2}\right)
},ir\sqrt{2\left(  1+\xi^{2}\right)  }\right]  $ is crossed ($r$ is a fixed
parameter), whereas $K_{\nu}(\omega_{2}r)$ changes discontinuously as the
negative real axis is crossed, and $K_{\nu}(\omega_{3}r)$ changes
discontinuously as the branch cut $\left[  -ir,ir\right]  $ is crossed on the
imaginary axis of $t$ plane. Note that the Bessel functions $K_{\nu}%
(\omega_{2}r)$ and $K_{\nu}\left(  tr\right)  $ are defined in $t$ plane cut
along the negative real axis.

\item[(3)] We also observe that all the integrands of the singular integrals,
for example, in $K_{c}^{B}$ and $K_{s}^{B}$ in Eqs. (\ref{KcB}) and
(\ref{KsB}) are even with respect to $t$.

\item[(4)] Moreover, for $0<\arg t<\pi$ we find%
\begin{equation}
-\left(  \frac{\pi}{2}-\delta\right)  <\arg\omega_{l}<\left(  \frac{\pi}%
{2}-\delta\right)  \quad\left(  \frac{\pi}{2}>\delta>0;\;l=1,2,3\right)  .
\label{47}%
\end{equation}
Therefore in the upper half of complex $t$ plane
\begin{equation}
\lim_{\left\vert t\right\vert \rightarrow\infty}K_{\nu}(\omega_{l}%
r)=\lim_{\left\vert t\right\vert \rightarrow\infty}\sqrt{\frac{\pi}%
{2\omega_{l}r}}e^{-\omega_{l}r}\rightarrow0. \label{48}%
\end{equation}

\item[(5)] Lastly, all the integrands in Eqs. (\ref{KcB})--(\ref{KsB}) have
simple poles on the real axis at
\begin{equation}
t=\pm\frac{1}{\sqrt{2}\xi}. \label{49}%
\end{equation}
There is also a branch cut between $t=-\frac{1}{\sqrt{2}\xi}$ and $t=\frac
{1}{\sqrt{2}\xi}$ because of the $\left(  1\pm R\right)  $ factor in the
integrals, but this particular branch cut associated with $\sqrt{1-2\xi
^{2}t^{2}}$ does not play a role in the contour integrals considered in the
present work, because the real axis is not crossed by the contours in
performing integrations. Therefore we may ignore this particular branch
cut.\medskip
\end{enumerate}

All these properties (1)--(5) together suggest it is possible to evaluate the
integrals by using methods of contour integration\cite{whittaker} along the
closed contours of a semicircle as depicted in Figs. 8--10. However, in this
approach the results obtained would not cover the entire region of the upper
positive quadrant of plane $(x,r)$. In the region outside the domain defined
by the inequalities, Ineq. (\ref{vc}) the Fourier transform integrals must be
computed numerically because it is the region where Jordan's
lemma\cite{whittaker} is violated; that is, the contour integral along the
circle $C_{\infty}$ does not vanish. In the exterior region their numerical
values are small and hence of no importance. The practical advantage of this
kind of contour integration method of evaluation is to isolate out the major
part of contributions to the integrals and discuss the connection with the
existing results where possible and with experimental data. It would be
convenient to decompose $K_{c}^{B}$ and $K_{s}^{B}$ into component integrals
as follows:%
\begin{align}
K_{c}^{B}\left(  x,r,\xi\right)   &  =K_{1}^{c}+K_{2}^{c}-4\xi^{2}K_{3}%
^{c},\label{AKcB}\\
K_{s}^{B}\left(  x,r,\xi\right)   &  =K_{1}^{s}+K_{2}^{s}-2K_{3}^{s},
\label{AKsB}%
\end{align}
where $K_{l}^{c}$ and $K_{l}^{s}$ are in the order of their appearance in Eqs.
(\ref{KcB}) and (\ref{KsB}).

Since methods of integration will be similar for the integrals involved in
$K_{l}^{c}$ and $K_{l}^{s}$ $\left(  l=1,2,3\right)  $ we will illustrate them
with the examples of integrals in $K_{1}^{c}$ and $K_{1}^{s}$ in the
following. The results for the rest of integrals can be similarly obtained.

\subsubsection{Contour integrations of $K_{1}^{c}$ and $K_{1}^{s}$}

As prototypes of contour integrals appearing in the axial velocity formula,
integrals $K_{1}^{c}$ and $K_{1}^{s}$ are explicitly evaluated below; see Eq.
(\ref{Kvx})--Eq. (\ref{Ks4}). Integrals $K_{1}^{c}$ and $K_{1}^{s}$ both have
simple poles at $t=\pm\left(  \sqrt{2}\xi\right)  ^{-1}$. There is a branch
cut along the imaginary axis between $t=-i\sqrt{2\left(  1+\xi^{2}\right)  }$
and $+i\sqrt{2\left(  1+\xi^{2}\right)  }$ and also a branch cut on the real
axis between $t=-1/\sqrt{2}\xi$ and $t=+1/\sqrt{2}\xi$, but the latter branch
cut plays no role in integration since the path of integration stays above the
cut. For this reason the latter branch cut is not shown in Figs. 11--13. For
evaluation of both $K_{1}^{c}$ and $K_{1}^{s}$ the contour in Fig. 12 is used.

Consider the contour integral denoted by $\mathcal{C}_{1}K_{1}^{c}$ along the
contour $\mathcal{C}_{1}$ in complex plane $z$ depicted in Fig. 9:%
\begin{equation}
\mathcal{C}_{1}K_{1}^{c}\equiv\int_{\mathcal{C}_{1}}dze^{ixz}\frac{\omega
_{1}^{2}}{1-2\xi^{2}z^{2}}K_{0}(\omega_{1}r), \label{A2}%
\end{equation}
where
\begin{equation}
\omega_{1}=\left[  1+z^{2}+\sqrt{1-2\xi^{2}z^{2}}\right]  ^{1/2}. \label{A3}%
\end{equation}
Since there is no singularity enclosed by contour $\mathcal{C}_{1}$, this
contour integral $\mathcal{C}_{1}K_{1}^{c}$ is clearly equal to zero. Integral
$\mathcal{C}_{1}K_{1}^{c}$ can be decomposed into integrals along the paths
$C_{-}$, $C_{+}$, $C$, $C_{\infty}$, and along the real axis $t$ in
$\mathcal{C}_{1}$. We thus may write it as%
\begin{align}
\mathcal{C}_{1}K_{1}^{c}  &  =\int_{-\infty}^{\infty}dt\frac{\left(
1+t^{2}+\sqrt{1-2\xi^{2}t^{2}}\right)  }{1-2\xi^{2}t^{2}}e^{ixt}K_{0}%
(\omega_{1}\left(  t\right)  r)\nonumber\\
&  \quad+\int_{C_{-}}dz\frac{\left(  1+z^{2}+\sqrt{1-2\xi^{2}z^{2}}\right)
}{1-2\xi^{2}z^{2}}e^{ixz}K_{0}(\omega_{1}\left(  z\right)  r)\nonumber\\
&  \quad+\int_{C_{+}}dz\frac{\left(  1+z^{2}+\sqrt{1-2\xi^{2}z^{2}}\right)
}{1-2\xi^{2}z^{2}}e^{ixz}K_{0}(\omega_{1}\left(  z\right)  r)\nonumber\\
&  \quad+\int_{C}dz\frac{\left(  1+z^{2}+\sqrt{1-2\xi^{2}z^{2}}\right)
}{1-2\xi^{2}z^{2}}e^{ixz}K_{0}(\omega_{1}\left(  z\right)  r)\nonumber\\
&  \quad+\int_{C_{\infty}}dz\frac{\left(  1+z^{2}+\sqrt{1-2\xi^{2}z^{2}%
}\right)  }{1-2\xi^{2}z^{2}}e^{ixz}K_{0}(\omega_{1}\left(  z\right)
r)\nonumber\\
&  =0. \label{A4}%
\end{align}
The first integral on the right, the integral along the real axis, can be
shown to be equal to $2K_{1}^{c}$:%
\begin{align*}
\int_{-\infty}^{\infty}dt  &  \frac{\left(  1+t^{2}+\sqrt{1-2\xi^{2}t^{2}%
}\right)  }{1-2\xi^{2}t^{2}}e^{ixt}K_{0}(\omega_{1}\left(  t\right)  r)\\
&  =2\int_{0}^{\infty}dt\cos\left(  t\overline{x}\right)  \frac{\omega_{1}%
^{2}}{\left(  1-2\xi^{2}t^{2}\right)  }K_{0}(\omega_{1}r)\\
&  =2K_{1}^{c}.
\end{align*}
The remaining integrals along contours $C_{-}$, $C_{+}$, $C$, $C_{\infty}$
will be denoted by $C_{-}K_{1}^{c}$, $C_{+}K_{1}^{c}$, $CK_{1}^{c}$,
$C_{o}K_{1}^{c}$, respectively. By the theorem of residues\cite{whittaker} the
integral $C_{-}K_{1}^{c}$ gives $\pi i$ times the residue of $C_{-}K_{1}^{c}$
at $t=-1/\sqrt{2}\xi$:%
\[
C_{-}K_{1}^{c}=i\frac{\sqrt{2}\pi}{8\xi^{3}}\left(  2\xi^{2}+1\right)
e^{-ix/\sqrt{2}\xi}K_{0}(\overline{\omega}_{1}r),
\]
where%
\begin{equation}
\overline{\omega}=\frac{\sqrt{1+2\xi^{2}}}{\sqrt{2}\xi}. \label{A6}%
\end{equation}
Similarly, we obtain%
\[
C_{+}K_{1}^{c}=-\pi i\frac{\sqrt{2}}{8\xi^{3}}\left(  2\xi^{2}+1\right)
e^{ix/\sqrt{2}\xi}K_{0}(\overline{\omega}_{1}r).
\]
Thus combining the results for $C_{-}K_{1}^{c}$ and $C_{+}K_{1}^{c}$, we
obtain%
\begin{equation}
C_{-}K_{1}^{c}+C_{+}K_{1}^{c}=\frac{\sqrt{2}\pi\left(  1+2\xi^{2}\right)
}{4\xi^{3}}\sin\left(  \frac{x}{\sqrt{2}\xi}\right)  K_{0}\left(
\overline{\omega}r\right)  . \label{A5}%
\end{equation}

To transform the contour integral $CK_{1}^{c}$ around the branch cut along the
imaginary axis we observe that if the phase of the argument of $K_{0}\left(
\omega_{1}r\right)  $ on the right hand lip of the cut is chosen equal to
zero, the phase of the argument on the left hand lip is $\pi i$, so that the
argument has the form $e^{\pi i}\omega_{1}r$ for the Bessel function on the
left side of contour $C$. In this connection, it must be recalled that only
the relative phase across the branch cut is of importance. When traced along
$C$ from the left to the right side of the cut, the Bessel function must be
continued from the left side of the cut to the right side by the following
continuation formula\cite{watson,abramowitz}%
\begin{equation}
K_{0}\left(  e^{i\pi}z\right)  =K_{0}\left(  z\right)  -\pi iI_{0}\left(
z\right)  , \label{A7}%
\end{equation}
where $I_{0}\left(  z\right)  $ is the regular solution for the second kind of
the Bessel function of order $0$; $K_{0}(z)$ is irregular in contrast to
$I_{0}\left(  z\right)  $. The irregular Bessel function $K_{\nu}(z)$ $\left(
\nu\geq0\right)  $ diverges logarithmically as $z\rightarrow0$, whereas in
series representation the Bessel function $I_{0}\left(  z\right)  $ is regular
and given by the formula\cite{watson,abramowitz}%
\begin{equation}
I_{0}\left(  z\right)  =\sum_{m=0}^{\infty}\frac{\left(  \frac{1}{2}z\right)
^{2m}}{\left(  m!\right)  ^{2}}. \label{A8}%
\end{equation}
This function is finite at $z=0$, but it behaves asymptotically as%
\begin{equation}
I_{0}\left(  z\right)  \sim\left(  2\pi z\right)  ^{-1/2}e^{z}\left[
1+O(z^{-1})\right]  \quad\left(  \left\vert \arg z\right\vert <\frac{1}{2}%
\pi\right)  . \label{A8a}%
\end{equation}
Using formula (\ref{A7}) and changing variable from $iy$ to $y$, we obtain the
integral along contour $C$:%
\begin{equation}
CK_{1}^{c}=\pi\int_{0}^{\sqrt{2\left(  1+\xi^{2}\right)  }}dy\frac
{e^{-xy}\left(  1-y^{2}+\sqrt{1+2\xi^{2}y^{2}}\right)  }{1+2\xi^{2}y^{2}}%
I_{0}\left(  \overline{\omega}_{1}r\right)  , \label{A9}%
\end{equation}
where%
\begin{equation}
\overline{\omega}_{1}=\left[  1-y^{2}+\sqrt{1+2\xi^{2}y^{2}}\right]  ^{1/2}.
\label{A9a}%
\end{equation}
If the series form for $I_{0}(z)$ in Eq. (\ref{A8}) is used, $CK_{1}^{c}$ can
be computed in terms of quadratures of elementary functions---in fact,
incomplete Laplace transforms. It should be recalled that this integral
(\ref{A9}) is subject to the condition (\ref{vc}) for the relation of $x$ to
$r$ that is deducible from the Jordan lemma\cite{whittaker}. To satisfy this
lemma the integrands of integrals in $K_{c}^{B}$ and $K_{s}^{B}$ must satisfy
the condition
\begin{equation}
x\operatorname{Im}t+r\operatorname{Re}\omega_{k}>0\quad(k=1,2,3) \label{A9J}%
\end{equation}
for $x,r>0$. Thus values of $x$ and $r$ in the $(x,r)$ are limited to the
region satisfying this condition plane, assuring the contour integrals along
the curve $C_{\infty}$ vanishes as $x$ and $r$ tend to infinity. In the case
of integral $K_{1}^{c}$ the condition implies the inequality
\begin{equation}
x\sqrt{1+\xi^{2}}>r, \label{A49c}%
\end{equation}
which in fact assures that the integral vanishes as $x$ and $r$ tend to
infinity. Outside this region the contour integration method is not
applicable. Therefore, integral (\ref{A9a}) does not hold and the Fourier
transform integral $K_{1}^{c}$ must be evaluated numerically. However, the
numerical values of the integral in the exterior region gets diminishingly
smaller as $x$ and $r$ increase to infinity. The contour integral along the
outer semicircle $C_{\infty}$ does vanish in the region satisfying Jordan's lemma.

Collecting the results for the contour integrals obtained above into Eq.
(\ref{A4}), we obtain the integral $K_{1}^{c}$ in the form:%
\begin{align}
K_{1}^{c}  &  =-\frac{\sqrt{2}\pi\left(  1+2\xi^{2}\right)  }{8\xi^{3}}%
\sin\left(  \frac{x}{\sqrt{2}\xi}\right)  K_{0}(\overline{\omega}r)\nonumber\\
&  \quad-\frac{\pi}{2}\int_{0}^{\sqrt{2\left(  1+\xi^{2}\right)  }}%
dy\frac{e^{-xy}\left(  1-y^{2}+\sqrt{1+2\xi^{2}y^{2}}\right)  }{1+2\xi
^{2}y^{2}}I_{0}(\overline{\omega}_{1}r). \label{50c}%
\end{align}

The procedure of evaluating integrals $K_{1}^{s}$ with the contour in Fig. 13
is entirely parallel to the one presented above for $K_{1}^{c}$. The result
for the integral $K_{1}^{s}$ is
\begin{align}
K_{1}^{s}  &  =\frac{\pi\left(  1+2\xi^{2}\right)  }{8\xi^{4}}\cos\left(
xt\right)  K_{0}(\overline{\omega}r)\nonumber\\
&  \quad-\frac{\pi}{2}\int_{0}^{\sqrt{2\left(  1+\xi^{2}\right)  }}%
dy\frac{e^{-xy}y\left(  1-y^{2}+\sqrt{1+2\xi^{2}y^{2}}\right)  }{\left(
1+2\xi^{2}y^{2}\right)  \left(  1+\sqrt{1+2\xi^{2}y^{2}}\right)  }%
I_{0}(\overline{\omega}_{1}r). \label{50s}%
\end{align}
This integral also is subject to condition (\ref{A49c}) and in the exterior
region the Fourier transform integral must be evaluated numerically.
Therefore, collecting results for $K_{1}^{c}$ and $K_{1}^{s}$, we obtain
\begin{align}
\frac{1}{2}K_{1}^{c}+\frac{\xi}{\sqrt{2}}K_{1}^{s}  &  =-\frac{\pi\left(
1+2\xi^{2}\right)  }{8\sqrt{2}\xi^{3}}\left[  \sin\left(  \frac{x}{\sqrt{2}%
\xi}\right)  -\cos\left(  xt\right)  K_{0}(\overline{\omega}r)\right]
K_{0}(\overline{\omega}r)\nonumber\\
&  \quad-\frac{\pi}{4}\int_{0}^{\sqrt{2\left(  1+\xi^{2}\right)  }}%
dy\frac{e^{-xy}\left(  1-y^{2}+\sqrt{1+2\xi^{2}y^{2}}\right)  }{1+2\xi
^{2}y^{2}}\times\nonumber\\
&  \qquad\qquad\qquad\left[  1-\frac{\sqrt{2}\xi y}{\left(  1+\sqrt{1+2\xi
^{2}y^{2}}\right)  }\right]  I_{0}(\overline{\omega}_{1}r). \label{51c}%
\end{align}
The first line involving sine and cosine functions in this result will be seen
canceled by similar terms in the results for integrals $K_{2}^{c}$\textbf{,
}$K_{2}^{s}$, $K_{3}^{c}$\textbf{, }and $K_{3}^{s}$.

\subsubsection{$K_{2}^{c}$\textbf{ and }$K_{2}^{s}$}

Evaluation of these integrals proceeds similarly to that of $K_{1}^{c}$ and
$K_{1}^{s}$ with the contour given in Fig. 12 except that since the integrand
does not have a branch cut on the imaginary axis, there is no integral along
the imaginary axis. There are only contributions from the residues at the
singularities. They give rise to the following results:%
\begin{align}
K_{2}^{c}  &  =-\frac{\sqrt{2}\pi\left(  1+2\xi^{2}\right)  }{8\xi^{3}}%
\sin\left(  \frac{x}{\sqrt{2}\xi}\right)  K_{0}(\overline{\omega
}r),\label{K2c}\\
K_{2}^{s}  &  =\frac{\pi\left(  1+2\xi^{2}\right)  }{8\xi^{4}}\cos\left(
\frac{x}{\sqrt{2}\xi}\right)  K_{0}(\overline{\omega}r). \label{K2s}%
\end{align}
Therefore we obtain%
\begin{equation}
\frac{1}{2}K_{2}^{c}+\frac{\xi}{\sqrt{2}}K_{2}^{s}=-\frac{\pi\left(
1+2\xi^{2}\right)  }{8\sqrt{2}\xi^{3}}\left[  \sin\left(  \frac{x}{\sqrt{2}%
\xi}\right)  -\cos\left(  \frac{x}{\sqrt{2}\xi}\right)  \right]
K_{0}(\overline{\omega}r). \label{K2cs}%
\end{equation}

\subsubsection{$K_{3}^{c}$\textbf{ and }$K_{3}^{s}$}

In the present cases, the integrands involve a branch cut along the imaginary
axis from $z=-i$ to $+i$. Therefore the appropriate contour to use is
$\mathcal{C}_{3}$ depicted in Fig. 13. Evaluation of integrals $K_{3}^{c}%
$\textbf{ }and\textbf{ }$K_{3}^{s}$ is entirely parallel to those of integrals
$K_{1}^{c}$\textbf{ }and\textbf{ }$K_{1}^{s}$. The results of their evaluation
are as follows:%
\begin{align}
K_{3}^{c}  &  =-\frac{\sqrt{2}\pi}{16\xi^{5}}\left(  1+2\xi^{2}\right)
K_{0}(\overline{\omega}r)\sin\left(  \frac{x}{\sqrt{2}\xi}\right)  +\frac{\pi
}{2}\int_{0}^{1}dy\frac{e^{-xy}y^{2}\left(  1-y^{2}\right)  }{1+2\xi^{2}y^{2}%
}I_{0}(\overline{\omega}_{3}r),\label{K3c}\\
K_{3}^{s}  &  =\frac{\pi}{8\xi^{4}}\left(  1+2\xi^{2}\right)  K_{0}%
(\overline{\omega}r)\cos\left(  \frac{x}{\sqrt{2}\xi}\right)  -\frac{\pi}%
{2}\int_{0}^{1}dy\frac{e^{-xy}y\left(  1-y^{2}\right)  }{1+2\xi^{2}y^{2}}%
I_{0}(\overline{\omega}_{3}r). \label{K3s}%
\end{align}
Therefore we find%
\begin{align}
-2\xi^{2}K_{3}^{c}-\sqrt{2}\xi K_{3}^{s}  &  =\frac{\pi}{4\sqrt{2}\xi^{3}%
}\left(  1+2\xi^{2}\right)  K_{0}(\overline{\omega}r)\left[  \sin\left(
\frac{x}{\sqrt{2}\xi}\right)  -\cos\left(  \frac{x}{\sqrt{2}\xi}\right)
\right] \nonumber\\
&  \quad-\frac{\pi}{2}\int_{0}^{1}dy\frac{e^{-xy}\sqrt{2}\xi y\left(
1-y^{2}\right)  \left(  \sqrt{2}\xi y-1\right)  }{1+2\xi^{2}y^{2}}%
I_{0}(\overline{\omega}_{3}r). \label{K3cs}%
\end{align}
The integrals in Eqs. (\ref{K3c}) and (\ref{K3s}) are subject to the condition
deduced from the Jordan lemma, namely, $x>r$.

\subsubsection{Summary for the Reduced Axial Velocity}

Collecting the results presented earlier, we obtain the reduced axial velocity%
\begin{align}
\widehat{\mathbf{v}}_{x}\left(  x,r,\xi\right)   &  =-\frac{\pi}{2\sqrt{2}\xi
}\left(  \widehat{v}_{x}\right)  _{me}\left(  x,r,\xi\right) \nonumber\\
&  \quad\,-\frac{\pi}{4}\left[  \mathfrak{C}_{1}\left(  x,r,\xi\right)
-\mathfrak{S}_{1}\left(  x,r,\xi\right)  \right]  -\frac{\pi}{2}\left[
\mathfrak{C}_{2}\left(  x,r,\xi\right)  -\mathfrak{S}_{2}\left(
x,r,\xi\right)  \right]  , \label{rvax}%
\end{align}
where various symbols are defined by%
\begin{align}
\left(  \widehat{v}_{x}\right)  _{\text{me}}  &  =\frac{x}{\left(  x^{2}%
+r^{2}\right)  ^{3/2}}+\sqrt{2}\xi\left[  \!\frac{1}{\left(  x^{2}%
+r^{2}\right)  ^{\frac{1}{2}}}-\frac{r^{2}}{2\left(  x^{2}+r^{2}\right)
^{\frac{3}{2}}}-\frac{2x^{2}-r^{2}}{\left(  x^{2}+r^{2}\right)  ^{\frac{5}{2}%
}}\!\right]  ,\label{phi}\\
\mathfrak{C}_{1}\left(  x,r;\xi\right)   &  =\int_{0}^{\sqrt{2\left(
1+\xi^{2}\right)  }}dy\frac{\left(  1-y^{2}+\sqrt{1+2\xi^{2}y^{2}}\right)
}{1+2\xi^{2}y^{2}}e^{-xy}I_{0}(\overline{\omega}_{1}r),\label{C1}\\
\mathfrak{C}_{2}\left(  x,r;\xi\right)   &  =\int_{0}^{1}dy\frac{2\xi^{2}%
y^{2}\left(  1-y^{2}\right)  }{1+2\xi^{2}y^{2}}e^{-xy}I_{0}(\overline{\omega
}_{3}r),\label{C2}\\
\mathfrak{S}_{1}\left(  x,r;\xi\right)   &  =\int_{0}^{\sqrt{2\left(
1+\xi^{2}\right)  }}dy\frac{\sqrt{2}\xi y\left(  1-y^{2}+\sqrt{1+2\xi^{2}%
y^{2}}\right)  }{\left(  1+2\xi^{2}y^{2}\right)  \left(  1+\sqrt{1+2\xi
^{2}y^{2}}\right)  }e^{-xy}I_{0}(\overline{\omega}_{1}r),\label{S1}\\
\mathfrak{S}_{2}\left(  x,r;\xi\right)   &  =\int_{0}^{1}dy\frac{\sqrt{2}\xi
y\left(  1-y^{2}\right)  }{1+2\xi^{2}y^{2}}e^{-xy}I_{0}(\overline{\omega}%
_{3}r). \label{S2}%
\end{align}
As noted earlier, the terms made up of trigonometric functions in Eqs.
(\ref{51c}), (\ref{K2cs}), and (\ref{K3cs}) indeed cancel each other out. This
velocity formula (\ref{rvax}) is the axial velocity profile of the
countercurrent of the ion and its ion atmosphere in the coordinate system
fixed at the center ion of the ion atmosphere, both of which are pulled by the
external electric field. The first four terms making up $\left(  \widehat
{v}_{x}\right)  _{\text{me}}\left(  x,r,\xi\right)  $ represent the
\textquotedblleft deterministic\textquotedblright\ part of the velocity
$\widehat{\mathbf{v}}_{x}$, and the integrals $\mathfrak{C}_{1}$,
$\mathfrak{C}_{2}$, $\mathfrak{S}_{1}$, and $\mathfrak{S}_{2}$ involving the
Bessel functions $I_{0}(\overline{\omega}_{1}r)$ and $I_{0}(\overline{\omega
}_{3}r)$ stem from the Brownian motion part of the mean local force---namely,
the dressed-up part of the local body force arising from the interaction of
the center ion, its ion atmosphere, and their interaction with the external
electric field, which distorts the ion atmosphere to an asymmetric form. This
velocity formula obtained in Eq. (\ref{rvax}) is in a convenient form to
analyze Wilson's result, further examine the cause of divergence, and find a
way to avoid the divergence difficulty in the evaluation of the
electrophoretic coefficient. This aspect is discussed in the main text.

\subsection{Distribution Functions and Potentials}

The same methods of contour integration can be employed for the distribution
functions $f_{ji}(\mathbf{r})$ representing the nonequilibrium ionic liquid
structure (pair distribution function) and the mean ionic potential $\psi
_{j}\left(  \mathbf{r}\right)  $. They are summarized below:%
\begin{align}
f_{ij}  &  =f_{ji}\left(  \pm\mathbf{r}\right) \nonumber\\
&  =n^{2}+\frac{\sqrt{2}\kappa zn^{2}e^{2}}{\pi Dk_{B}T}\left\{  -\frac{\pi
}{2\sqrt{2}}\int_{0}^{\sqrt{2\left(  1+\xi^{2}\right)  }}dye^{-yx}\left(
1+\frac{1}{\sqrt{1+2\xi^{2}y^{2}}}\right)  I_{0}\left(  \overline{\omega}%
_{1}r\right)  \right. \nonumber\\
&  \qquad\mp\frac{\pi\xi}{2\sqrt{2}}\int_{0}^{\sqrt{2\left(  1+\xi^{2}\right)
}}dy\frac{e^{-xy}y\left(  1+\sqrt{1+2\xi^{2}y^{2}}\right)  }{\left(
1+2\xi^{2}y^{2}\right)  }I_{0}\left(  \overline{\omega}_{1}r\right)
\nonumber\\
&  \qquad\left.  \pm\frac{\pi\xi}{\sqrt{2}}\int_{0}^{1}dy\frac{ye^{-xy}%
}{1+2\xi^{2}y^{2}}I_{0}\left(  \overline{\omega}_{3}r\right)  \right\}
\label{FS1}%
\end{align}
and%
\begin{align}
\psi_{j}\left(  \pm\mathbf{r}\right)   &  =-\psi_{i}\left(  \mp\mathbf{r}%
\right) \nonumber\\
&  =\frac{\kappa ze}{\sqrt{2}\pi D}\left\{  \left[  -\frac{\pi}{2}\int
_{0}^{\sqrt{2\left(  1+\xi^{2}\right)  }}dy\frac{e^{-xy}\left(  1+\sqrt
{1+2\xi^{2}y^{2}}\right)  }{1+2\xi^{2}y^{2}}I_{0}\left(  \overline{\omega}%
_{1}r\right)  \right.  \right. \nonumber\\
&  \qquad\qquad\qquad\left.  +2\pi\xi^{2}\int_{0}^{1}dy\frac{e^{-xy}y^{2}%
}{1+2\xi^{2}y^{2}}I_{0}\left(  \overline{\omega}_{3}r\right)  \right]
\nonumber\\
&  \qquad\qquad\quad\mp\left[  \frac{\pi}{\sqrt{2}}\xi\int_{0}^{\sqrt{2\left(
1+\xi^{2}\right)  }}dy\frac{ye^{-xy}}{1+2\xi^{2}y^{2}}I_{0}\left(
\overline{\omega}_{1}r\right)  \right. \nonumber\\
&  \qquad\qquad\qquad\left.  \left.  -\sqrt{2}\pi\int_{0}^{1}dy\frac{ye^{-xy}%
}{1+2\xi^{2}y^{2}}I_{0}\left(  \overline{\omega}_{3}r\right)  \right]
\right\}  \label{PS1}%
\end{align}
in the region of $(x,r)$ in the upper plane where Jordan's lemma is satisfied.
These results can be easily obtained by using the contour integration method
described earlier in this Appendix. We notice that the nonequilibrium pair
distribution functions and potentials do not contain mechanical contributions,
but only the Brownian motion contributions. The reason is that $f_{ji}%
(\mathbf{r)}$ and $\psi_{j}\left(  \mathbf{r}\right)  $ are solutions of the
OF equations and Poisson equations for the nonequilibrium part described by
the Brownian motion model.

\newpage Figure Captions\bigskip

Fig. 1\quad The cylindrical coordinate system employed. The $x$ axis is
parallel to the external electric field.

Fig. 2\quad Nonequilibrium part of the distribution function $\Delta
\overline{f}_{ij}\left(  +\mathbf{r}\right)  $ is plotted in ($x,r$) plane at
$\xi=1$. Here $\Delta\overline{f}_{ij}\left(  +\mathbf{r}\right)  =\left[
f_{ij}\left(  +\mathbf{r}\right)  -n^{2}\right]  /\left(  \sqrt{2}\kappa
ze^{2}/\pi Dk_{B}T\right)  $. $\Delta\overline{f}_{ij}\left(  +\mathbf{r}%
\right)  $ is computed with the contour integration methods within the range
defined by Ineq. (\ref{vc}) and, outside this range, by means of the method of
principal values for singular integrals.

Fig. 3\quad Nonequilibrium part of the potential $\Delta\overline{\psi}%
_{j}\left(  +\mathbf{r}\right)  $ is plotted in ($x,r$) plane at $\xi=1$. Here
$\Delta\overline{\psi}_{j}\left(  +\mathbf{r}\right)  =\psi_{j}\left(
+\mathbf{r}\right)  /\left(  \kappa ze/\sqrt{2}\pi D\right)  $ with $\psi
_{j}^{0}$ denoting the Debye--H\"{u}ckel potential. In Eq. (\ref{34p})
$\psi_{j}^{0}$ is not explicitly put in since $\psi_{j}\left(  \pm
\mathbf{r}\right)  $ is the nonequilibrium part of the potential in the
external field. Therefore $\Delta\overline{\psi}_{j}\left(  +\mathbf{r}%
\right)  $ should be understood as $\Delta\overline{\psi}_{j}\left(
+\mathbf{r}\right)  =\left[  \psi_{j}\left(  +\mathbf{r}\right)  -\psi_{j}%
^{0}\right]  /\left(  \kappa ze/\sqrt{2}\pi D\right)  $. Within the range of
$x$ and $r$ satisfying Ineq. (\ref{33}) [also see Ineq. (\ref{A49c})] the
contour integration method is used and outside the region the method of
principal value integration is used for computation.

Fig. 4\quad The reduced axial velocity profile $\widehat{\mathbf{v}}%
_{x}\left(  x,r,\xi\right)  $ is plotted in $(x,r)$ plane at $\xi=1$. Within
the range of $x$ and $r$ satisfying Ineq. (\ref{vc}) the contour integration
method is used and outside the region the method of principal\ value
integration is used for computation. The axial velocity profile is
directional, being positive the positive $x$ direction parallel to the
external field before vanishing to zero at large distance whereas being
negative in the transversal (radial) direction before vanishing to zero as $r$
increases. Thus the boundary conditions are satisfied in both $x$ and $r$
directions. This figure indicates the mode of behaviors of the counterflow of
the medium to the ionic movement when the external field is turned on.

Fig. 5\quad The electrophoretic factor $\mathfrak{f}\left(  x,r,\xi\right)  $
is plotted in 3D in a similar color coding to Fig. 4 in the case of $\xi=1$.

Fig. 6\quad The projection of surface $\mathfrak{f}\left(  x,r,\xi\right)  $
onto $(x,r)$ plane. There are two sets of quasi-elliptical level curves; one
with the major axis on the $x$ axis and the other on the $r$ axis. The former
corresponds to the contours of the negative part of the $\mathfrak{f}\left(
x,r,\xi\right)  $ surface projected onto ($x,r$) plane, and the latter to the
contours of the positive part projected onto ($x,r$) plane. The outermost
level curve $C_{p}$ is the locus of $\mathfrak{f}\left(  x,r,\xi\right)  =0$.
This level curve $C_{P}$ depicts the moving ion atmosphere distorted by the
external electric field from the spherical form assumed by the ion atmosphere
at $\xi=0$. This moving ion atmosphere is seen polarized toward the field direction.

Fig. 7\quad The distorted ion atmosphere is seen to have the center at
($x_{c},0$) on the $x$ axis. The field dependence of the center of the ion
atmosphere $\left(  x_{c},0\right)  $ describes the trajectory of its motion.
The trajectory is shown in this figure. The curve indicates the mode of
migration for the center from the origin of the coordinate system where the
center is located when $\xi=0$, as the field strength is increased. It
decreases to a plateau after reaching a maximum as $\xi$ increases.

Fig. 8\quad Plot of an example for $\mathfrak{f}\left(  x,r,\xi\right)  $ at
$x=r=0.5$ as a function of $\xi$ and its comparison with Wilson's
electrophoretic coefficient $f(\xi)$. The solid line, the present theory; the
dotted line, the OW theory.

Fig. 9\quad A $3-D$ relaxation time coefficient $\mathfrak{g}\left(
x,r;\xi\right)  $. A combination of the contour integration results and the
method of principal integration is used to construct the surface.

Fig. 10\quad Plot of and example for $\mathfrak{g}\left(  x,r,\xi\right)  $ at
$x=r=0.5$ as a function of $\xi$ and its comparison with Wilson's
electrophoretic coefficient $g(\xi)$. The solid line, the present theory; the
dotted line, the OW theory.

Fig. 11\quad Contour $\mathcal{C}_{1}$ for integrals $K_{1}^{c}$ and
$K_{1}^{s}$. This contour also applies to integrals $J_{1}^{c}$ and $J_{1}%
^{s}$ and $P_{1}^{c}$ and $P_{1}^{s}$. The bold line denotes the branch cut.

Fig. 12\quad Contour $\mathcal{C}_{3}$ for integrals $K_{3}^{c}$ and
$K_{3}^{s}$. This contour also applies to integrals $J_{3}^{c}$ and $J_{3}%
^{s}$ and $P_{3}^{c}$ and $P_{3}^{s}$. The bold line denotes the branch cut.

Fig. 13\quad Contour $\mathcal{C}_{2}$ for integrals $K_{2}^{c}$ and
$K_{2}^{s}$. This contour also applies to integrals $J_{2}^{c}$ and $J_{2}%
^{s}$ and $P_{2}^{c}$ and $P_{2}^{s}$. The bold line denotes the branch cut on
the negative real axis.\newpage%

\begin{figure}
[ptb]
\begin{center}
\includegraphics[
natheight=5.391800in,
natwidth=7.311400in,
height=2.7607in,
width=3.7341in
]%
{../New Submission/Final Figures/Fig 1.png}%
\caption{ }%
\label{Fig1}%
\end{center}
\end{figure}
%

\begin{figure}
[ptb]
\begin{center}
\includegraphics[
natheight=4.571700in,
natwidth=6.886200in,
height=2.4879in,
width=3.7332in
]%
{../New Submission/Final Figures/Fig 2.png}%
\caption{ }%
\label{Fig2}%
\end{center}
\end{figure}
%

\begin{figure}
[ptb]
\begin{center}
\includegraphics[
natheight=4.058900in,
natwidth=6.886200in,
height=2.2115in,
width=3.7332in
]%
{../New Submission/Final Figures/Fig 3.png}%
\caption{ }%
\label{Fig3}%
\end{center}
\end{figure}
%

\begin{figure}
[ptb]
\begin{center}
\includegraphics[
natheight=4.410200in,
natwidth=6.886200in,
height=2.4012in,
width=3.7332in
]%
{../New Submission/Final Figures/Fig 4.png}%
\caption{ }%
\label{Fig4}%
\end{center}
\end{figure}
%

\begin{figure}
[ptb]
\begin{center}
\includegraphics[
natheight=5.318800in,
natwidth=8.718200in,
height=2.289in,
width=3.7341in
]%
{../New Submission/Final Figures/Fig 5.png}%
\caption{ }%
\label{Fig5}%
\end{center}
\end{figure}
%

\begin{figure}
[ptb]
\begin{center}
\includegraphics[
natheight=6.637200in,
natwidth=8.409800in,
height=2.9523in,
width=3.7341in
]%
{../New Submission/Final Figures/Fig 6.png}%
\caption{ }%
\label{Fig6}%
\end{center}
\end{figure}
%

\begin{figure}
[ptb]
\begin{center}
\includegraphics[
natheight=4.600900in,
natwidth=6.784000in,
height=2.5399in,
width=3.7341in
]%
{../New Submission/Final Figures/Fig 7.png}%
\caption{ }%
\label{Fig7}%
\end{center}
\end{figure}
%

\begin{figure}
[ptb]
\begin{center}
\includegraphics[
natheight=4.439400in,
natwidth=5.611700in,
height=2.9596in,
width=3.7332in
]%
{../New Submission/Final Figures/Fig 8.png}%
\caption{ }%
\label{Fig8}%
\end{center}
\end{figure}
%

\begin{figure}
[ptb]
\begin{center}
\includegraphics[
natheight=4.937500in,
natwidth=6.886200in,
height=2.685in,
width=3.7341in
]%
{../New Submission/Final Figures/Fig 9.png}%
\caption{ }%
\label{Fig9}%
\end{center}
\end{figure}
%

\begin{figure}
[ptb]
\begin{center}
\includegraphics[
natheight=4.439400in,
natwidth=5.611700in,
height=2.9596in,
width=3.7332in
]%
{../New Submission/Final Figures/Fig 10.png}%
\caption{ }%
\label{Fig10}%
\end{center}
\end{figure}
%

\begin{figure}
[ptb]
\begin{center}
\includegraphics[
natheight=5.421000in,
natwidth=7.399000in,
height=2.7424in,
width=3.7332in
]%
{../New Submission/Final Figures/Fig 11.png}%
\caption{ }%
\label{Fig11}%
\end{center}
\end{figure}
%

\begin{figure}
[ptb]
\begin{center}
\includegraphics[
natheight=5.406400in,
natwidth=7.442700in,
height=2.7187in,
width=3.7332in
]%
{../New Submission/Final Figures/Fig 12.png}%
\caption{ }%
\label{Fig12}%
\end{center}
\end{figure}
%

\begin{figure}
[ptb]
\begin{center}
\includegraphics[
natheight=5.406400in,
natwidth=7.238400in,
height=2.7954in,
width=3.7332in
]%
{../New Submission/Final Figures/Fig 13.png}%
\caption{ }%
\label{Fig13}%
\end{center}
\end{figure}
\newpage


\begin{thebibliography}{99}                                                                                               %


\bibitem {chapman}S. Chapman and T. G. Cowling, \textit{The Mathematical
Theory of Nonuniform Gases} (Cambridge U.P., London, 1970), third edition.

\bibitem {mason}E. A. Mason and E. W. McDaniel, \textit{Transport Properties
of Ions in Gases} (Wiley, New York, 1988).

\bibitem {ting}C. S. Ting, ed., \textit{Physics of Hot Electron Transport in
Semiconductors (World Scientific, Singapore, 1992).}

\bibitem {nag}B. R. Nag, \textit{Electron Transport in Compound
Semiconductors} (Springer, Berlin, 1980).

\bibitem {landsberg}P. T. Landsberg, \textit{Basic Properties of
Semiconductors,} Vol. 1 (North-Holland, Amsterdam, 1992).

\bibitem {wien}M. Wien, Ann. Physik. \textbf{85}, 795 (1928); Phys. Z.
\textbf{29}, 751(1928); Ann. Physik. [5] \textbf{1}, 400 (1929); Phys. Z.
\textbf{32}, 545 (1931); J. Malsch and M. Wien, Ann. Physik. [4] \textbf{83},
305 (1927).

\bibitem {harned}H. S. Harned and B. B. Owen, \textit{The Physical Chemistry
of Electrolytic Solutions} (Reinhold, New York, 1958).

\bibitem {accascina}R. M. Fuoss and F. Accascina, \textit{Electrolytic
Conductance} (Interscience, New York, 1959).

\bibitem {daily}J. W. Daily and M. M. Micci, J. Chem. Phys. \textbf{131},
094501 (2009).

\bibitem {shabanov}O. M. Shabanov, R. T. Kachaev, S. A. Dzhamalova, and A. A.
Iskakova, Russian J. Electrochem. \textbf{46}, 1390 (2010); O. M. Shabanov, S.
M. Gadzhiv, A. A. Iskakova, R. T. Kachaev, A. O. Magomedova, and S. I.
Suleimanov, ibid. \textbf{47}, 221 (2011).

\bibitem {friedman}Y. Wang, C. Li, W. Wang, J. Jiang, D. Zhou, R. Xu, and S.
P. Friedman, Soil Sci. Soc. Am. J. \textbf{73}, 569 (2008).

\bibitem {castner}E. W. Castner and J. F. Wishart, J. Chem. Phys.
\textbf{132}, 120901 (2010).

\bibitem {ionic}R. D. Rogers and K. R. Seddon, eds., \textit{Ionic Liquids
IIIA, Fundamentals, Progress, Challenges, and Opportunities. Properties and
Structure}, ACS Symposium Series Vol. 901; R. D. Rogers and K. R. Seddon,
eds., \textit{Ionic Liquids IIIB, Fundamentals, Progress, Challenges and
Opportunities. Transformations and Progress}. ACS Symposium Series Vol. 902
(American Chemical Society, Washington DC, 2005).

\bibitem {asymmetric}B. C. Eu, \textquotedblleft Brownian movement theory of
nonequilibrium statistical mechanics, transport, and hydrodynamics of strong
asymmetric electrolyte solutions in an electric field\textquotedblright\ (to
be sumbitted).

\bibitem {asymmetric-2}B. C. Eu and H. Xu, \textquotedblleft Wien effect on
ionic conductance of asymmetric strong electrolytes in an electric field. (to
be sumbitted).

\bibitem {onsager}L. Onsager, Phys. Z. \textbf{27}, 388 (1926); \textbf{28},
277 (1927).

\bibitem {debye}P. Debye and E. H\"{u}ckel, Physik. Z. \textbf{24}, 305 (1923).

\bibitem {stokes}G. G. Stokes, \textit{Mathematical and Physical Paper}s
(Cambridge U.P., London, 1880), Vol. 1, pp 36-43.

\bibitem {landau}L. D. Landau and E. M. Lifshitz, \textit{Fluid Mechanics}
(Pergamon. Oxford, 1958).

\bibitem {batchelor}G. K. Batchelor, \textit{Fluid Dynamics} (Cambridge U.P.,
London, 1967).

\bibitem {bird}R. B. Bird, W. E. Stewart, E. N. Lightfoot, \textit{Transport
Phenomena} (Wiley, New York, 1960), p. 132.

\bibitem {fuoss}L. Onsager and R. M. Fuoss, J. Phys. Chem. \textbf{36}, 2698 (1932).

\bibitem {smoluchowski}M. von Smoluchowski, Phys. Z. \textbf{17}, 557, 585 (1916).

\bibitem {jackson}J. D. Jackson, \textit{Classical Electrodynamics} (Wiley,
New York, 1975), second ed.

\bibitem {wilson}W. S. Wilson, \textit{The Theory of the Wien Effect for a
Binary Electrolyte}, PhD Thesis, Yale University, June, 1936.

\bibitem {kirkwood}J. G. Kirkwood, J. Chem. Phys. (1946).

\bibitem {eu87}B. C. Eu, J. Chem. Phys. \textbf{87}, 1238 (1985).

\bibitem {eu92}B. C. Eu, \textit{Kinetic Theory and Irreversible
Thermodynamics} (Wiley, New York, 1992).

\bibitem {eu98}B. C. Eu, \textit{Nonequilibrium Statistical Mechanics}
(Kluwer, Dordrecht, 1998).

\bibitem {whittaker}E. Whittaker and G. N. Watson, \textit{Modern Analysis}
(Cambridge U. P., London, 1952).

\bibitem {eurah2}B. C. Eu, H. Xu, and K. Rah, the following paper entitled
\textquotedblleft Wien effect on ionic conductance of binary strong
electrolyte solutions in a high external electric field\textquotedblright.

\bibitem {watson}G. N. Watson, \textit{Theory of Bessel Functions} (Cambridge
U. P., London, 1966).

\bibitem {abramowitz}M. Abramowitz and I. Stegun, \textit{Handbook of
Mathematical Functions} (National Bureau of Standards, Washington, DC, 1966).

\bibitem {singular}N. I. Muskhelishvili, \textit{Singular Integral Equations}
(P. Noordhoff, Groningen, 1953).

\bibitem {mazur}S. R. de Groot and P. Mazur, \textit{Nonequilibrium
Thermodynamics }(North-Holland, Amsterdam, 1962).

\bibitem {haase}R. Haase, \textit{Thermodynamics of Irreversible Processes}
(Dover, New York, 1969), Chapter 4 and, in particular, Sec. 4-16.

\bibitem {pressure}B. C. Eu, (to be submitted).

\bibitem {plasma}S. Ichimaru, \textit{Basic Principles of Plasma Physics}
(Benjamin/Cummings, reading, MA, 1973); E. A. Mason and E. W. McDaniel,
\textit{Transport Properties of Ions in Gases} (Wiley, New York, 1988).

\bibitem {benabdallah}N. Benabdallah, A. Arnold, P. Degond, I. M. Gamba, R. T.
Glassey, C. D. Levermore, C. Ringhofer, edts., \textit{Transport in Transition
Regimes} (Springer, Heidelberg, 2004).

\bibitem {micro-nano}D. K. Ferry and S. M. Goodnick, \textit{Transport in
Nanostructure} (Cambridge U. P., London, 1997); G. Em Karniadakis and A.
Beskok, \textit{Micro Flows} (Springer, Heidelberg, 2002);

\bibitem {tanaka}Y. Tanaka, ed., \textit{Ion Exchange Membranes: Fundamentals
and Applications} (Elsevier, Amsterdam, 2007)\textit{.}
\end{thebibliography}
\end{document}